\documentclass[a4paper,11pt]{article}
\usepackage{jheppub}
\usepackage{besphysics}
\usepackage{color}
\usepackage{amsmath}
\usepackage{epsfig}
\usepackage{multirow}
\usepackage{overpic}
\usepackage{colortbl}
\usepackage{enumitem}
\newsavebox{\tablebox}
\usepackage{pifont}
\usepackage{enumitem}
\usepackage{tikz}
\usepackage{etoolbox}
\usepackage{comment}
\usepackage[T1]{fontenc}
\usepackage{hyperref}
\usepackage{breakurl}
\usepackage{lineno}
\usepackage[figuresright]{rotating}

\def\chifc{\chi^2_{\rm 4C}}

\title{\boldmath Study of $\eta(1405)/\eta(1475)$ in $\jp\to\gamma\ksz\ksz\piz$ decay}

\collaboration{The BESIII Collaboration}

\collaborationImg{\includegraphics[height=30mm,angle=90]{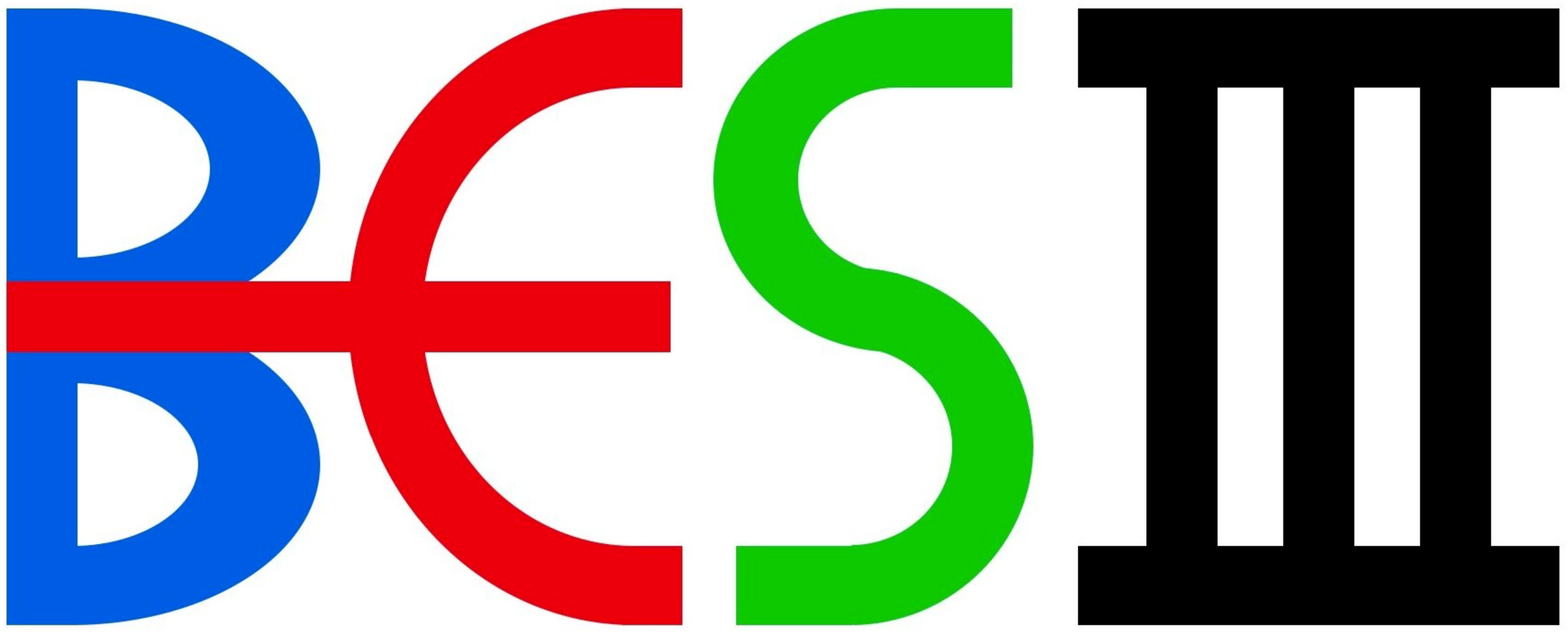}} 

\abstract{
Using a sample of $(10.09\pm0.04)\times10^{9}$ $\jpsi$ decays collected with the BESIII detector, partial wave analyses of the decay $\jpsi\to\gamma\ksz\ksz\piz$ are performed within the $\ksz\ksz\piz$ invariant mass region below 1.6$\gevcc$.   The covariant tensor amplitude method is used in both mass independent and mass dependent approaches. Both analysis approaches exhibit dominant pseudoscalar and axial vector components, and show good consistency for the other individual components. Furthermore, the mass dependent analysis reveals that the $\ksz\ksz\piz$ invariant mass spectrum for the pseudoscalar component can be well described with two isoscalar resonant states using relativistic Breit-Wigner model, {\it i.e.}, 
the $\eta(1405)$ with a mass of $1391.7\pm0.7_{-0.3}^{+11.3}\mevcc$ and a width of $60.8\pm1.2_{-12.0}^{+5.5}\mev$, 
and the $\eta(1475)$ with a mass of $1507.6\pm1.6_{-32.2}^{+15.5}\mevcc$ and a width of $115.8\pm2.4_{-10.9}^{+14.8}\mev$. The first and second uncertainties are statistical and systematic, respectively. 
Alternate models for the pseudoscalar component are also tested, but the description of the $\ksz\ksz\piz$ invariant mass spectrum deteriorates significantly.}


\begin{document}
\setcounter{tocdepth}{10}
\maketitle
\let\clearpage\relax
\bibliographystyle{JHEP}
\flushbottom
%\linenumbers

\section{INTRODUCTION}
The non-abelian nature of quantum chromodynamics (QCD) predicts the existence of exotic states, such as glueballs, hybrids, and multiquarks states. The experimental confirmation of these states would provide fundamental information about QCD in the confinement regime and would be a direct test of the QCD theory. The radiative decays of charmonium are  glue-rich processes and are considered to be excellent probes for the production of gluonic matter and light hadron structures. Much progress has been made in the past few decades, but many issues remain unresolved.

In the field of light hadron spectroscopy, one of the most interesting and disputed questions is the nature of the pseudoscalar structure with a mass around 1.4~$\gevcc$, the so-called "$\iota$" state. The "$\iota$" state was first observed in the $\jp$ radiative decay to the $K\bar{K}\pi$ final state in the early 1980s by the Crystal Barrel~\cite{1} and Mark II~\cite{2} Collaborations. 
The structure has been subsequently confirmed by different experiments~\cite{3,4,5,6,7,8,9,10,11}, and is often interpreted as the combination of two isoscalar resonant states, {\it i.e.}, the $\eta(1405)$ and $\eta(1475)$~\cite{13}. Pseudoscalar structures in the vicinity of 1.4~$\gevcc$ are also observed in the $\eta\pi\pi$, $\pi\pi\pi$ and $\gamma V$ (where $V$ refers to vector meson) final states of $\jp$ radiative decays, but with quite different lineshapes~\cite{ppe1,ppe2,ppe3,3pi,gv1,gv2}. Consequently, whether or not these observed pseudoscalar structures are from the same origin is controversial.

From the theoretical point of view, the pseudoscalar nonet of ground states is well established, and the $\eta(1295)$ and $\eta(1475)$ are generally assigned to be the first radial excitations of the ground states $\eta$ and $\etap$, taking into account their production and decay properties~\cite{13}. It is very difficult to accomodate another pseudoscalar state with a mass around 1.4$\gevcc$ in the quark model. Therefore, the possibility of an additional state with a non-$q\bar{q}$ nature~\cite{14,15,16,17,18,19,20} has been widely discussed. Meanwhile, there is still a long-standing controversy about whether or not the $\eta(1405)$ and $\eta(1475)$ are two separate states or just one pseudoscalar state, namely the $\eta(1440)$, in different decay modes. In refs.~\cite{tsm1,tsm2,tsm3,tsm4}, 
a coherent phenomenological analysis of $\eta(1405/1475)\to K\bar{K}\pi$, $\eta\pi\pi$ and $\pi\pi\pi$ is performed by considering a triangle singularity mechanism (TSM), and it is concluded that only one pseudoscalar state is needed in this mass vicinity, and the distorted lineshape and shifted peak position in the different decay modes could be due to dynamic features within the TSM.

To solve the $\eta(1405/1475)$ puzzle and to pin down their nature, a precise and comprehensive measurement is prerequisite. The BESIII experiment has collected $10.09\times10^{9}$ $\jp$ decays~\cite{njp}, which provides an excellent opportunity to explore the nature of the $\eta(1405/1475)$. Meanwhile, comparing to other decay modes, $\jp\to\gamma\ksz\ksz\piz$ is free of background from similar final states with one fewer or more photon, and can provide a much cleaner sample than the corresponding charged channels. This article presents a partial wave analysis (PWA) of the decay $\jp\to\gamma\ksz\ksz\piz$ based on the full $\jpsi$ sample collected by the BESIII collaboration, where both mass dependent and independent approaches are performed. Both $\ksz$ mesons are reconstructed from $\pip\pim$ and the $\piz$ meson is reconstructed from $\gamma\gamma$.

\section{BESIII DETECTOR AND MONTE CARLO SIMULATION}
The BESIII detector~\cite{23} records the final state particles produced in symmetric $e^{+}e^{-}$ collisions provided by the BEPCII storage ring~\cite{24} in the center-of-mass energy range from 2.0 to 4.95 GeV~\cite{25}, with a peak luminosity of $1\times10^{33}$cm$^{-2}$s$^{-1}$ achieved at $\sqrt{s} = 3.77\gev$. The cylindrical core of the BESIII detector covers 93\% of the full solid angle. From inner detectors to outer, it consists of a helium-based multilayer drift chamber (MDC), a plastic scintillator time-of-flight system (TOF), and a CsI(Tl) electromagnetic calorimeter (EMC), which are all enclosed in a superconducting solenoidal magnet providing a 1.0 T (0.9 T in 2012) magnetic field. The solenoid is supported by an octagonal flux-return yoke with resistive plate counter muon identification modules interleaved with steel. The charged-particle momentum resolution at 1$\gevc$ is 0.5\%, and the ${\rm d}E/{\rm d}x$ resolution is 6\% for electrons from Bhabha scattering. The EMC measures photon energies with a resolution of 2.5\% (5\%) at 1$\gev$ in the barrel (end cap) region. The time resolution in the TOF barrel region is 68 ps, while that in the end cap region was initially 110~ps. The end cap TOF system was upgraded in 2015 with multi-gap resistive plate chamber technology, improving the time resolution to be 60~ps~\cite{26,27}. 

Simulated data samples produced with a $\textsc{geant4}$-based Monte Carlo (MC) package~\cite{28}, which includes the geometric description of the BESIII detector~\cite{29,30} and the detector response, are used to optimize the event selection criteria, determine detection efficiencies and estimate backgrounds. The simulation models the beam energy spread and initial state radiation in the $e^{+}e^{-}$ annihilations with the generator $\textsc{kkmc}$~\cite{31,32}. The signal MC sample for $\jp\to\gamma\ksz\ksz\piz$, with subsequent decays $\ksz\to\pip\pim$ and $\piz\to\gamma\gamma$, is generated uniformly in phase space (PHSP). 

An inclusive MC sample with $10.01\times10^{9}$ $\jp$ decays is used to study background, where the production of the $\jp$ resonance and the continuum processes is incorporated in $\textsc{kkmc}$~\cite{31,32}. All particle decays are modeled with $\textsc{evtgen}$~\cite{33,34} using branching fractions either taken from the Particle Data Group (PDG)~\cite{13}, when available, or otherwise estimated with $\textsc{lundcharm}$~\cite{35,36}. Final state radiation from charged final state particles is incorporated using $\textsc{photos}$~\cite{37}.

\section{EVENT SELECTION}
Charged tracks are reconstructed using hits in the MDC. The point of closest approach of each charged track to the interaction point (IP) is required to be less than 20 cm along the $z$ axis, which is the symmetry axis of the MDC, because of the relatively long lifetime of $\ksz$. Charged tracks detected in the MDC are required to be within a polar angle ($\theta$) range of $\left|\costht\right|<0.93$, where $\theta$ is defined with respect to the $z$ axis. All charged tracks are reconstructed under the pion hypothesis.To reconstruct $\ksz$ candidates, a vertex fit and a subsequent secondary vertex fit are performed on each pair of charged tracks with opposite charges. When two vertex fits are performed successfully, the corresponding combination is regarded as an $\ksz$ candidate. Photon candidates are identified using showers in the EMC. The deposited energy of each shower must be more than 25$\mev$ in the EMC barrel region ($\left|\costht\right|<0.80$) and more than 50$\mev$ in the end cap region ($0.86<\left|\costht\right|<0.92$). To exclude those showers originating from charged tracks, the angle between the position of each shower in the EMC and the closest extrapolated charged track must be greater than $10^{\circ}$ as measured from the IP. To suppress electronic noise and energy deposits unrelated to the event, the difference between the EMC time and the event start time is required to be within $[0,700]$ ns.

\begin{figure}[htbp]
	\centering
	\begin{overpic}[width=0.49\textwidth]{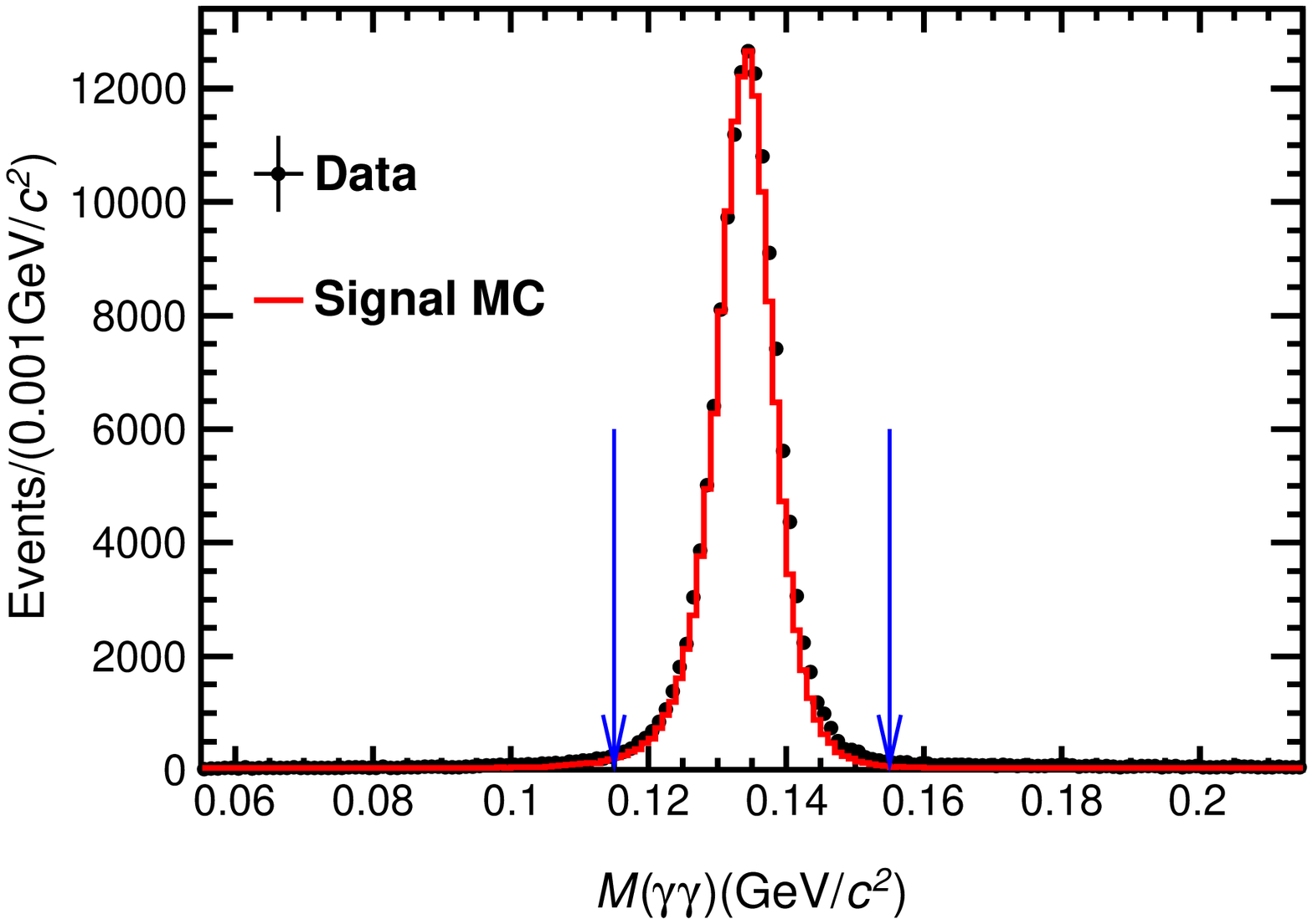}
		\put(20,60){$\bf (a)$}
	\end{overpic}
	\begin{overpic}[width=0.49\textwidth]{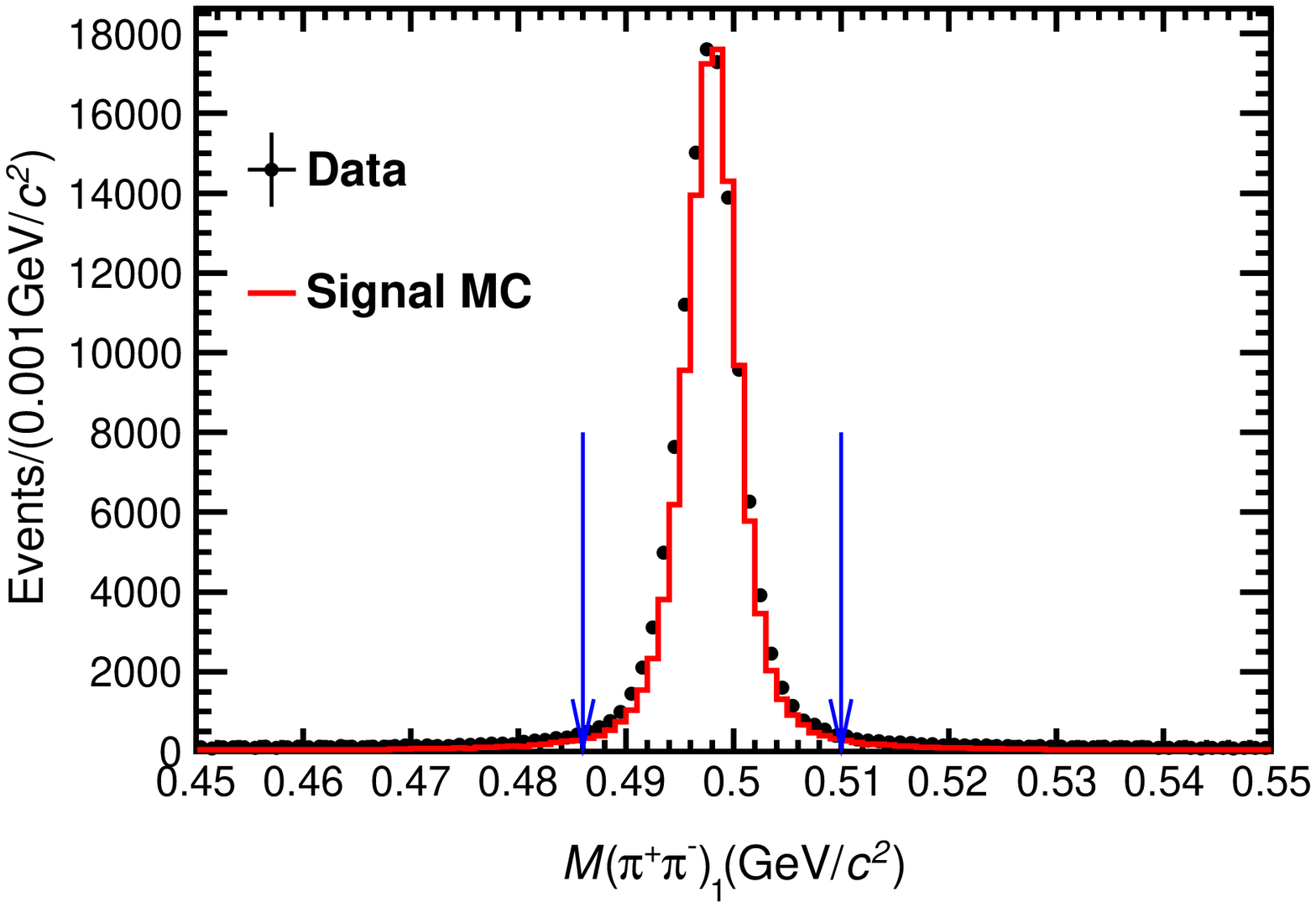}
		\put(20,60){$\bf (b)$}
	\end{overpic}
	\begin{overpic}[width=0.49\textwidth]{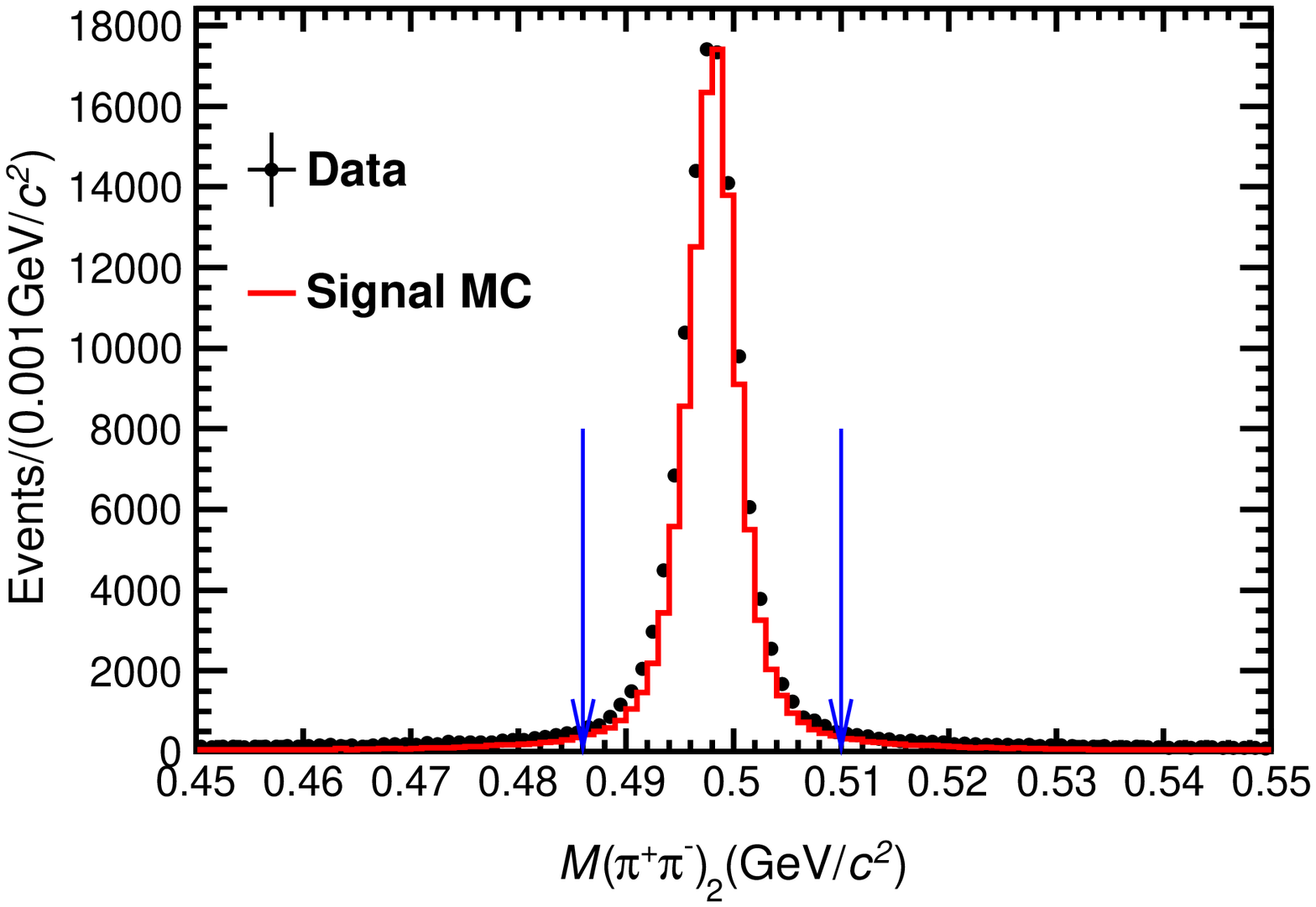}
		\put(20,60){$\bf (c)$}
	\end{overpic}
	\caption{Invariant mass distributions of (a) $\gamma\gamma$, (b) one $\pip\pim$ pair, (c) another $\pip\pim$ pair for candidate events with the 4C kinematic fit applied. Dots with error bars are data and the red solid line represents the signal MC sample (normalized by height).}	
	\label{f0}
\end{figure}

In $\jp\to\gamma\ksz\ksz\piz$ process, candidates are required to have four charged tracks with zero net charge and at least three photon candidates. The $\ksz$ candidates are required to satisfy $L/\sigma_{L}>2$, where $L$ and $\sigma_{L}$ are the distance of the common vertex of the $\pip\pim$ pair away from the IP, and the corresponding uncertainty, respectively. The $\gamma\gamma\gamma\ksz\ksz$ candidates are further subjected to a four-constraint (4C) kinematic fit, which ensures energy and momentum conservation. Only candidates which yield a $\chifc$ of the 4C kinematic fit of less than 40 are retained for further analysis. For events with more than three photon candidates, multiple signal candidates are possible, and only the one with the smallest $\chifc$ is retained for further analysis. Furthermore, the $\jp\to\gamma\ksz\ksz\piz$ candidates are required to have exactly one pair of photons with  invariant mass within the $\piz$ mass region ($\left|M(\gamma\gamma)-M(\piz)\right|<20\mevcc$)~\cite{13}, as shown in figure~\ref{f0}(a). MC simulation studies indicate this criterion significantly reduces the rate of miscombined photon pairs to be less than 0.3\%. Both $\ksz$ candidates are required to satisfy $\left|M(\pip\pim)-M(\ksz)\right|<12\mevcc$, where $M(\ksz)$ is the known $K^{0}$ mass~\cite{13}, as shown in figures~\ref{f0}(b)-(c). The miscombination of pion pairs to form a $\ksz$ signal is also studied and found to be negligible. To suppress backgrounds containing an $\eta$ signal, events with any photon pair of invariant mass within $\pm30$~\mevcc of the known $\eta$ mass ~\cite{13} are rejected, while the efficiency for real signal is about 90.5\%. The decay $\jp\to\omega\ksz\ksz$ with $\omega\to\gamma\piz$ is of the same final state as the signal decay $\jp\to\gamma\ksz\ksz\piz$. Therefore, events with $\gamma\piz$ invariant mass within $\pm40$~\mevcc of
the known $\omega$ mass~\cite{13} are rejected, while the efficiency for real signal is about 95.5\%. After all the above selection criteria, potential backgrounds are studied by subjecting the inclusive MC sample of $10.01\times10^{9}$ $\jp$ decays to the same selection criteria applied to data. No significant peaking background is identified in the invariant mass spectrum of $\ksz\ksz\piz$. The dominant background stemming from non-$\ksz\ksz$ processes is estimated to be 0.5\% in total. Non-$\ksz\ksz$ backgrounds are estimated using events from $\ksz\ksz$ two-dimensional sidebands from data, where the sideband regions are defined as $\left|M(\pip\pim)_{1,2}-M(\ksz)\pm35\mevcc\right|<12\mevcc$, and the subscripts 1 and 2 stand for the two sets of $\pip\pim$
corresponding to the two $\ksz$. The total contribution is estimated to be at a level of 0.2\%, which is neglected in the following analysis.

\begin{figure}[htbp]
	\flushleft
	\begin{overpic}[width=0.49\textwidth]{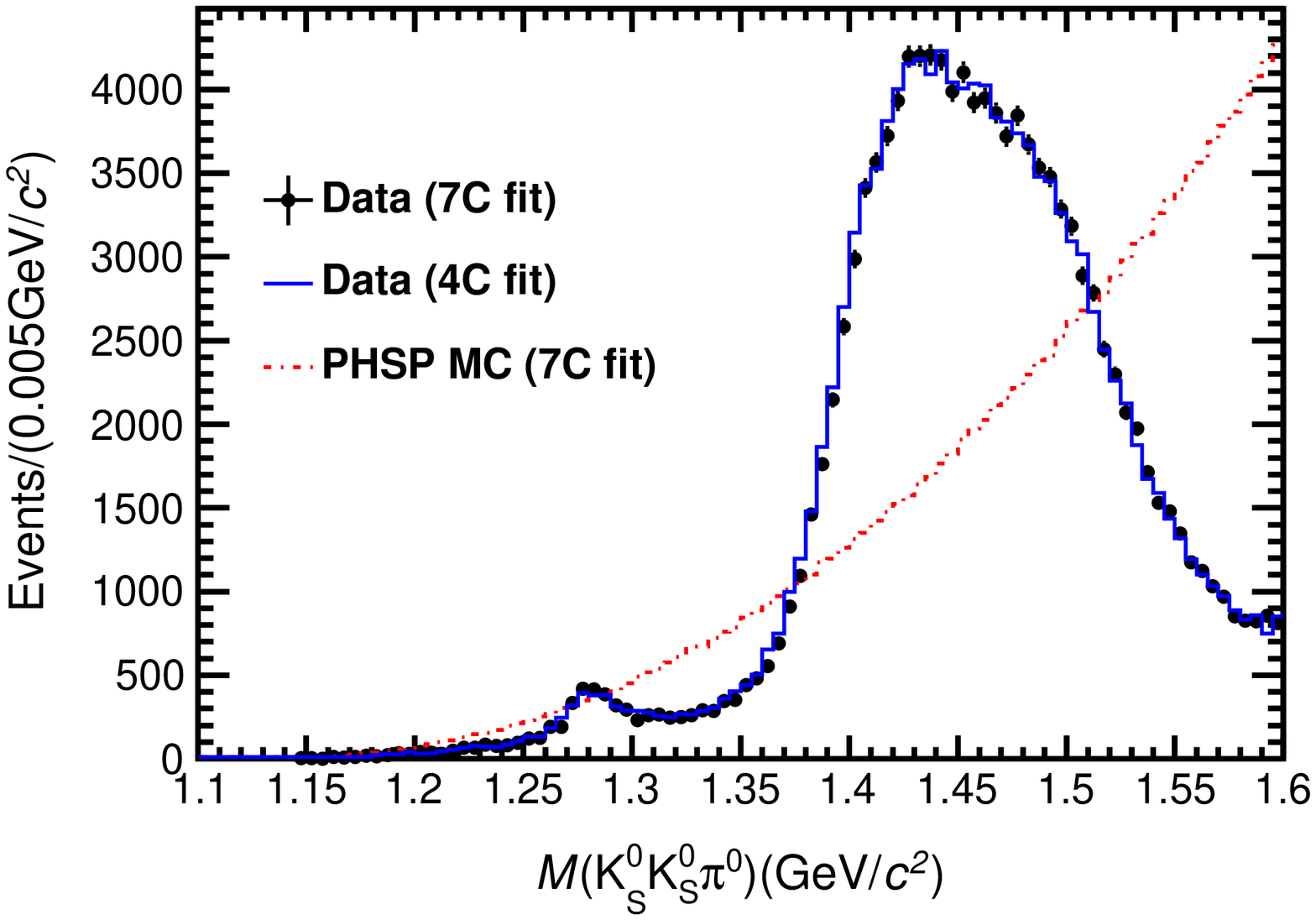}
		\put(20,60){\Large {$\bf (a)$}}
	\end{overpic}
	\begin{overpic}[width=0.49\textwidth]{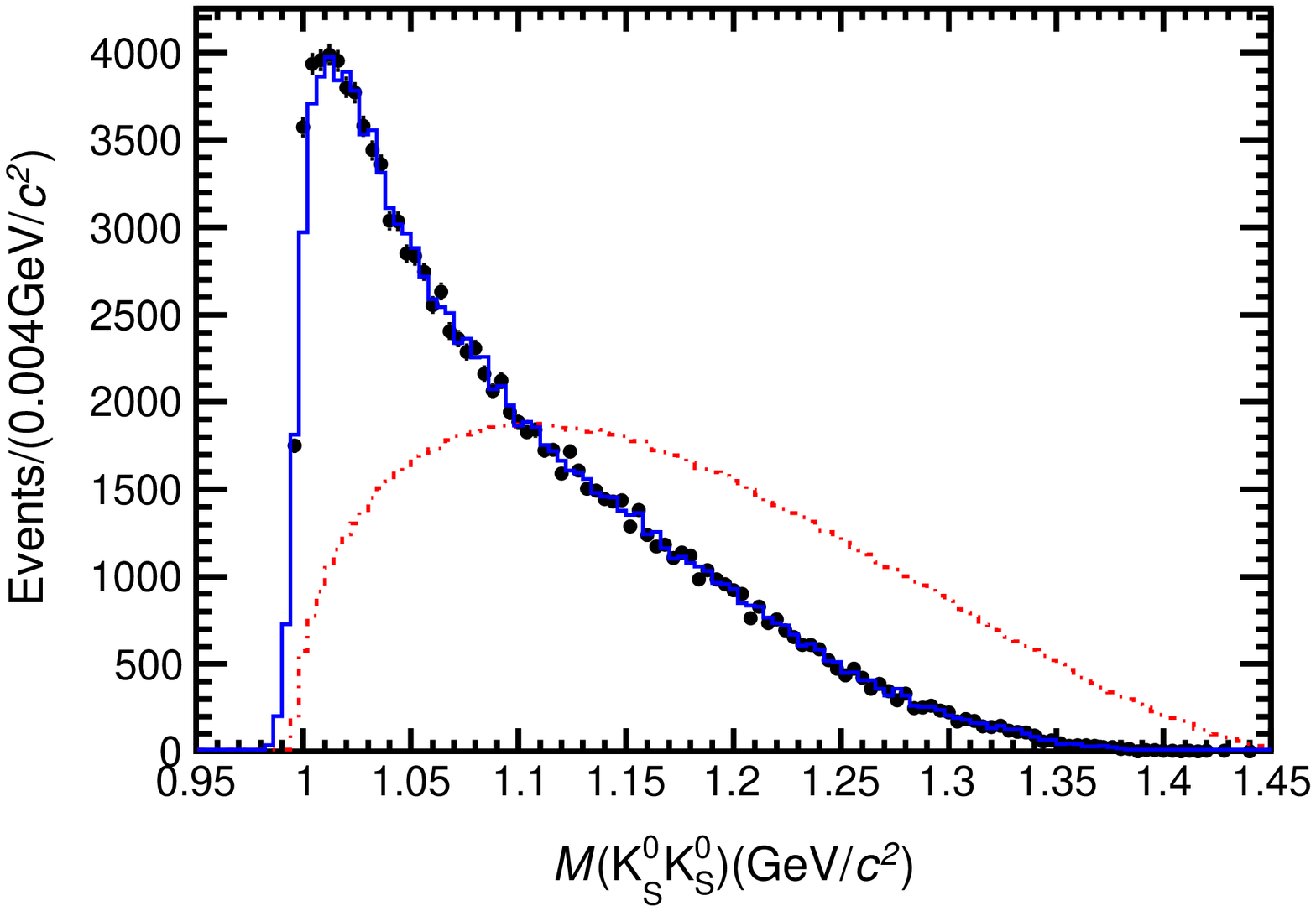}
		\put(18,60){$\bf (b)$}
	\end{overpic}
	\begin{overpic}[width=0.49\textwidth]{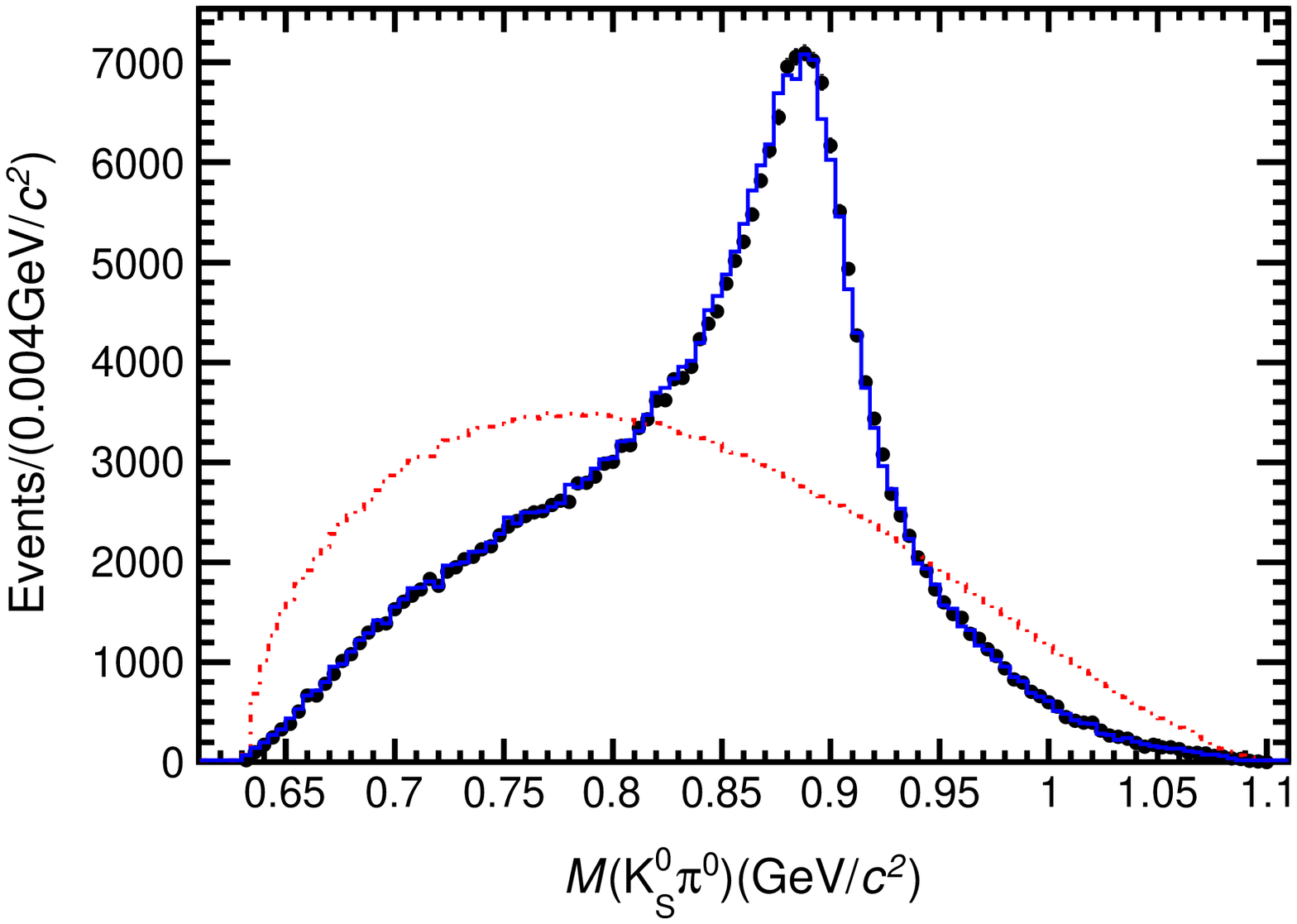}
		\put(20,60){$\bf (c)$}
	\end{overpic}
	\begin{overpic}[width=0.50\textwidth]{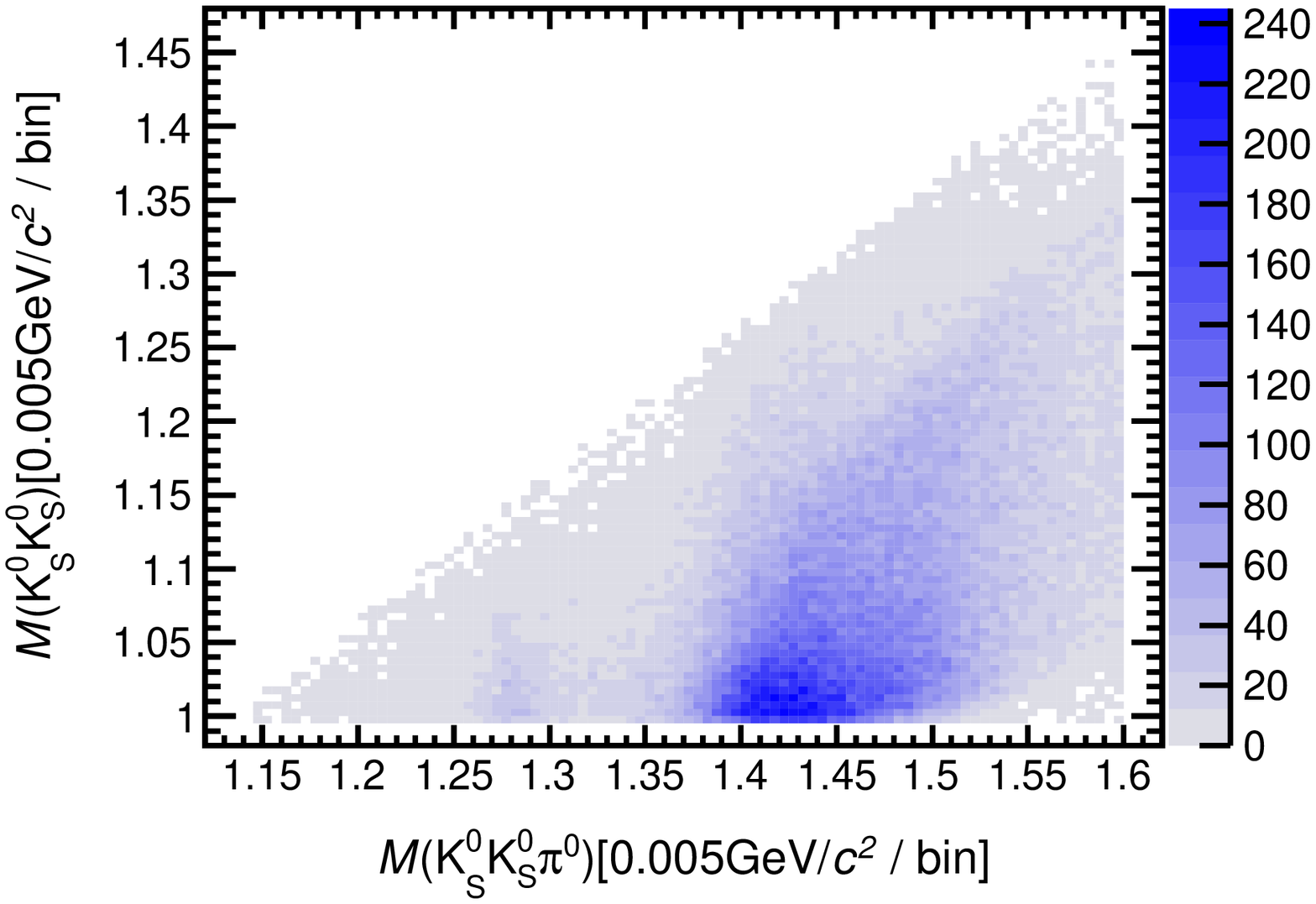}
		\put(20,60){$\bf (d)$}
	\end{overpic}
	\begin{overpic}[width=0.49\textwidth]{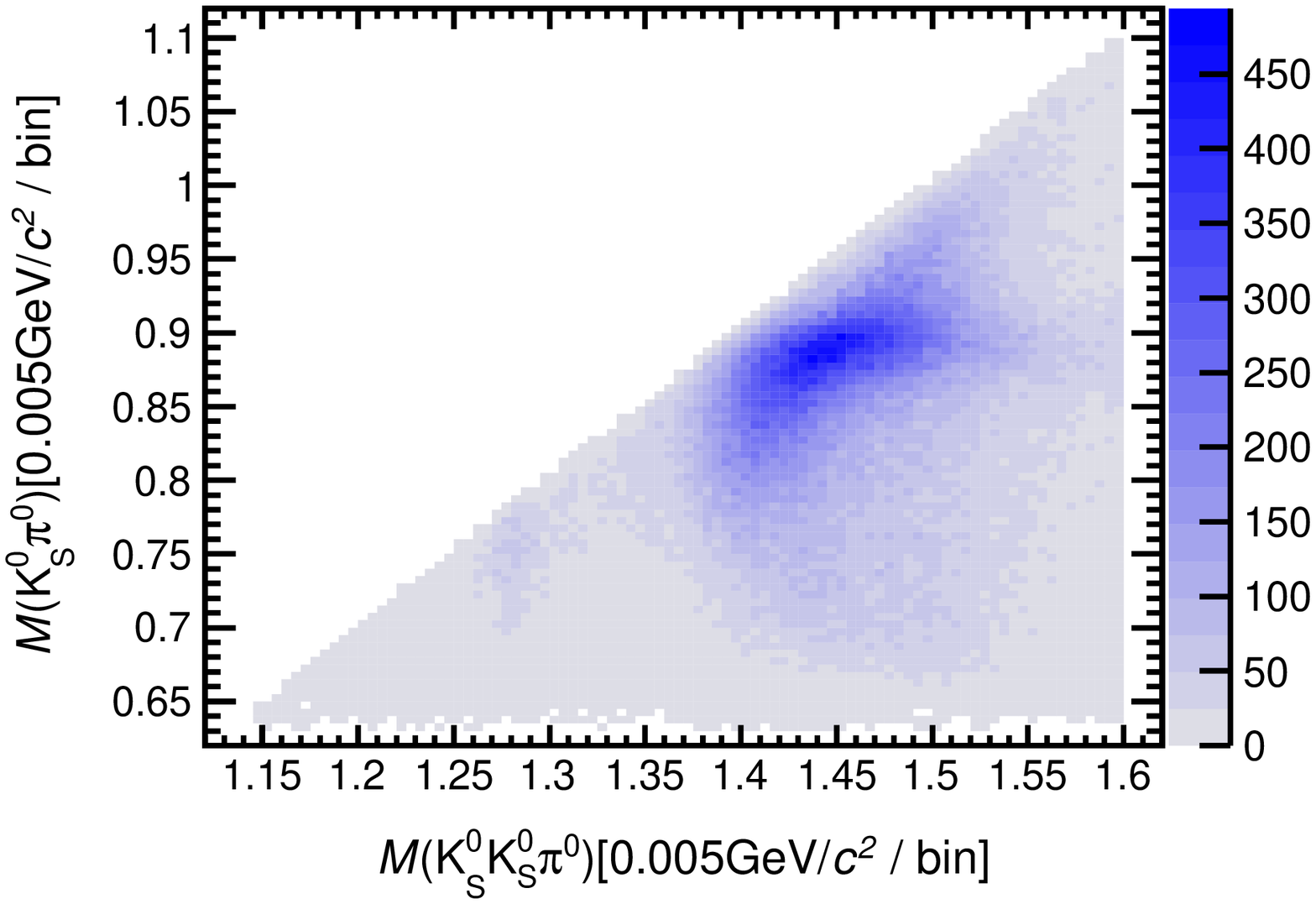}
		\put(20,60){$\bf (e)$}
	\end{overpic}
	\begin{overpic}[width=0.49\textwidth]{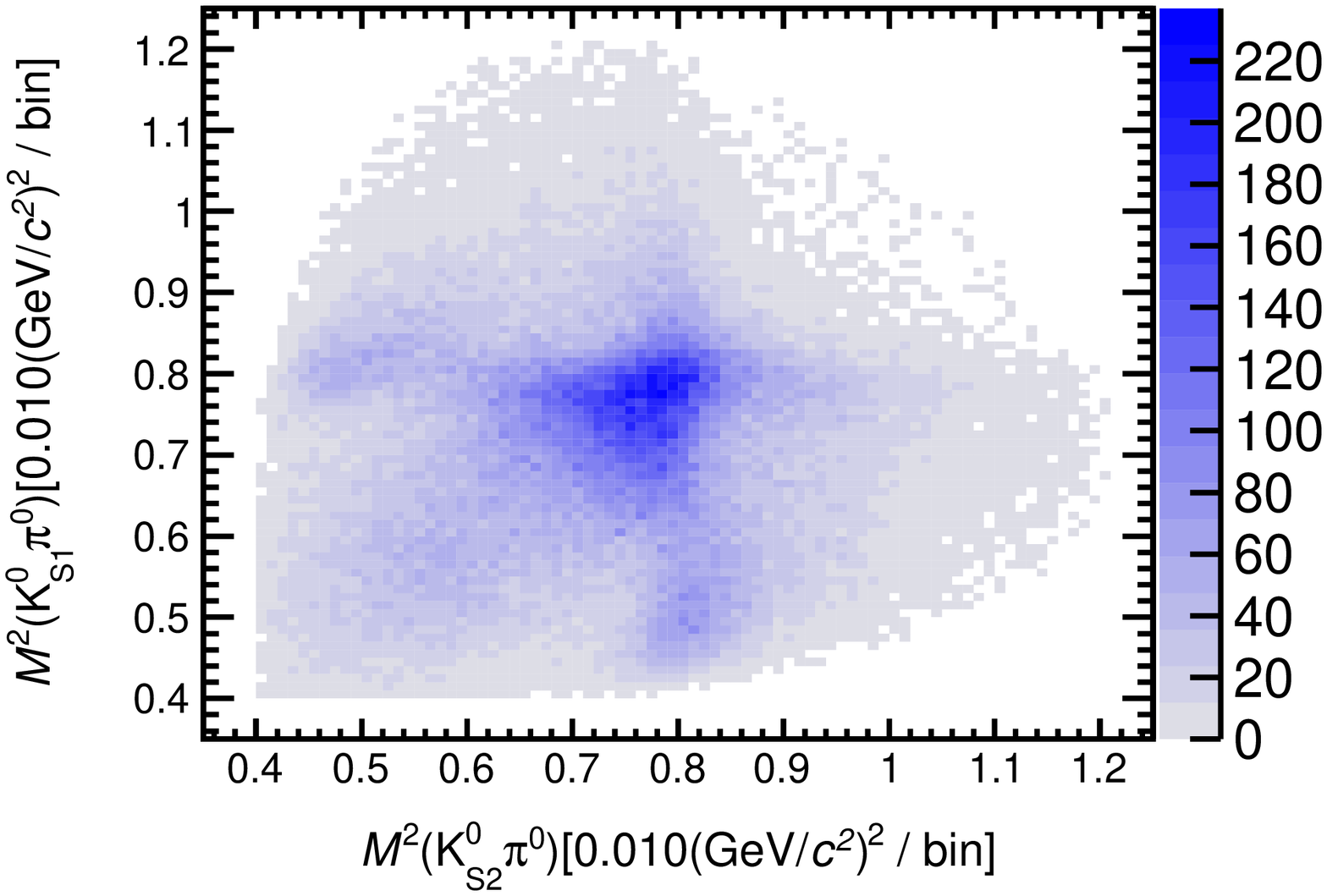}
		\put(20,60){$\bf (f)$}
	\end{overpic}
	\caption{Invariant mass spectra of (a) $\ksz\ksz\piz$, (b) $\ksz\ksz$, and (c) $\ksz\piz$ for events with the requirement $M(\ksz\ksz\piz)<1.6\gevcc$. The corresponding two-dimensional distributions of (d) $M(\ksz\ksz)$ versus $M(\ksz\ksz\piz)$ and (e) $M(\ksz\piz)$ versus $M(\ksz\ksz\piz)$. The corresponding Dalitz plot of (f) $M^{2}(K_{S1}^{0}\piz)$ versus $M^{2}(K_{S2}^{0}\piz)$. Dots with error bars are the data with the 7C kinematic fit applied. The blue solid line and the red dotted line represent the data with the 4C kinematic fit applied and the signal PHSP MC samples with the 7C kinematic fit applied (normalized by integral), respectively.}	
	\label{f1}
\end{figure}

To improve the resolution of kinematic variables, the remaining $\jpsi\to\gamma\ksz\ksz\piz$ events are subjected to a seven-constraint (7C) kinematic fit, which, in addition to imposing energy and momentum conservation, further constrains the one $\piz$ and two $\ksz$ masses to their known values quoted in the PDG~\cite{13}. Detailed MC simulation studies show that the 7C kinematic fit improves the resolution of the $\ksz\ksz\piz$ invariant mass from 8.4$\mevcc$ to 4.9$\mevcc$ at 1.45$\gevcc$, while the efficiency is more than 99.5\%. After all the above criteria are imposed, the invariant mass spectrum of $\ksz\ksz\piz$ up to 1.6$\gevcc$ is shown in figure~\ref{f1}(a) for data and the signal PHSP MC sample. A prominent structure around 1.45$\gevcc$ as well as a clear bump around 1.28$\gevcc$ are observed. The corresponding spectra of $\ksz\ksz$ and $\ksz\piz$ invariant mass are presented in figures~\ref{f1}(b) and (c), respectively. The spectra reveal strong enhancements near the $\ksz\ksz$ mass threshold and around 0.90$\gevcc$ in the $\ksz\piz$ mass spectrum. The two-dimensional distributions of the invariant mass of $\ksz\ksz\piz$ versus that of $\ksz\ksz$ and $\ksz\piz$ are also shown in figures~\ref{f1}(d) and (e), respectively. The Dalitz plot of $M^{2}(K_{S1}^{0}\piz)$ versus $M^{2}(K_{S2}^{0}\piz)$ for the selected candidates is shown in figure~\ref{f1}(f).

\section{PARTIAL WAVE ANALYSIS}
\subsection{Analysis method}
Using the GPUPWA framework~\cite{38}, a PWA is performed on the remaining 126,436 events in the region $M(\ksz\ksz\piz)<1.6\gevcc$ in order to disentangle the structures present in the $\ksz\ksz\piz$ invariant mass spectrum. The quasi-two-body decay amplitudes in the decay process $\jpsi\to\gamma R_{1}$ with two sequential decays $R_{1}\to\ksz R_{2}\to\ksz\ksz\piz$ and $R_{1}\to R_{2} \piz \to\ksz\ksz\piz$ are considered and constructed using the covariant tensor amplitudes described in ref.~\cite{39}, 
where $R_1$ and $R_2$ are the intermediate states that can decay to $\ksz\ksz\piz$, $\ksz\piz$ or $\ksz\ksz$, respectively. 

Let $A_X$ be the amplitude for a decay mode $X$, which has $N_{W_X}$ independent partial wave amplitudes. For the $\jp$ radiative decay, the general form of the $A_X$ decay amplitude is:
\begin{equation}\label{e1}
A_{X}=\psi_{\mu}(m_1)e_{\nu}^{*}(m_2)\sum_{j=1}^{N_{W_X}}\Lambda_{j}U^{\mu\nu}_{j},
\end{equation}
according to ref.~\cite{39}, where $\psi_{\mu}(m_{1})$ is the polarization four-vector for the $\jp$; $e_{\nu}(m_2)$ is the polarization four-vector for the photon; $m_1$ and $m_2$ are the spin projections of the $\jp$ and photon, respectively; $U^{\mu\nu}_{j}$ is the $j^{\textrm{th}}$ independent partial wave amplitude of $\jp$ radiative decay by the mode $X$ with coupling strength determined by a complex parameter $\Lambda_{j}$. The partial wave amplitudes $U^{\mu\nu}_{j}$ used in the analysis are constructed with the four-momenta of the particles in the final state, and their specific expressions are given in ref.~\cite{39}.

The ordinary intermediate resonance is parametrized by a relativistic Breit-Wigner (BW) propagator with a constant-width
\begin{equation}
BW(s)=\dfrac{1}{M^{2}-s-iM\Gamma},
\end{equation}
where $s$ is the invariant mass squared of resonances, $M$ and
$\Gamma$ are the corresponding mass and width. 

For $a_{0}(980)^{0}$ with mass near $K\bar{K}$ threshold, we use dispersion integrals to describe its lineshape, following the prescription given in ref.~\cite{dif}. The $a_{0}(980)^{0}$ amplitude is constructed using the following denominator: 
\begin{equation}
	D_{\alpha}(s)=m_{0}^{2}-s-\sum_{ch}\Pi_{ch}(s),
\end{equation}
where $m_{0}$ is the $a_{0}(980)$ mass and $\Pi_{ch}(s)$ in the sum over channels is a complex function, with imaginary part
\begin{equation}
	\textrm{Im}\Pi_{ch}(s)=g^{2}_{ch}\rho_{ch}(s)F_{ch}(s),
\end{equation}
while real parts are given by principal value integrals,
\begin{equation}\label{e2}
	\textrm{Re}\Pi_{ch}(s)=\dfrac{1}{\pi}P\int_{s_{ch}}^{\infty}\dfrac{\textrm{Im}\Pi_{ch}(s^{'})ds^{'}}{(s^{'}-s)}.
\end{equation}
In the above expressions $\rho_{ch}(s)$ is the available phase space for a given channel, obtained from the corresponding decay momentum $q_{ch}(s)$. The integral in eq.~(\ref{e2}) is divergent when $s\to\infty$, so the phase space is modified by a form factor $F_{ch}(s)=e^{-\beta q^{2}_{ch}(s)}$, where the parameter $\beta$ is related to the root-mean-square size of an emitting source~\cite{dif}. We consider three $a_{0}(980)^{0}$ decay channels here, the $\pi\eta$, $K\bar{K}$, and $\etap\pi$, with corresponding coupling constants and the value of $a_{0}(980)^{0}$ mass quoted from ref.~\cite{41}.

The complex coefficients of the amplitudes and resonance parameters are determined by an unbinned maximum likelihood fit to the data. The likelihood function is constructed following a method similar to that used in ref.~\cite{42}. 

The probability to observe the $i^{\textrm{th}}$ event characterized by
the measurement $\xi_{i}$, {\it i.e.}, the measured four-momenta of the particles in the final state, is:
\begin{equation}
P(\xi_{i})=\dfrac{|M(\xi_i)|^{2}\epsilon(\xi_i)\Phi(\xi_i)}{\sigma^{'}},
\end{equation}
where $\epsilon(\xi_{i})$ is the detection efficiency, $\Phi(\xi_{i})$ is the standard element of phase space, and $M(\xi_{i})=\sum_{X}A_{X}(\xi_{i})$ is the matrix element describing the decay processes from the $\jp$ to
the final state $\gamma\ksz\ksz\piz$. $A_{X}(\xi_{i})$ is the amplitude corresponding to decay mode $X$ as defined in eq.~(\ref{e1}) and $\sigma^{'}\equiv\int |M(\xi)|^{2}\epsilon(\xi)\Phi(\xi)d\xi$ is the normalization integral.

The joint probability for observing N events in the data sample is:
\begin{equation}
\mathcal{L}=\prod_{i=1}^{N}P(\xi_{i})=\prod_{i=1}^{N}\dfrac{|M(\xi_i)|^{2}\epsilon(\xi_i)\Phi(\xi_i)}{\sigma^{'}}.
\end{equation}
For technical reasons, rather than maximizing $\mathcal{L}$, $\mathcal{S}=-\ln\mathcal{L}$ is minimized, with
\begin{equation}
\mathcal{S}=-\ln\mathcal{L}=-\sum_{i=1}^{N}\ln|M(\xi_i)|^{2}+N \ln\sigma^{'}-\sum_{i=1}^{N}\ln\epsilon({\xi_{i}})\Phi(\xi_{i}),
\end{equation}
For a given data set. The third term is a constant and has no impact on the determination of the parameters of the amplitudes or on the relative changes of $-\ln\mathcal{L}$ values. In the fitting, the third term will not be considered.

The free parameters are optimized by MINUIT~\cite{43}. The normalization integral $\sigma^{'}$ is evaluated using MC techniques. A signal PHSP MC sample of $N_\textrm{gen}$ events is generated. These events are put through the detector simulation, subjected to the selection criteria and yield a sample of $N_\textrm{acc}$ accepted events. The normalization integral is computed as:
\begin{equation}
\sigma^{'}=\int|M(\xi)|^{2}\epsilon(\xi)\Phi(\xi)d\xi\to\dfrac{1}{N_\textrm{gen}}\sum_{j=1}^{N_\textrm{acc}}|M(\xi_j)|^{2},
\end{equation}
where the constant value of the phase space integral $\int\Phi(\xi)d\xi$ is ignored.

The number of the fitted events $N_{X}$ for a decay mode $X$, which has $N_{W_{X}}$ independent decay amplitudes, is defined as:
\begin{equation}
N_{X}=\dfrac{\sigma_{X}}{\sigma^{'}}\cdot N,
\end{equation}
where $N$ is the number of selected events after background subtraction, and
\begin{equation}
\sigma_{X}=\dfrac{1}{N_\textrm{gen}}\sum_{j=1}^{N_\textrm{acc}}|A_{X}(\xi_j)|^{2},
\end{equation}
is calculated with the same MC sample as the normalization integral $\sigma^{'}$.

The branching fraction of a decay mode $X$ is calculated as:
\begin{equation}
\mathcal{B}=\dfrac{N_{X}}{N_{\jp}\cdot \epsilon_{X}\cdot \mathcal{B}^{2}_{\ksz\to\pip\pim}\cdot \mathcal{B}_{\piz\to\gamma\gamma}},
\end{equation}
where the detection efficiency $\epsilon_{X}$ for the decay mode $X$ is obtained by the partial wave amplitude weighted MC sample,
\begin{equation}
\epsilon_{X}=\dfrac{\sigma_{X}}{\sigma_{X}^\textrm{gen}}=\dfrac{\sum_{k=1}^{N_\textrm{acc}}|A_{X}(\xi_{k})|^{2}}{\sum_{j=1}^{N_\textrm{gen}}|A_{X}(\xi_{j})|^{2}},
\end{equation}
$N_{X}$ is the number of the fitted events for the decay mode $X$; $N_{\jp}=(10.09\pm0.04)\times10^{9}$ is the total number of $\jp$ events~\cite{njp}; $\mathcal{B}_{\ksz\to\pip\pim}=(69.2\pm 0.05)\%$ and $\mathcal{B}_{\piz\to\gamma\gamma}=(98.823\pm 0.034)\%$ are the branching fractions of $\ksz\to\pip\pim$ and $\piz\to\gamma\gamma$ quoted from ref.~\cite{13}, respectively. Additionally, the branching fractions are defined as the coherent sum over the $R_1\to K^{0}_{S1} R_2\to K^{0}_{S1} K^{0}_{S2}\piz$ and $R_1\to K^{0}_{S2} R_2\to K^{0}_{S2}K^{0}_{S1}\piz$ contributions  for $\jpsi\to\gamma R_1$,$R_1\to\ksz R_2\to\ksz\ksz\piz$ processes in this analysis.

\subsection{Mass independent PWA}
To investigate the contributions from different components, a mass independent (MI) PWA is performed, which is mandatory in order to explore the lineshape of the $\ksz\ksz\piz$ invariant mass distribution for the different decay modes, and to minimize the bias from a particular model for the dynamics of the intermediate states.

\begin{table}[htbp]
	\begin{center}
		\renewcommand\arraystretch{1.2}
		\footnotesize
		\begin{tabular}{c|l|l|l|l|l}
			\hline
			\hline
			$J^{PC}$ &$0^{-+}$ &$1^{++}$ &$1^{-+}$  &$2^{++}$ &$2^{-+}$ \\
			\hline
			
			\multirow{5}{*}{for $\ksz\piz$} &$K^{*}(892)^{0}\ksz$ &$K^{*}(892)^{0}\ksz$ &$K^{*}(892)^{0}\ksz$ &$K^{*}(892)^{0}\ksz$  &$K^{*}(892)^{0}\ksz$   \\
			
			&$(\ksz\piz)_\textrm{P-phsp}\ksz$ &$(\ksz\piz)_\textrm{P-phsp}\ksz$ &$(\ksz\piz)_\textrm{P-phsp}\ksz$ &$(\ksz\piz)_\textrm{P-phsp}\ksz$  &$(\ksz\piz)_\textrm{P-phsp}\ksz$  \\
			
			&$(\ksz\piz)_\textrm{D-phsp}\ksz$ &$(\ksz\piz)_\textrm{D-phsp}\ksz$ &$(\ksz\piz)_\textrm{D-phsp}\ksz$ &$(\ksz\piz)_\textrm{D-phsp}\ksz$ &$(\ksz\piz)_\textrm{D-phsp}\ksz$ \\
			
			&$K^{*}_{0}(700)^{0}\ksz$ &$K^{*}_{0}(700)^{0}\ksz$ &  &  &$K^{*}_{0}(700)^{0}\ksz$  \\
			
			&$(\ksz\piz)_\textrm{S-phsp}\ksz$ &$(\ksz\piz)_\textrm{S-phsp}\ksz$ &  &  &$(\ksz\piz)_\textrm{S-phsp}\ksz$  \\
			\hline
			
			\multirow{5}{*}{for $\ksz\ksz$}
			&$a_{2}(1320)^{0}\piz$ &$a_{2}(1320)^{0}\piz$ &$a_{2}(1320)^{0} \piz$ &$a_{2}(1320)^{0}\piz$ &$a_{2}(1320)^{0}\piz$
			\\			
			
			&$(\ksz\ksz)_\textrm{D-phsp} \piz$ &$(\ksz\ksz)_\textrm{D-phsp} \piz$ &$(\ksz\ksz)_\textrm{D-phsp} \piz$ &$(\ksz\ksz)_\textrm{D-phsp} \piz$ &$(\ksz\ksz)_\textrm{D-phsp} \piz$ \\
			
			&$a_{0}(980)^{0}\piz$ &$a_{0}(980)^{0}\piz$ &  &  &$a_{0}(980)^{0}\piz$  \\
			
			&$a_{0}(1450)^{0}\piz$ &$a_{0}(1450)^{0}\piz$ &  &  &$a_{0}(1450)^{0}\piz$  \\
			
			&$(\ksz\ksz)_\textrm{S-phsp} \piz$ &$(\ksz\ksz)_\textrm{S-phsp} \piz$ &  &  &$(\ksz\ksz)_\textrm{S-phsp} \piz$  \\
			
			\hline
			\hline			
		\end{tabular}
	\end{center}
	\caption{The set of all possible decay mode candidates evaluated in the MI PWA.}
	\label{t1}
\end{table}

In practice, a MI PWA where the intermediate states in the $\ksz\ksz\piz$ invariant mass spectrum are parameterized by a separate complex constant for each of 24 bins of 15$\mevcc$ width, while the part of the amplitude that describes the dynamical function is constant over a small range of $s$, is performed in the region $M(\ksz\ksz\piz)<1.6\gevcc$ to extract the contribution of components with different decay modes, this method also has been described as "Bin-by-bin analysis" in ref.~\cite{44}. Taking into account the spin-parity, charge conjugation and isospin conservation, all possible decay mode candidates are evaluated, as listed in table~\ref{t1}, where "S", "P" and "D" represent the mother resonance decaying with orbital angular momentum equal to 0, 1 and 2, respectively. The "phsp" refers to the direct decay process without an intermediate resonance, and is modeled by the corresponding PHSP distribution without a propagator. Only the partial wave amplitudes with smallest orbital angular momentum are considered in the analysis because of the strong suppression on the magnitude from the centrifugal barrier~\cite{39} for large orbital angular momentum waves. In the fit, all resonance parameters are fixed to the PDG values~\cite{13} or published results~\cite{41}. The changes on the negative log-likelihood (NLL) value and the number of free parameters in the fits with and without a component are used to evaluate its statistical significance. The nominal solution, in which the significance of each component is greater than $5\sigma$, is retained as the baseline in each bin.

\begin{sidewaystable}[htbp]
	\begin{center}
		\resizebox{\textwidth}{10mm}{
			\begin{tabular}{l|l|l|l|l|l|l|l|l|l|l|l}
				\hline\hline
				1.240$-$1.255$\gevcc$ &1.255$-$1.270$\gevcc$ &1.270$-$1.285$\gevcc$ &1.285$-$1.300$\gevcc$ &1.300$-$1.315$\gevcc$ &1.315$-$1.330$\gevcc$ &1.330$-$1.345$\gevcc$ &1.345$-$1.360$\gevcc$ &1.360$-$1.375$\gevcc$ &1.375$-$1.390$\gevcc$ &1.390$-$1.405$\gevcc$ &1.405$-$1.420$\gevcc$ \\
				\hline
				$0^{-+}\to a_{0}(980)^{0}\piz$ &$0^{-+}\to a_{0}(980)^{0}\piz$ &$0^{-+}\to a_{0}(980)^{0}\piz$ &$0^{-+}\to a_{0}(980)^{0}\piz$ &$0^{-+}\to a_{0}(980)^{0}\piz$ &$0^{-+}\to a_{0}(980)^{0}\piz$ &$0^{-+}\to a_{0}(980)^{0}\piz$ &$0^{-+}\to a_{0}(980)^{0}\piz$ &$0^{-+}\to a_{0}(980)^{0}\piz$ &$0^{-+}\to a_{0}(980)^{0}\piz$ &$0^{-+}\to a_{0}(980)^{0}\piz$ &$0^{-+}\to a_{0}(980)^{0}\piz$\\
				
				$0^{-+}\to K^{*}(892)^{0}\ksz$ &$0^{-+}\to K^{*}(892)^{0}\ksz$ &$0^{-+}\to K^{*}(892)^{0}\ksz$ &$0^{-+}\to K^{*}(892)^{0}\ksz$ 
				&$0^{-+}\to K^{*}(892)^{0}\ksz$ &$0^{-+}\to K^{*}(892)^{0}\ksz$ &$0^{-+}\to K^{*}(892)^{0}\ksz$ &$0^{-+}\to K^{*}(892)^{0}\ksz$ &$0^{-+}\to K^{*}(892)^{0}\ksz$ &$0^{-+}\to K^{*}(892)^{0}\ksz$ &$0^{-+}\to K^{*}(892)^{0}\ksz$ &$0^{-+}\to K^{*}(892)^{0}\ksz$\\
				
				$1^{++}\to a_{0}(980)^{0}\piz$ &$1^{++}\to a_{0}(980)^{0}\piz$ &$1^{++}\to a_{0}(980)^{0}\piz$ &$1^{++}\to a_{0}(980)^{0}\piz$ &$1^{++}\to a_{0}(980)^{0}\piz$ &$1^{++}\to a_{0}(980)^{0}\piz$ &$1^{++}\to a_{0}(980)^{0}\piz$  &$1^{++}\to a_{0}(980)^{0}\piz$  &$0^{-+}\to a_{2}(1320)^{0}\piz$ &$0^{-+}\to a_{2}(1320)^{0}\piz$ &$0^{-+}\to a_{2}(1320)^{0}\piz$  &$0^{-+}\to a_{2}(1320)^{0}\piz$\\
				
				& & & & & & & &$0^{-+}\to(\ksz\piz)_\textrm{P-phsp}\ksz$ &$0^{-+}\to(\ksz\piz)_\textrm{P-phsp}\ksz$ &$0^{-+}\to(\ksz\piz)_\textrm{P-phsp}\ksz$ &$0^{-+}\to(\ksz\piz)_\textrm{P-phsp}\ksz$ \\
				
				& & & & & & & &$0^{-+}\to(\ksz\ksz)_\textrm{S-phsp}\piz$ &$0^{-+}\to(\ksz\ksz)_\textrm{S-phsp}\piz$ &$0^{-+}\to(\ksz\ksz)_\textrm{S-phsp}\piz$ &$0^{-+}\to(\ksz\ksz)_\textrm{S-phsp}\piz$ \\	
				
				& & & & & & & &$1^{++}\to K^{*}(892)^{0}\ksz$ &$1^{++}\to K^{*}(892)^{0}\ksz$ &$1^{++}\to K^{*}(892)^{0}\ksz$ &$1^{++}\to K^{*}(892)^{0}\ksz$ \\		
				\hline\hline
		\end{tabular}}
	\end{center}
	\begin{center}
		\resizebox{\textwidth}{12mm}{
			\begin{tabular}{l|l|l|l|l|l|l|l|l|l|l|l}
				\hline\hline
				1.420$-$1.435$\gevcc$ &1.435$-$1.450$\gevcc$ &1.450$-$1.465$\gevcc$ &1.465$-$1.480$\gevcc$ &1.480$-$1.495$\gevcc$ &1.495$-$1.510$\gevcc$ &1.510$-$1.525$\gevcc$ &1.525$-$1.540$\gevcc$ &1.540$-$1.555$\gevcc$ &1.555$-$1.570$\gevcc$ &1.570$-$1.585$\gevcc$ &1.585$-$1.600$\gevcc$ \\
				\hline
				$0^{-+}\to a_{0}(980)^{0}\piz$ &$0^{-+}\to a_{0}(980)^{0}\piz$ &$0^{-+}\to a_{0}(980)^{0}\piz$ &$0^{-+}\to a_{0}(980)^{0}\piz$ &$0^{-+}\to a_{0}(980)^{0}\piz$ &$0^{-+}\to a_{0}(980)^{0}\piz$ &$0^{-+}\to a_{0}(980)^{0}\piz$ &$0^{-+}\to a_{0}(980)^{0}\piz$ &$0^{-+}\to a_{0}(980)^{0}\piz$ &$0^{-+}\to a_{0}(980)^{0}\piz$ &$0^{-+}\to a_{0}(980)^{0}\piz$ &$0^{-+}\to a_{0}(980)^{0}\piz$\\
				
				$0^{-+}\to K^{*}(892)^{0}\ksz$ &$0^{-+}\to K^{*}(892)^{0}\ksz$ &$0^{-+}\to K^{*}(892)^{0}\ksz$ &$0^{-+}\to K^{*}(892)^{0}\ksz$ 
				&$0^{-+}\to K^{*}(892)^{0}\ksz$ &$0^{-+}\to K^{*}(892)^{0}\ksz$ &$0^{-+}\to K^{*}(892)^{0}\ksz$ &$0^{-+}\to K^{*}(892)^{0}\ksz$ &$0^{-+}\to K^{*}(892)^{0}\ksz$ &$0^{-+}\to K^{*}(892)^{0}\ksz$ &$0^{-+}\to K^{*}(892)^{0}\ksz$ &$0^{-+}\to K^{*}(892)^{0}\ksz$\\
				
				$0^{-+}\to a_{2}(1320)^{0}\piz$ &$0^{-+}\to a_{2}(1320)^{0}\piz$ &$0^{-+}\to a_{2}(1320)^{0}\piz$  &$0^{-+}\to a_{2}(1320)^{0}\piz$ &$0^{-+}\to a_{2}(1320)^{0}\piz$ &$0^{-+}\to a_{2}(1320)^{0}\piz$ &$0^{-+}\to a_{2}(1320)^{0}\piz$  &$0^{-+}\to a_{2}(1320)^{0}\piz$ &$0^{-+}\to a_{2}(1320)^{0}\piz$ &$0^{-+}\to a_{2}(1320)^{0}\piz$ &$0^{-+}\to a_{2}(1320)^{0}\piz$  &$0^{-+}\to a_{2}(1320)^{0}\piz$ \\
				
				$0^{-+}\to(\ksz\piz)_\textrm{P-phsp}\ksz$ &$0^{-+}\to(\ksz\piz)_\textrm{P-phsp}\ksz$ &$0^{-+}\to(\ksz\piz)_\textrm{P-phsp}\ksz$ &$0^{-+}\to(\ksz\piz)_\textrm{P-phsp}\ksz$ &$0^{-+}\to(\ksz\piz)_\textrm{P-phsp}\ksz$ &$0^{-+}\to(\ksz\piz)_\textrm{P-phsp}\ksz$ &$0^{-+}\to(\ksz\piz)_\textrm{P-phsp}\ksz$ &$0^{-+}\to(\ksz\piz)_\textrm{P-phsp}\ksz$ &$0^{-+}\to(\ksz\piz)_\textrm{P-phsp}\ksz$ &$0^{-+}\to(\ksz\piz)_\textrm{P-phsp}\ksz$ &$0^{-+}\to(\ksz\piz)_\textrm{P-phsp}\ksz$ &$0^{-+}\to(\ksz\piz)_\textrm{P-phsp}\ksz$ \\
				
				$0^{-+}\to(\ksz\ksz)_\textrm{S-phsp}\piz$ &$0^{-+}\to(\ksz\ksz)_\textrm{S-phsp}\piz$ &$0^{-+}\to(\ksz\ksz)_\textrm{S-phsp}\piz$ &$0^{-+}\to(\ksz\ksz)_\textrm{S-phsp}\piz$
				&$1^{++}\to a_{0}(980)^{0}\piz$ &$1^{++}\to a_{0}(980)^{0}\piz$ &$1^{++}\to a_{0}(980)^{0}\piz$ &$1^{++}\to a_{0}(980)^{0}\piz$ &$1^{++}\to a_{0}(980)^{0}\piz$ &$1^{++}\to a_{0}(980)^{0}\piz$ &$1^{++}\to a_{0}(980)^{0}\piz$ &$1^{++}\to a_{0}(980)^{0}\piz$ \\
				
				$1^{++}\to K^{*}(892)^{0}\ksz$ &$1^{++}\to K^{*}(892)^{0}\ksz$ &$1^{++}\to K^{*}(892)^{0}\ksz$ &$1^{++}\to K^{*}(892)^{0}\ksz$ &$1^{++}\to K^{*}(892)^{0}\ksz$ &$1^{++}\to K^{*}(892)^{0}\ksz$ &$1^{++}\to K^{*}(892)^{0}\ksz$ &$1^{++}\to K^{*}(892)^{0}\ksz$ &$1^{++}\to K^{*}(892)^{0}\ksz$ &$1^{++}\to K^{*}(892)^{0}\ksz$ &$1^{++}\to K^{*}(892)^{0}\ksz$ &$1^{++}\to K^{*}(892)^{0}\ksz$ \\
				
				& & & &$2^{++}\to K^{*}(892)^{0}\ksz$ &$2^{++}\to K^{*}(892)^{0}\ksz$ &$2^{++}\to K^{*}(892)^{0}\ksz$ &$2^{++}\to K^{*}(892)^{0}\ksz$ &$2^{++}\to K^{*}(892)^{0}\ksz$ &$2^{++}\to K^{*}(892)^{0}\ksz$ &$2^{++}\to K^{*}(892)^{0}\ksz$ &$2^{++}\to K^{*}(892)^{0}\ksz$ \\
				
				& & & &$2^{-+}\to a_{0}(980)^{0}\piz$ &$2^{-+}\to a_{0}(980)^{0}\piz$ &$2^{-+}\to a_{0}(980)^{0}\piz$ &$2^{-+}\to a_{0}(980)^{0}\piz$ &$2^{-+}\to a_{0}(980)^{0}\piz$ &$2^{-+}\to a_{0}(980)^{0}\piz$ &$2^{-+}\to a_{0}(980)^{0}\piz$ &$2^{-+}\to a_{0}(980)^{0}\piz$ \\
				\hline\hline
		\end{tabular}}
	\end{center}
	\caption{The fit components of the nominal solutions for 24 bins of 15$\mevcc$ width in the $\ksz\ksz\piz$ invariant mass spectrum.}
	\label{t1b}		
\end{sidewaystable}

\begin{figure}[htbp]
	\centering
	\begin{overpic}[width=0.88\textwidth]{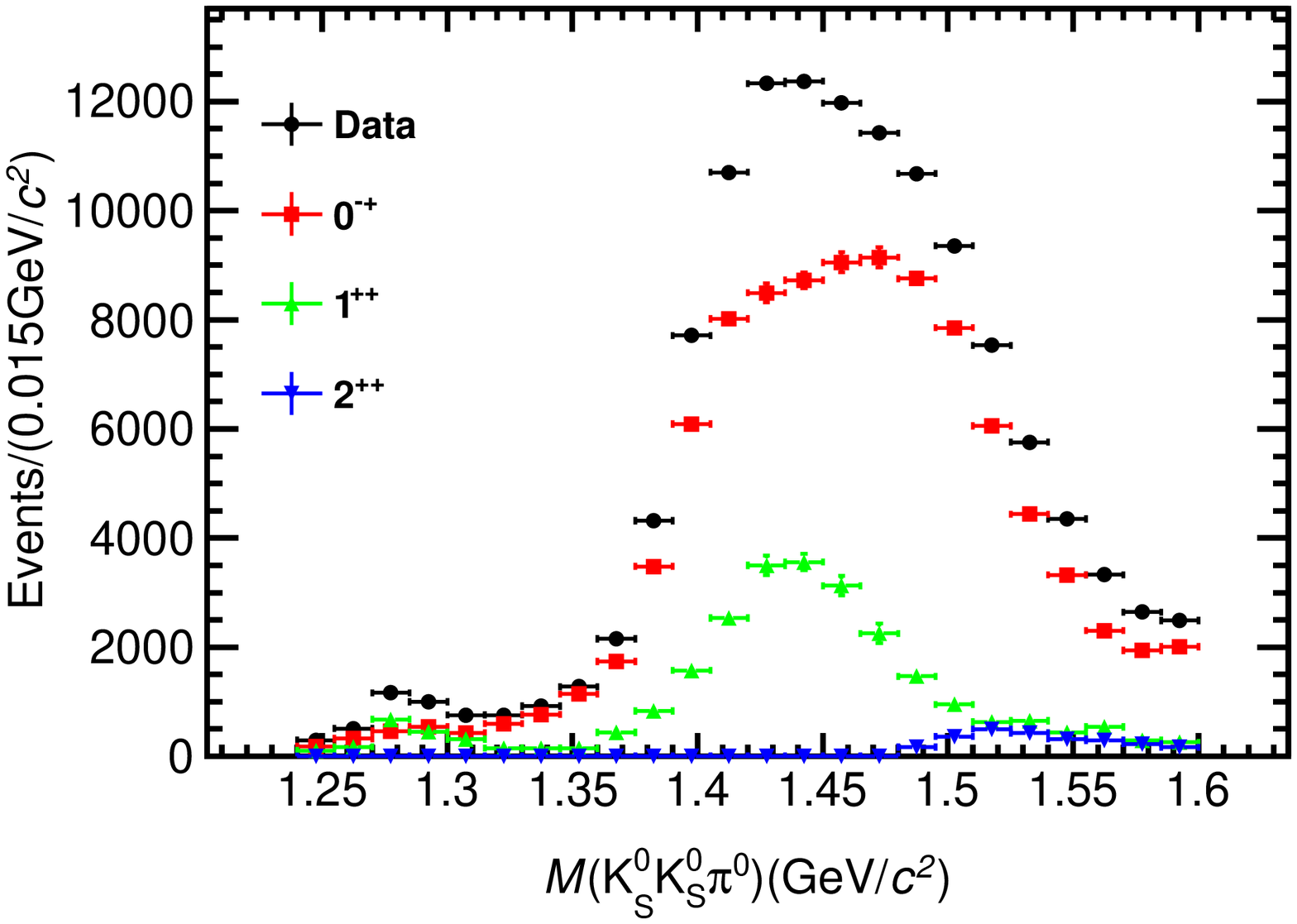}
		\put(80,60){\Large {$\bf (a)$}}
	\end{overpic}
	\begin{overpic}[width=0.88\textwidth]{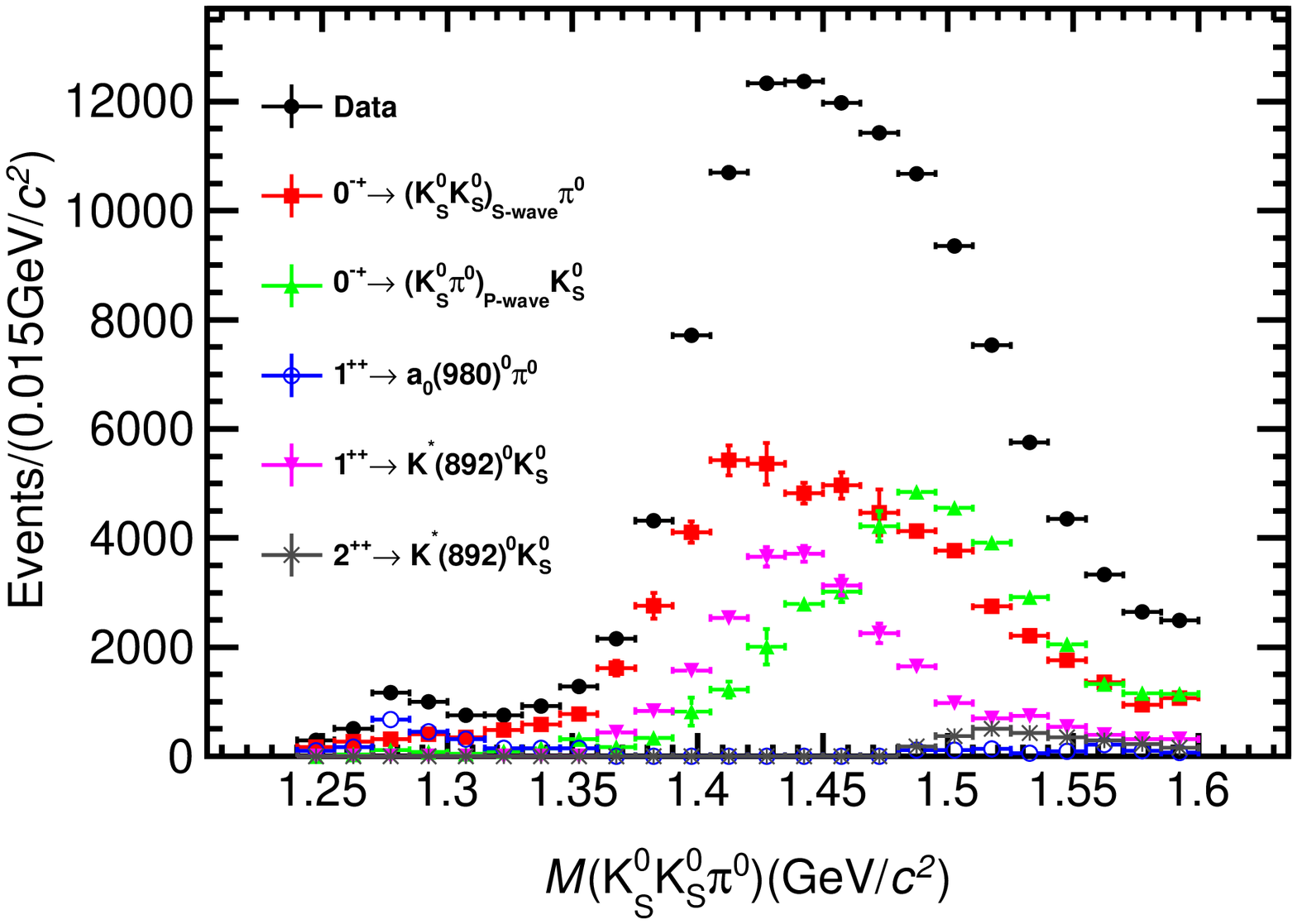}
		\put(80,60){\Large {$\bf (b)$}}
	\end{overpic}
	\caption{The $\ksz\ksz\piz$ intensity spectra for the (a) dominant spin-parity components and (b) dominant decay modes for different spin-parity components obtained from the MI PWA results. The uncertainty on the total data is statistical, and the uncertainties on the others are obtained from the PWA in each bin.}	
	\label{f2}
\end{figure} 

In the MI PWA results, the fit components of the nominal solutions for 24 bins are shown in table~\ref{t1b}. The $\ksz\ksz\piz$ intensity spectra obtained from the MI PWA results are shown in figure~\ref{f2}, where $(\ksz\ksz)_\textrm{S-wave}\piz$ represents the combination of $a_{0}(980)^{0}\piz$ and $(\ksz\ksz)_\textrm{S-phsp}\piz$ waves, and $(\ksz\piz)_\textrm{P-wave}\ksz$ represents the combination of $K^{*}(892)^{0}\ksz$ and $(\ksz\piz)_\textrm{P-phsp}\ksz$ waves. The fit results confirm that the pseudoscalar component is the dominant contribution. However, unlike a general resonance, the lineshape of the total pseudoscalar component does not exhibit an obvious peak position. It is instead relatively constant between 1.4$\gevcc$ and 1.5$\gevcc$. Further analysis shows that the pseudoscalar component decays into the $(\ksz\ksz)_\textrm{S-wave}\piz$ and $(\ksz\piz)_\textrm{P-wave}\ksz$ partial waves are of comparable magnitudes, but with different lineshapes and peaks. 
Axial vector components also contribute prominently, peaking at 1.28$\gevcc$ and 1.42$\gevcc$, respectively, and the corresponding component peaking at 1.28$\gevcc$ decays into $a_{0}(980)^{0}\piz$, while that peaking at 1.42$\gevcc$ into $K^{*}(892)^{0}\ksz$. There is also a small tensor component decaying into $K^{*}(892)^{0}\ksz$ around 1.52$\gevcc$, which is observed in the decay of $\jpsi\to\gamma K\bar{K}\pi$ for the first time.

\subsection{Mass dependent PWA}
A mass dependent (MD) PWA is also performed. All possible intermediate states listed in the PDG~\cite{13} with mass less than 1.6$\gevcc$, and satisfying spin-parity, charge conjugation and isospin conservation, {\it e.g.}, the $\eta(1295)$, $\eta(1405)$, $\eta(1475)$, $f_{1}(1285)$, $f_{1}(1420)$, $f_{1}(1510)$, $f_{2}(1270)$, $f_{2}(1525)$ and $f_{2}(1565)$, are considered as candidates in the analysis. Additionally, non-resonant processes with different spin-parities are also taken into consideration when searching for the nominal solution. These kinds of non-resonant components can mimic the contributions of the tails from higher mass resonances with different spin-parities and decay modes as well. The masses and widths of intermediate states from $\jpsi$ radiative decay are floating in the fit, while those of the resonances decaying into $\ksz\ksz$ or $\ksz\piz$ final states are fixed to the PDG values~\cite{13} or published results~\cite{41} as done in the MI PWA. All possible intermediate states and decay modes are tried in the fits, and only those with significance larger than $5\sigma$ are retained.

\begin{sidewaystable}[htbp]
	\normalsize
	\renewcommand\arraystretch{1.2}
	\begin{center}
		\resizebox{\textwidth}{30mm}{
			\begin{tabular}{l|cccccccccccccc}
				\hline
				\hline
				Components &$(1)$  &$(2)$  &$(3)$  &$(4)$  &$(5)$ &$(6)$ &$(7)$  &$(8)$  &$(9)$ &$(10)$  &$(11)$  &$(12)$ &$(13)$ &$(14)$ \\
				\hline
				$(1).$ $\jp\to\gamma$PHSP$(0^{-+})\to\gamma K^{*}(892)^{0}\ksz\to\gamma\ksz\ksz\piz$   &$3.4\pm0.1$ &$0.9\pm0.1$ &$9.0\pm0.1$ &$-2.7\pm0.1$ &$-1.2\pm0.1$ &$0.0\pm0.0$ &$0.3\pm0.1$ &$0.0\pm0.0$ &$1.4\pm0.1$ &$1.4\pm0.1$ &$0.0\pm0.0$ &$0.0\pm0.0$ &$0.2\pm0.1$ &$0.0\pm0.0$ \\
				
				$(2).$ $\jp\to\gamma$PHSP$(1^{++})\to\gamma K^{*}(892)^{0}\ksz\to\gamma\ksz\ksz\piz$   &  &$2.9\pm0.1$ &$1.2\pm0.1$ &$0.3\pm0.1$ &$-1.0\pm0.2$ &$0.0\pm0.0$ &$0.0\pm0.0$ &$0.0\pm0.0$ &$0.2\pm0.1$ &$0.4\pm0.1$ &$1.4\pm0.1$ &$1.0\pm0.2$ &$0.0\pm0.1$ &$0.0\pm0.0$ \\
				
				$(3).$ $\jp\to\gamma\eta(1405)\to\gamma\ksz(\ksz\piz)_\textrm{P-wave}\to\gamma\ksz\ksz\piz$   &  &  &$9.5\pm0.2$ &$-1.0\pm0.1$ &$-1.6\pm0.1$ &$0.0\pm0.0$ &$-1.7\pm0.1$ &$0.0\pm0.1$ &$1.1\pm0.1$ &$2.4\pm0.1$ &$0.0\pm0.0$ &$0.0\pm0.1$ &$1.0\pm0.1$ &$0.0\pm0.0$ \\
				
				$(4).$ $\jp\to\gamma\eta(1475)\to\gamma\ksz(\ksz\piz)_\textrm{P-wave}\to\gamma\ksz\ksz\piz$   &  &  &  &$11.2\pm0.2$ &$2.0\pm0.1$ &$0.0\pm0.0$ &$-2.0\pm0.1$ &$0.0\pm0.0$ &$-1.9\pm0.1$ &$-1.0\pm0.1$ &$0.0\pm0.0$ &$0.0\pm0.0$ &$-0.8\pm0.1$ &$0.3\pm0.1$ \\
				
				$(5).$ $\jp\to\gamma f_{1}(1420)\to\gamma K^{*}(892)^{0}\ksz\to\gamma\ksz\ksz\piz$   &  &  &  &  &$18.9\pm0.3$ &$-0.1\pm0.1$ &$-0.7\pm0.1$ &$0.0\pm0.0$ &$-0.6\pm0.1$ &$0.0\pm0.0$ &$1.6\pm0.1$ &$-5.8\pm0.3$ &$-0.2\pm0.1$ &$0.0\pm0.0$ \\
				
				$(6).$ $\jp\to\gamma f_{2}(1525)\to\gamma K^{*}(892)^{0}\ksz\to\gamma\ksz\ksz\piz$   &  &  &  &  &  &$2.2\pm0.1$ &$0.0\pm0.0$ &$0.0\pm0.1$ &$0.0\pm0.1$ &$0.0\pm0.0$ &$0.0\pm0.2$ &$0.0\pm0.1$ &$0.0\pm0.0$ &$0.0\pm0.1$ \\
				
				$(7).$ $\jp\to\gamma$PHSP$(0^{-+})\to\gamma a_{0}(980)^{0}\piz\to\gamma\ksz\ksz\piz$   &  &  &  &  &  &  &$22.9\pm0.3$ &$0.0\pm0.0$ &$9.1\pm0.1$ &$-2.1\pm0.2$ &$0.1\pm0.1$ &$0.0\pm0.0$ &$0.0\pm0.0$ &$0.0\pm0.0$ \\
				
				$(8).$ $\jp\to\gamma$PHSP$(2^{-+})\to\gamma a_{0}(980)^{0}\piz\to\gamma\ksz\ksz\piz$   &  &  &  &  &  &  &  &$0.2\pm0.1$ &$0.0\pm0.0$ &$0.0\pm0.0$ &$0.0\pm0.0$ &$0.0\pm0.0$ &$0.0\pm0.0$ &$0.0\pm0.0$ \\
				
				$(9).$ $\jp\to\gamma \eta(1405)\to\gamma (\ksz\ksz)_\textrm{S-wave}\piz\to\gamma\ksz\ksz\piz$   &  &  &  &  &  &  &  &  &$7.2\pm0.1$ &$-0.9\pm0.1$ &$-0.1\pm0.1$ &$0.0\pm0.0$ &$0.0\pm0.0$ &$0.0\pm0.0$ \\
				
				$(10).$ $\jp\to\gamma \eta(1475)\to\gamma (\ksz\ksz)_\textrm{S-wave}\piz\to\gamma\ksz\ksz\piz$   &  &  &  &  &  &  &  &  &  &$8.6\pm0.2$ &$0.1\pm0.1$ &$0.1\pm0.1$ &$0.0\pm0.0$ &$0.0\pm0.0$ \\
				
				$(11).$ $\jp\to\gamma f_{1}(1285)\to\gamma a_{0}(980)^{0}\piz\to\gamma\ksz\ksz\piz$   &  &  &  &  &  &  &  &  &  &  &$2.1\pm0.1$ &$-0.8\pm0.1$ &$0.0\pm0.0$ &$0.0\pm0.0$ \\
				
				$(12).$ $\jp\to\gamma f_{1}(1420)\to\gamma a_{0}(980)^{0}\piz\to\gamma\ksz\ksz\piz$   &  &  &  &  &  &  &  &  &  &  &  &$1.3\pm0.1$ &$0.0\pm0.0$ &$0.0\pm0.0$ \\
				
				$(13).$ $\jp\to\gamma \eta(1405)\to\gamma a_{2}(1320)^{0}\piz\to\gamma\ksz\ksz\piz$   &  &  &  &  &  &  &  &  &  &  &  &  &$0.9\pm0.1$ &$-0.4\pm0.1$ \\
				
				$(14).$ $\jp\to\gamma \eta(1475)\to\gamma a_{2}(1320)^{0}\piz\to\gamma\ksz\ksz\piz$   &  &  &  &  &  &  &  &  &  &  &  &  & &$0.1\pm0.1$ \\
				\hline
				\hline
		\end{tabular}}
	\end{center}
	\caption{Fraction of each component and interference fractions between two components (\%) in the MD PWA nominal solution. The uncertainties are statistical only.}
	\label{t2}
\end{sidewaystable}

\begin{sidewaystable}[htbp]
	\begin{center}
		\normalsize
		\renewcommand\arraystretch{1.3}
		\resizebox{\textwidth}{30mm}{\begin{tabular}{ccclcc}
				\hline
				\hline
				Resonance &$M(\textrm{MeV}/{c}^2)$ &$\Gamma(\textrm{MeV})$  &Decay Mode &B.F. &Sig.$(\sigma)$ \\
				\hline
				\multirow{2}{*}{$\eta(1405)$}  &\multirow{2}{*}{$1391.7\pm0.7_{-0.3}^{+11.3}$} &\multirow{2}{*}{$60.8\pm1.2_{-12.0}^{+5.5}$}  &$\jp\to\gamma\eta(1405)\to\gamma\ksz(\ksz\piz)_\textrm{P-wave}\to\gamma\ksz\ksz\piz$ &$(5.84\pm0.12^{+2.03}_{-3.36})\times 10^{-5}$  &$\gg 35$ \\
				\cline{4-6}
				& & &$\jp\to\gamma\eta(1405)\to\gamma(\ksz\ksz)_\textrm{S-wave}\piz\to\gamma\ksz\ksz\piz$ &$(2.88\pm0.04^{+1.64}_{-0.38})\times 10^{-5}$ &$18.4$ \\
				\hline
				
				\multirow{2}{*}{$\eta(1475)$}  &\multirow{2}{*}{$1507.6\pm1.6_{-32.2}^{+15.5}$} &\multirow{2}{*}{$115.8\pm2.4_{-10.9}^{+14.8}$}   &$\jp\to\gamma\eta(1475)\to\gamma\ksz(\ksz\piz)_\textrm{P-wave}\to\gamma\ksz\ksz\piz$ &$(6.58\pm0.12^{+3.98}_{-2.82})\times 10^{-5}$ &$\gg 35$ \\
				\cline{4-6}
				& & &$\jp\to\gamma\eta(1475)\to\gamma(\ksz\ksz)_\textrm{S-wave}\piz\to\gamma\ksz\ksz\piz$ &$(3.99\pm0.09^{+0.41}_{-0.66})\times 10^{-5}$  &$\gg 35$  \\
				\hline
				
				$f_{1}(1285)$  &$1280.2\pm0.6_{-1.5}^{+1.2}$ &$28.2\pm1.1_{-2.9}^{+5.5}$ &$\jp\to\gamma f_{1}(1285)\to\gamma a_{0}(980)^{0}\piz\to\gamma\ksz\ksz\piz$ &$(8.55\pm0.41^{+3.42}_{-1.04})\times 10^{-6}$  &$\gg 35$ \\
				\hline
				
				\multirow{2}{*}{$f_{1}(1420)$}  &\multirow{2}{*}{$1433.5\pm1.1_{-0.7}^{+27.9}$} &\multirow{2}{*}{$95.9\pm2.3_{-10.9}^{+13.6}$} &$\jp\to\gamma f_{1}(1420)\to\gamma K^{*}(892)^{0}\ksz\to\gamma\ksz\ksz\piz$ &$(7.25\pm0.12^{+0.73}_{-1.25})\times 10^{-5}$  & $\gg 35$ \\
				\cline{4-6}
				& & &$\jp\to\gamma f_{1}(1420)\to\gamma a_{0}(980)^{0}\piz\to\gamma\ksz\ksz\piz$ &$(4.62\pm0.36^{+2.36}_{-1.94})\times 10^{-6}$  &$17.8$ \\
				\hline
				
				$f_{2}(1525)$ &$1515.4\pm2.5_{-7.6}^{+3.2}$ &$64.0\pm4.3_{-6.1}^{+2.0}$ &$\jp\to\gamma f_{2}(1525)\to\gamma K^{*}(892)^{0}\ksz\to\gamma\ksz\ksz\piz$ &$(9.47\pm0.43^{+1.51}_{-0.66})\times 10^{-6}$  &$23.8$  \\
				\hline
				\hline							
		\end{tabular}}
	\end{center}
	\caption{Masses ($M$), widths ($\Gamma$), branching fractions (B.F.) and significances (Sig.) of predominant components in the MD PWA nominal solution. The first uncertainties are statistical and the second ones are systematic.}
	\label{t3}
\end{sidewaystable}

Finally, the nominal solution includes two pseudoscalar states, the $\eta(1405)$ and $\eta(1475)$, two axial vector states, the $f_1(1285)$ and $f_1(1420)$ and a tensor state, the $f_2(1525)$. In total, 14 components are considered, and the fractions of the various components, as well as the interference fractions between any two components are presented in table~\ref{t2}. The calculation of the fractions for individual components involves the PHSP MC truth information without detector acceptance or resolution effects. The fraction for the $i^{\textrm{th}}$ component is defined as:
\begin{equation}
	F_{i}=\dfrac{\sum_{k=1}^{N_{\textrm{truth}}}|A_{i}|^{2}_{k}}{\sum_{k=1}^{N_{\textrm{truth}}}|A|^{2}_{k}},
\end{equation}
and the interference between the $i^{\textrm{th}}$ and $j^{\textrm{th}}$ components is defined as (where $I_{ij} = I_{ji}$):
\begin{equation}
	I_{ij}=\dfrac{\sum_{k=1}^{N_{\textrm{truth}}}2Re[A_{i}A_{j}^{*}]_{k}}{\sum_{k=1}^{N_{\textrm{truth}}}|A|^{2}_{k}},
\end{equation}
where $N_{\textrm{truth}}$ is the number of PHSP MC events at the generator level; $A_{i}$ and $A_j$ are the amplitudes corresponding to the $i^{\textrm{th}}$ and $j^{\textrm{th}}$ components in the MD PWA nominal solution as defined in eq.~(\ref{e1}), respectively; $A=\sum_{i}A_{i}$ is the total amplitude describing the decay processes from the $\jp$ to the final state $\gamma\ksz\ksz\piz$ in the MD PWA nominal solution. These fractions will not sum to unity if there is net constructive or destructive interference. In order to determine the statistical uncertainties of the fractions, the amplitude coefficients are randomly sampled by a Gaussian-distributed amount set by the covariance matrix. Then the distribution of each fraction is fitted with a Gaussian function and the width of the Gaussian function is defined as the uncertainty of the fraction. The optimized masses and widths of the intermediate states, 
product branching fractions of $\jpsi\to\gamma R_1$ with $R_1\to\ksz R_2\to\ksz\ksz\piz$  or $R_1\to\piz R_2\to\ksz\ksz\piz$, the statistical significances of the predominant components are calculated accordingly and summarized in table~\ref{t3}, where the first uncertainties are statistical and the second are systematic.

\begin{figure}[htbp]
	\centering
	\begin{overpic}[width=0.68\textwidth]{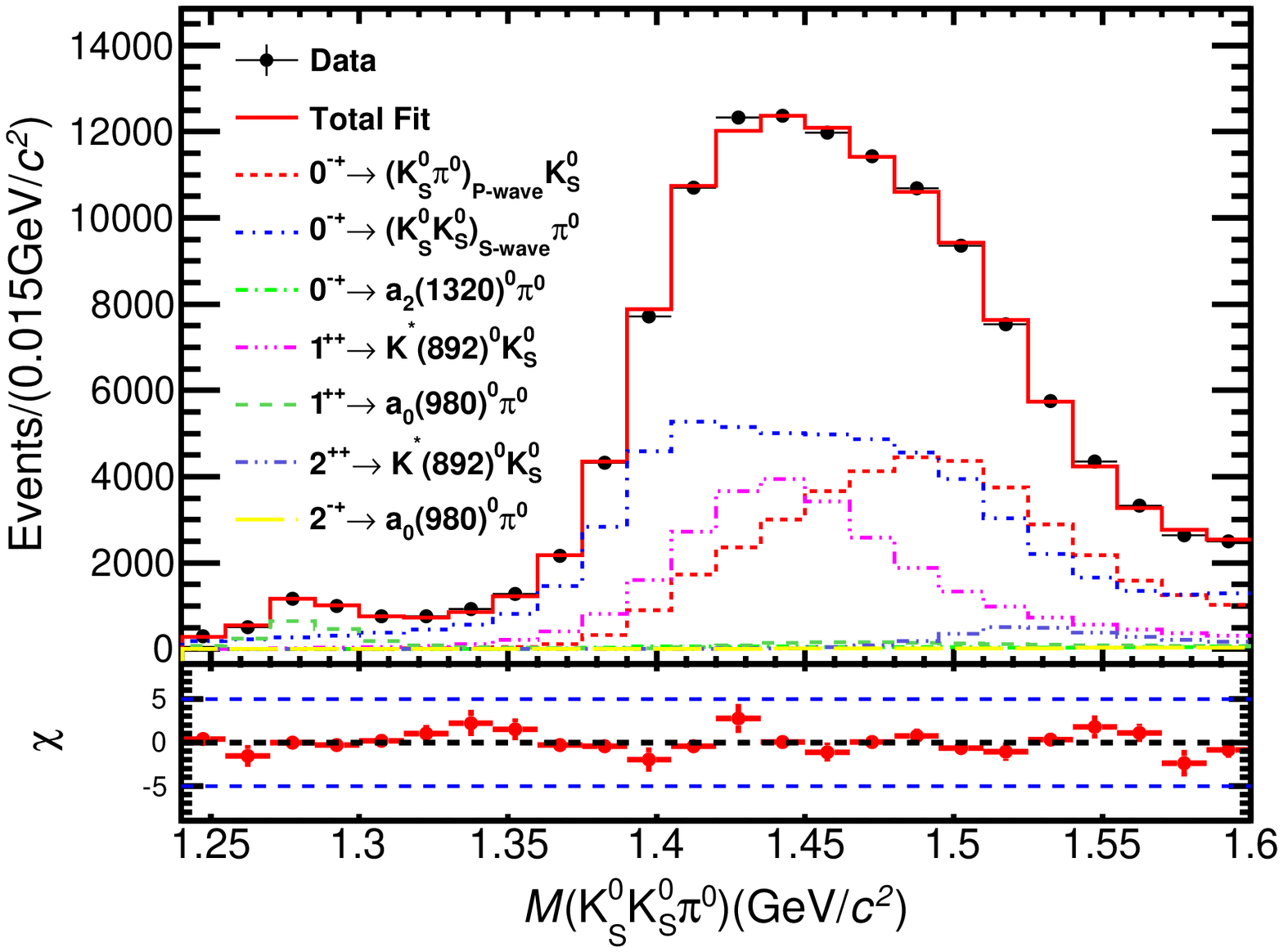}
		\put(80,60){$\bf (a)$}
	\end{overpic}
	\begin{overpic}[width=0.45\textwidth]{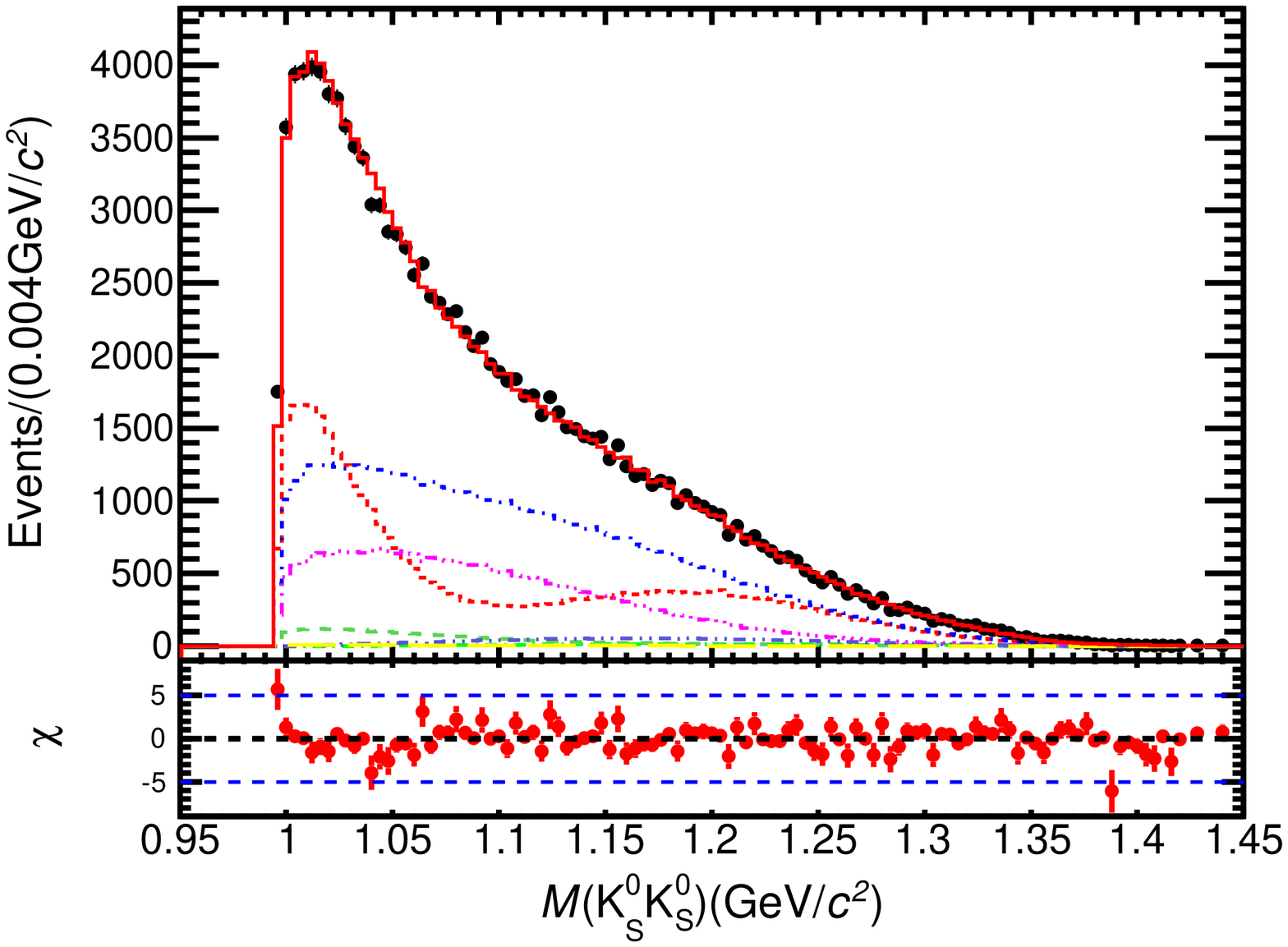}
		\put(75,58){$\bf (b)$}
	\end{overpic}
	\begin{overpic}[width=0.45\textwidth]{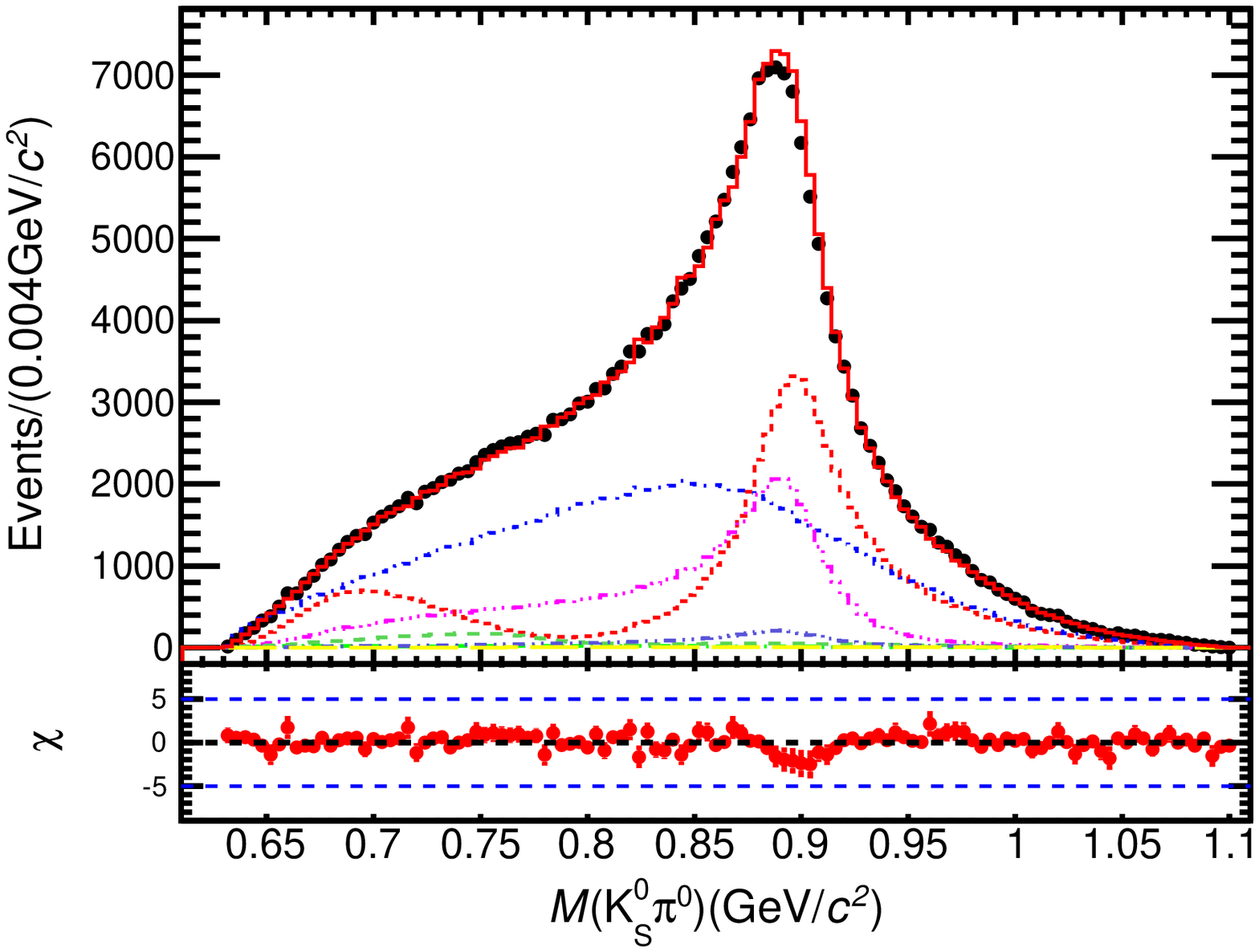}
		\put(75,58){$\bf (c)$}
	\end{overpic}
	\begin{overpic}[width=0.45\textwidth]{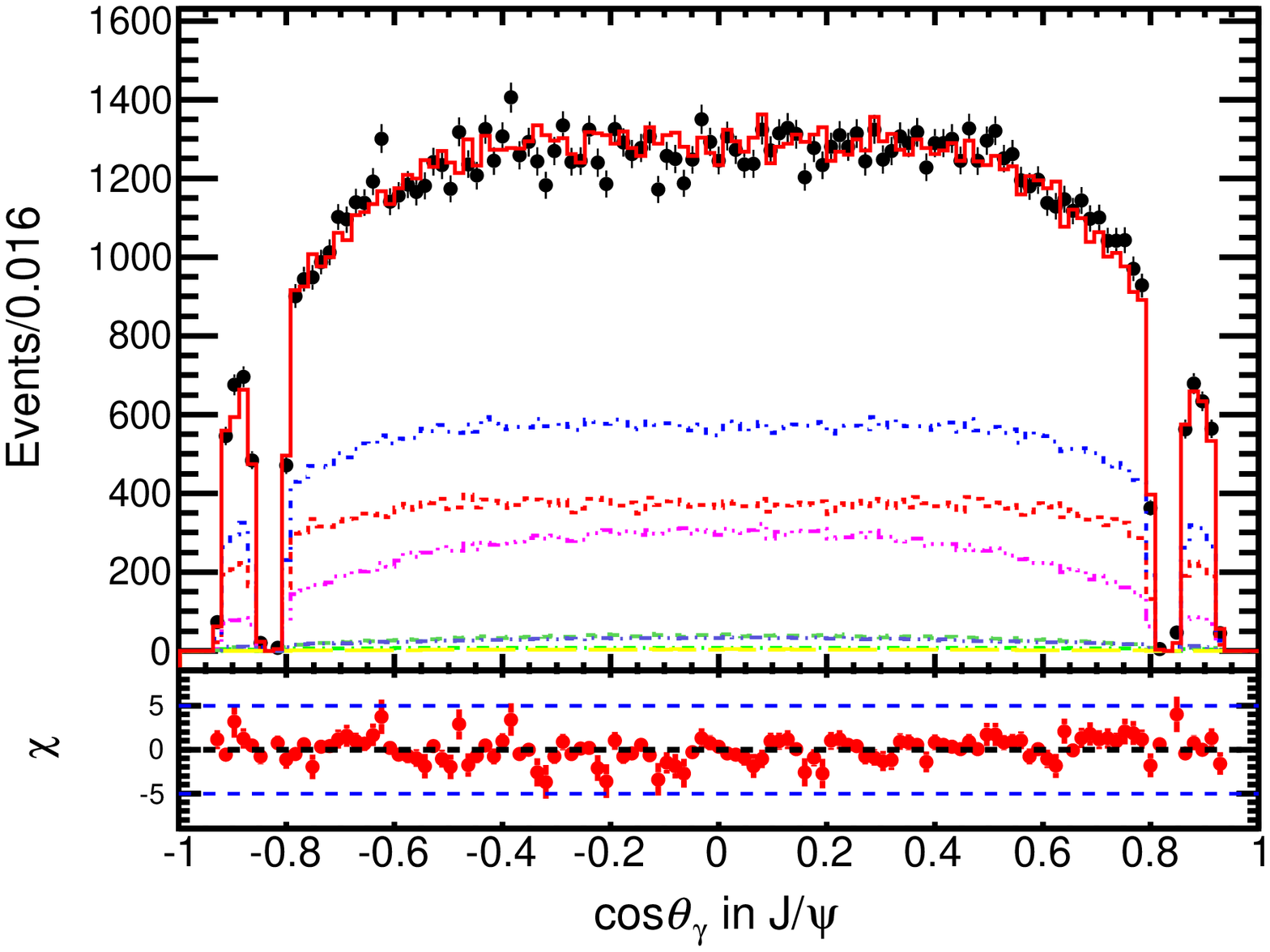}
		\put(75,45){$\bf (d)$}
	\end{overpic}
	\begin{overpic}[width=0.45\textwidth]{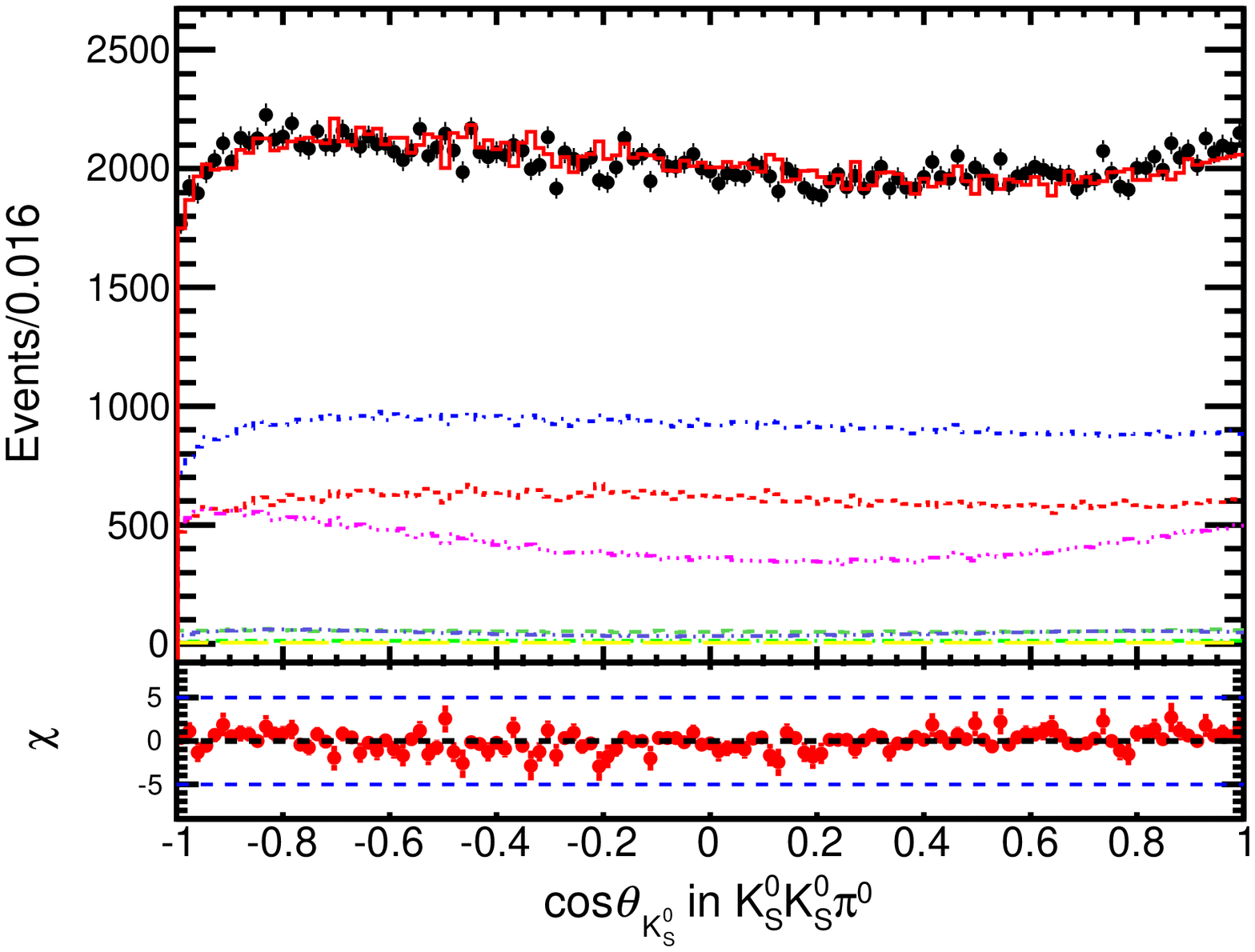}
		\put(75,45){$\bf (e)$}
	\end{overpic}
	\begin{overpic}[width=0.45\textwidth]{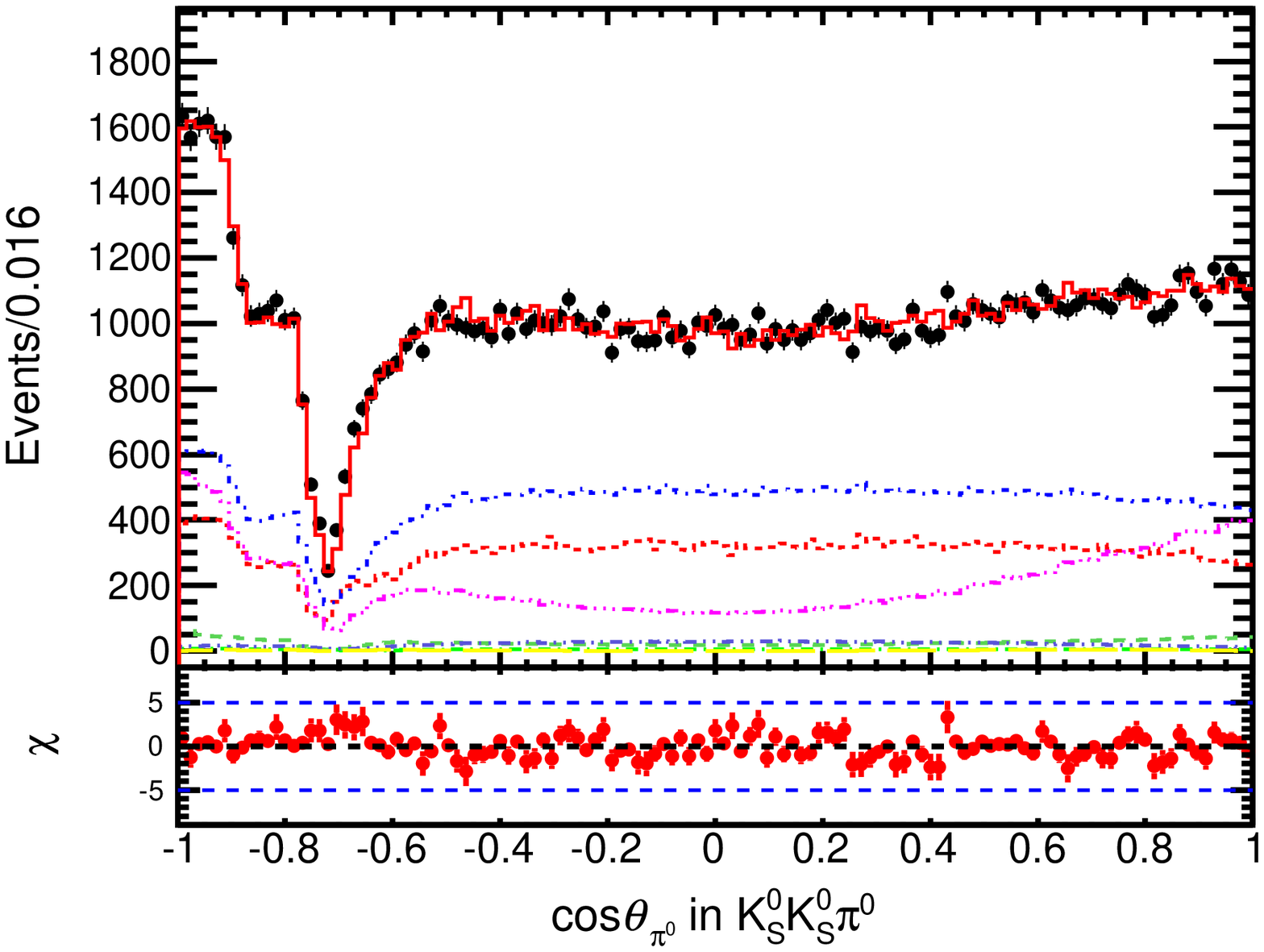}
		\put(75,55){$\bf (f)$}
	\end{overpic}
	\begin{overpic}[width=0.45\textwidth]{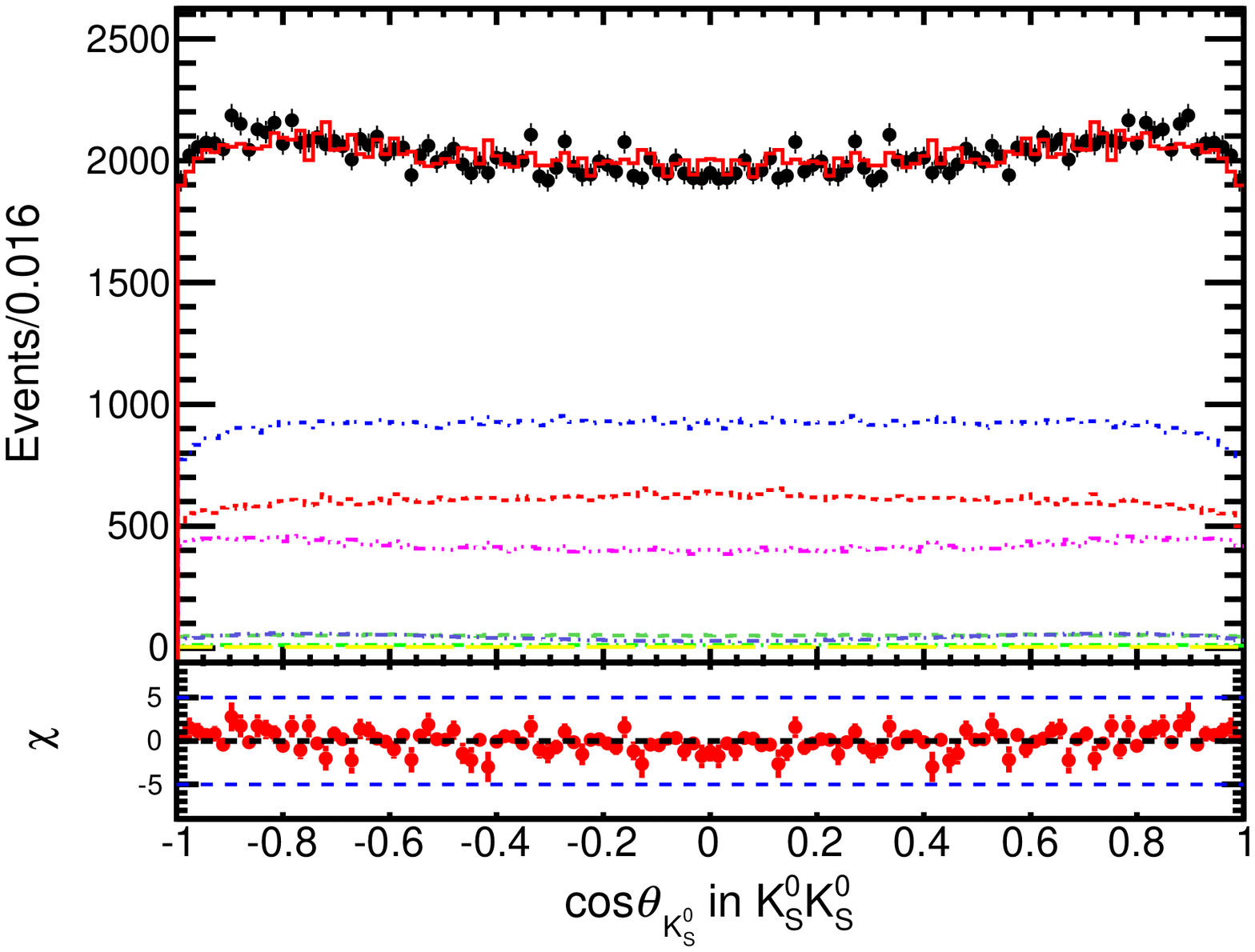}
		\put(75,45){$\bf (g)$}
	\end{overpic}
	\caption{Superposition of data and the MD PWA fit projections for invariant mass distributions of (a) $\ksz\ksz\piz$, (b) $\ksz\ksz$, and (c) $\ksz\piz$. The $\cos\theta$ distributions of (d) $\gamma$ in $\jp$ helicity frame, (e) $\ksz$ and (f) $\piz$ in $\ksz\ksz\piz$ system helicity frame, (g) $\ksz$ in $\ksz\ksz$ system helicity frame. The pull projection of residual is shown beneath each distribution correspondingly.}	
	\label{f3}
\end{figure}

\begin{figure}[htbp]
	\flushleft
	\begin{overpic}[width=0.36\textwidth]{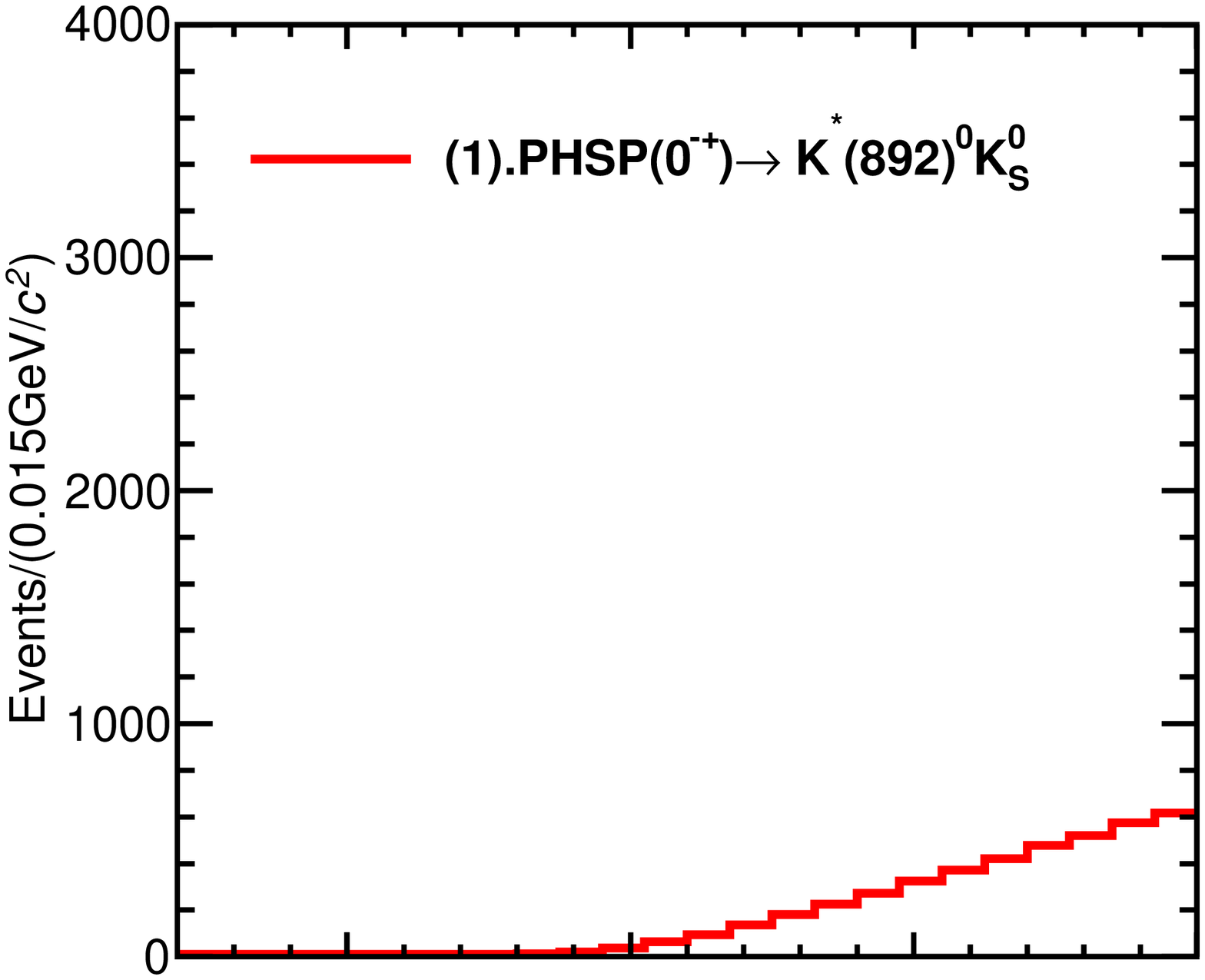}
		\put(28,34){$(a)$}
	\end{overpic}
	\begin{overpic}[width=0.30\textwidth]{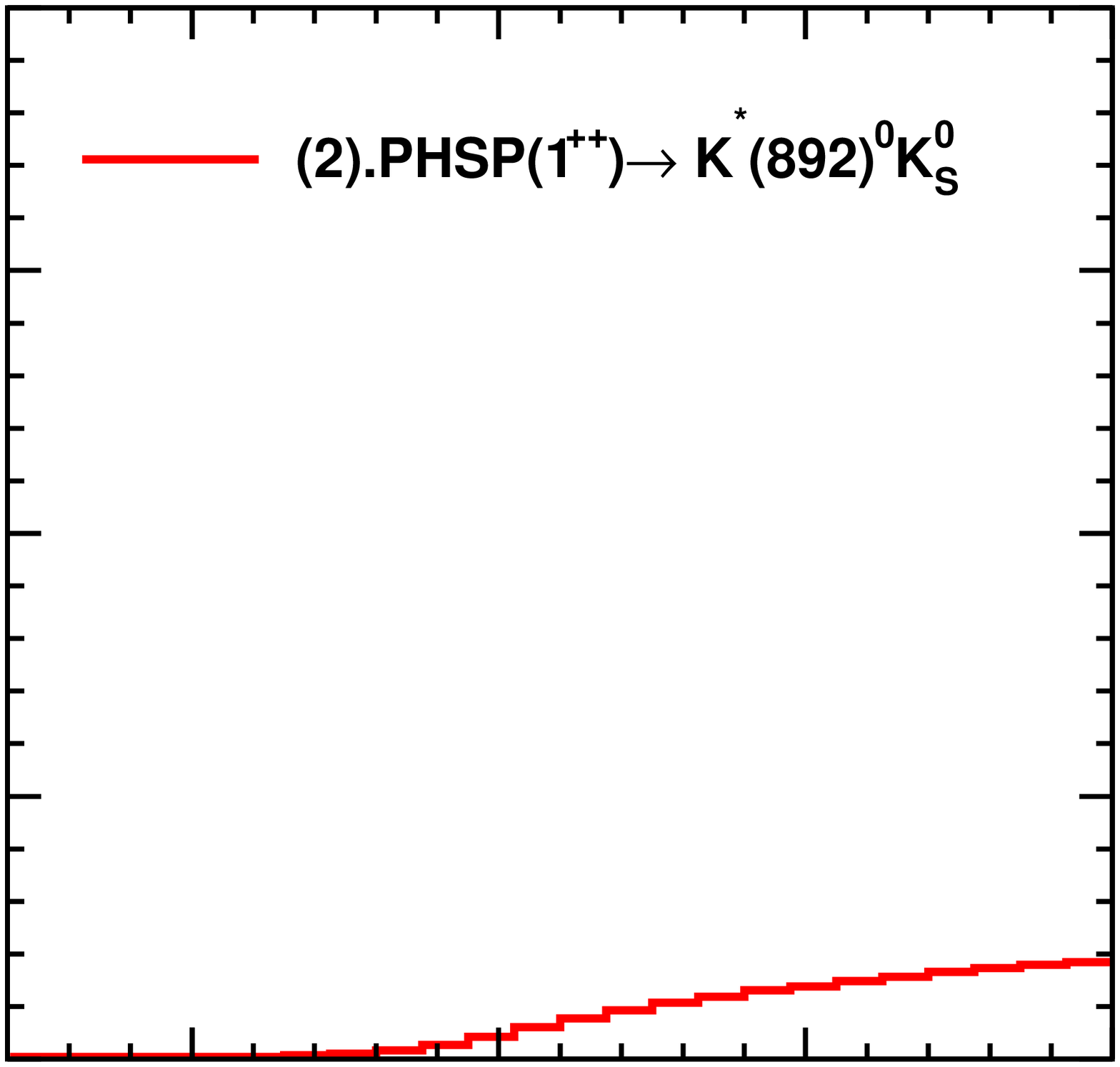}
		\put(18,42){$(b)$}
	\end{overpic}
	\begin{overpic}[width=0.30\textwidth]{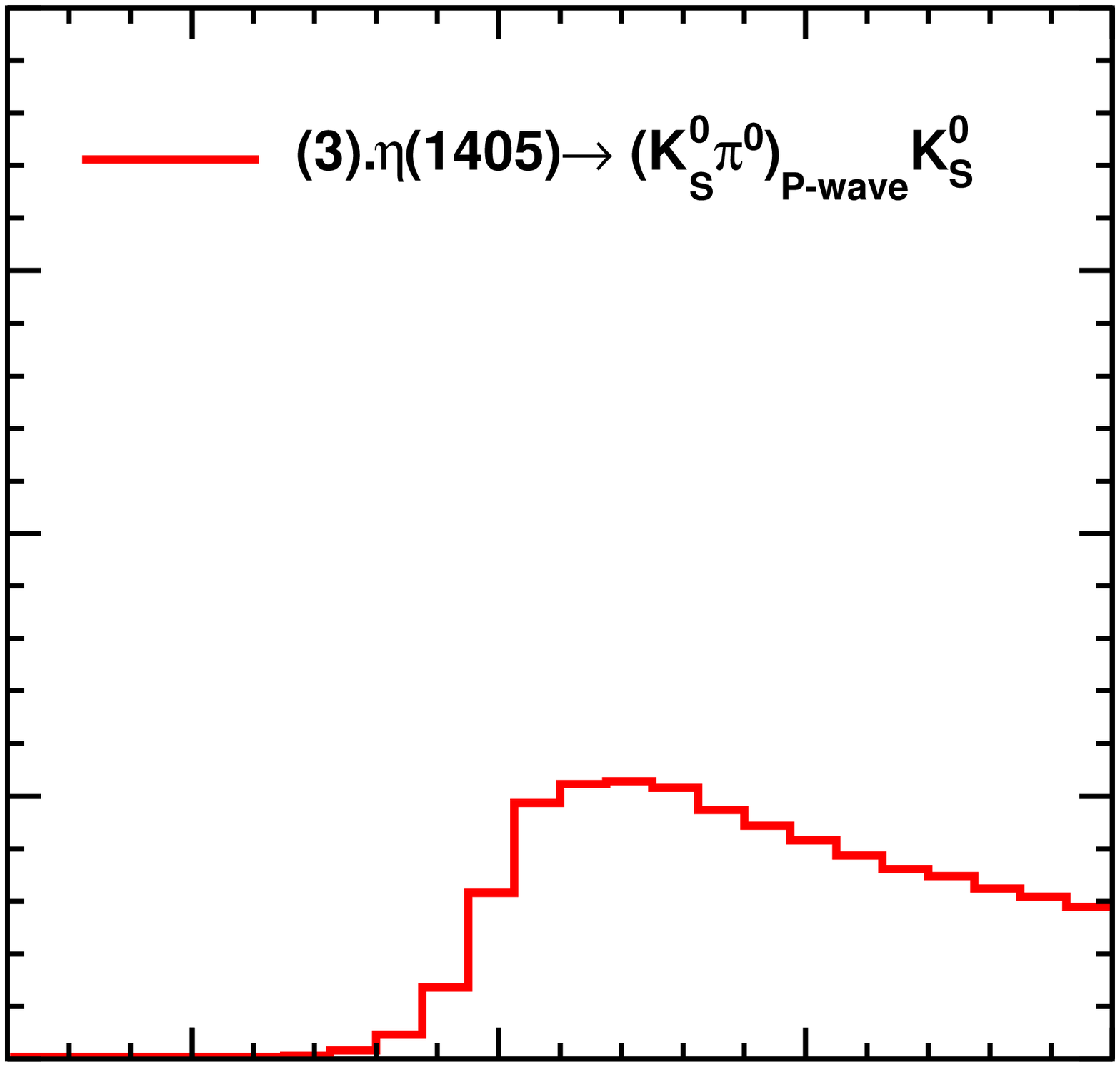}
		\put(18,42){$(c)$}
	\end{overpic}
	\begin{overpic}[width=0.36\textwidth]{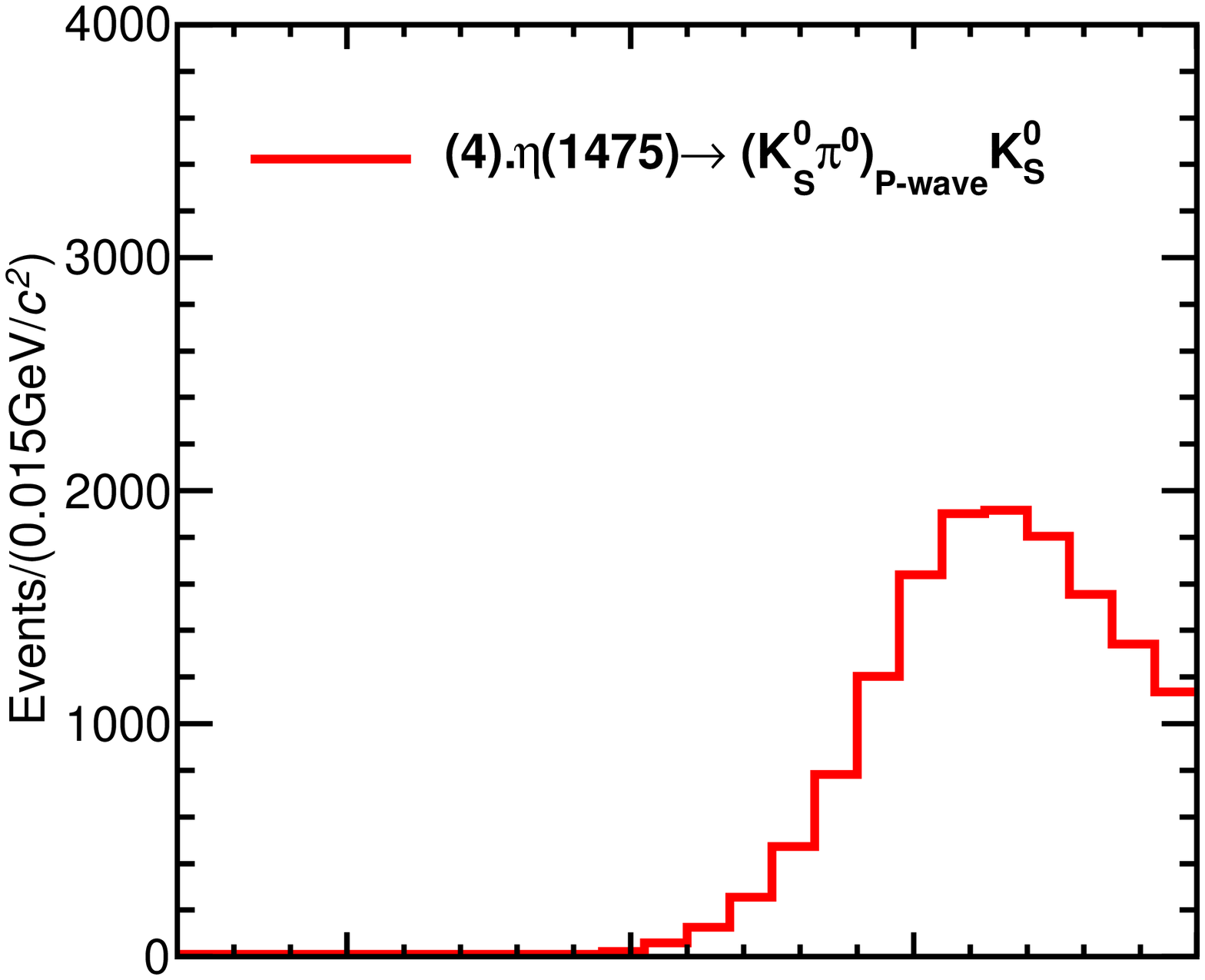}
		\put(28,34){$(d)$}
	\end{overpic}
	\begin{overpic}[width=0.30\textwidth]{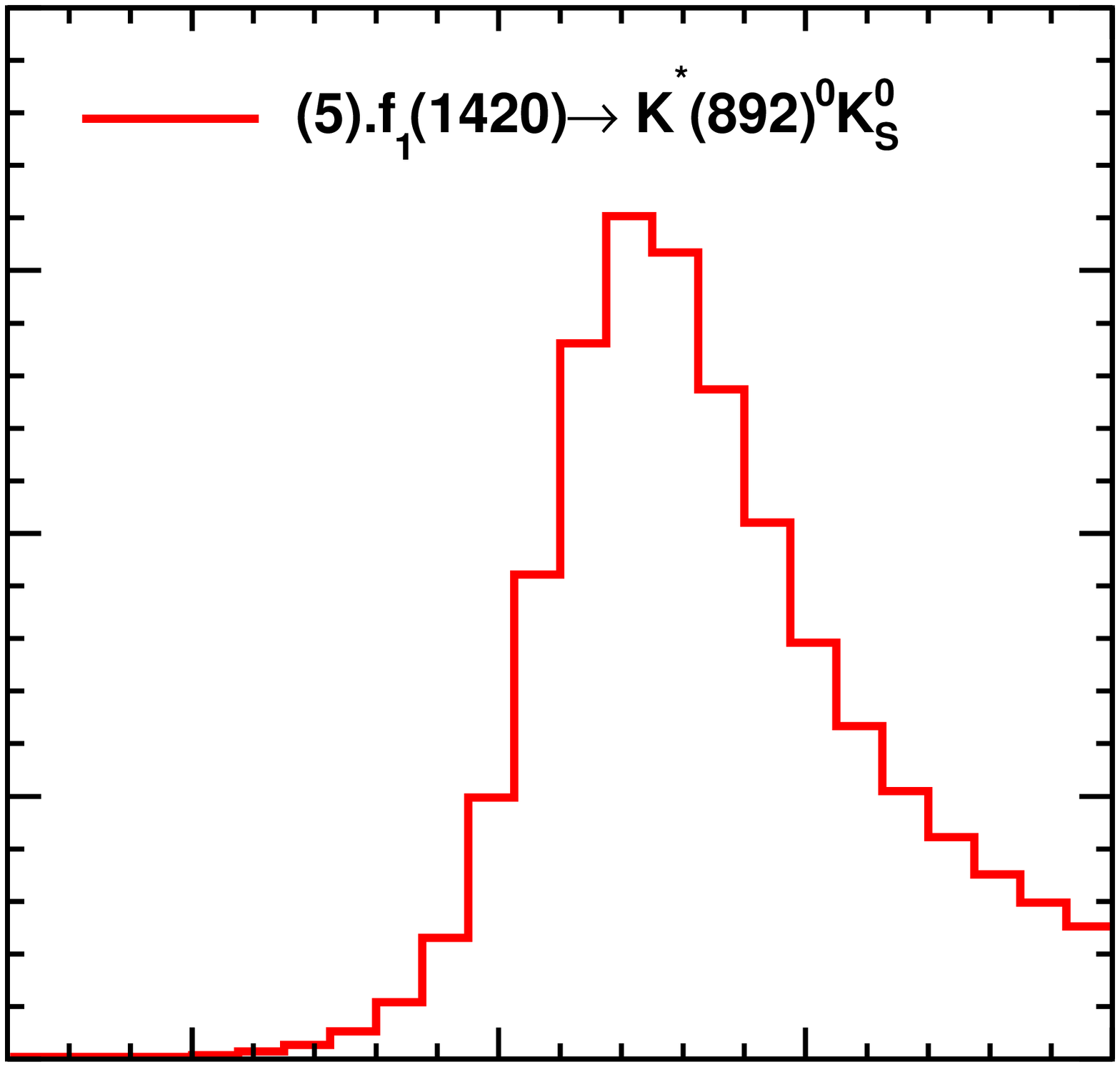}
		\put(18,42){$(e)$}
	\end{overpic}
	\begin{overpic}[width=0.30\textwidth]{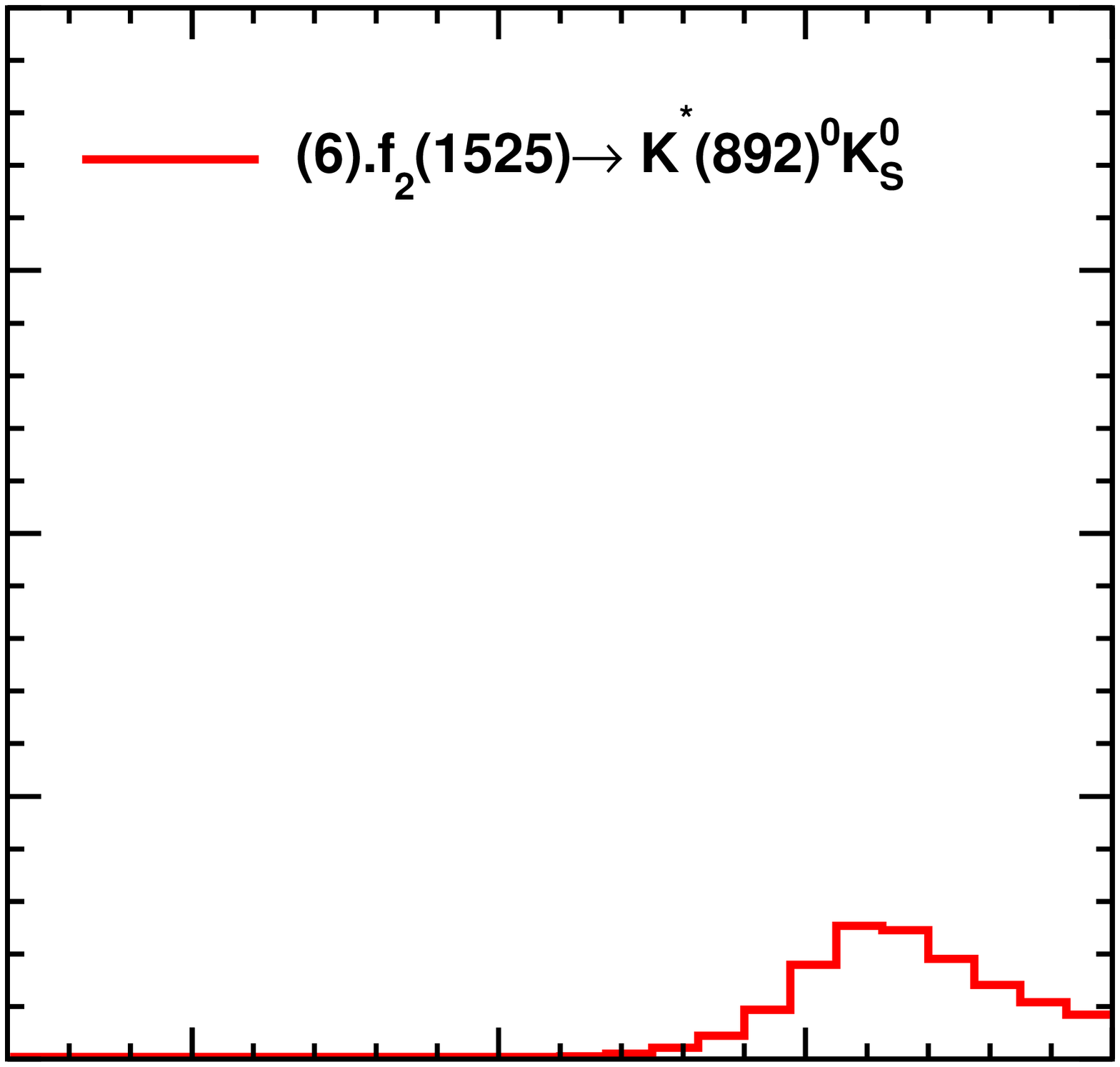}
		\put(18,42){$(f)$}
	\end{overpic}
	\begin{overpic}[width=0.36\textwidth]{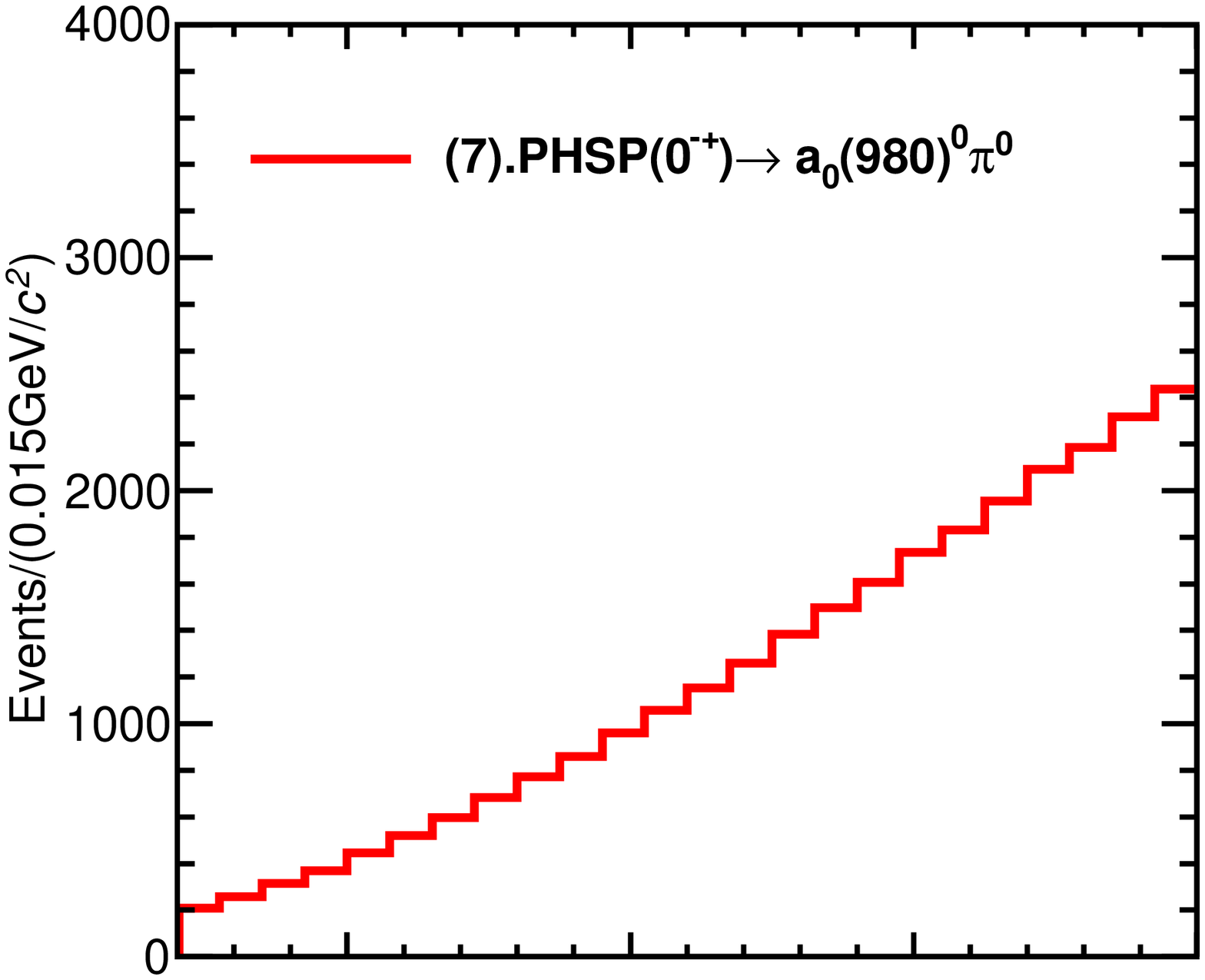}
		\put(28,34){$(g)$}
	\end{overpic}
	\begin{overpic}[width=0.30\textwidth]{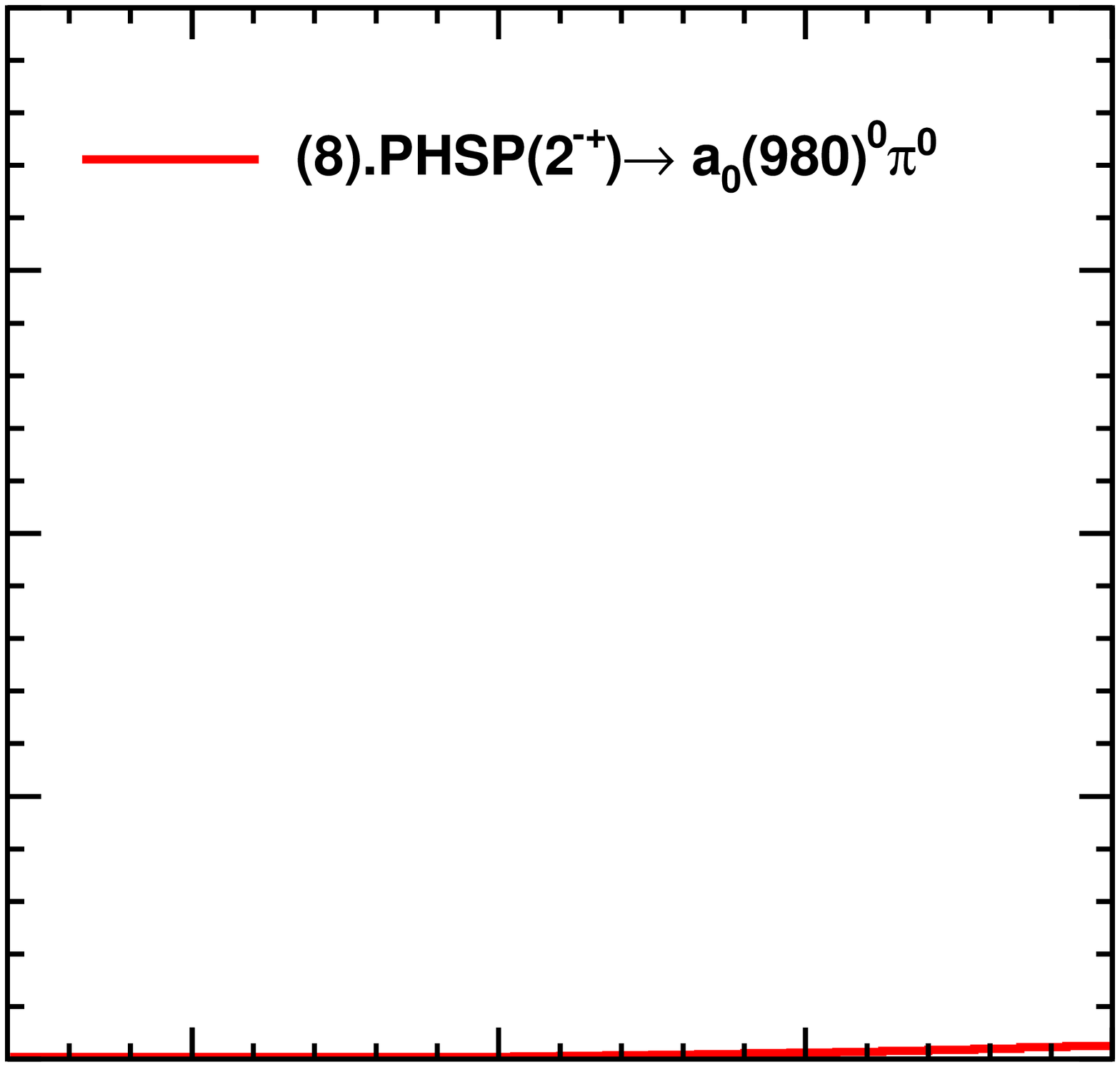}
		\put(18,42){$(h)$}
	\end{overpic}
	\begin{overpic}[width=0.30\textwidth]{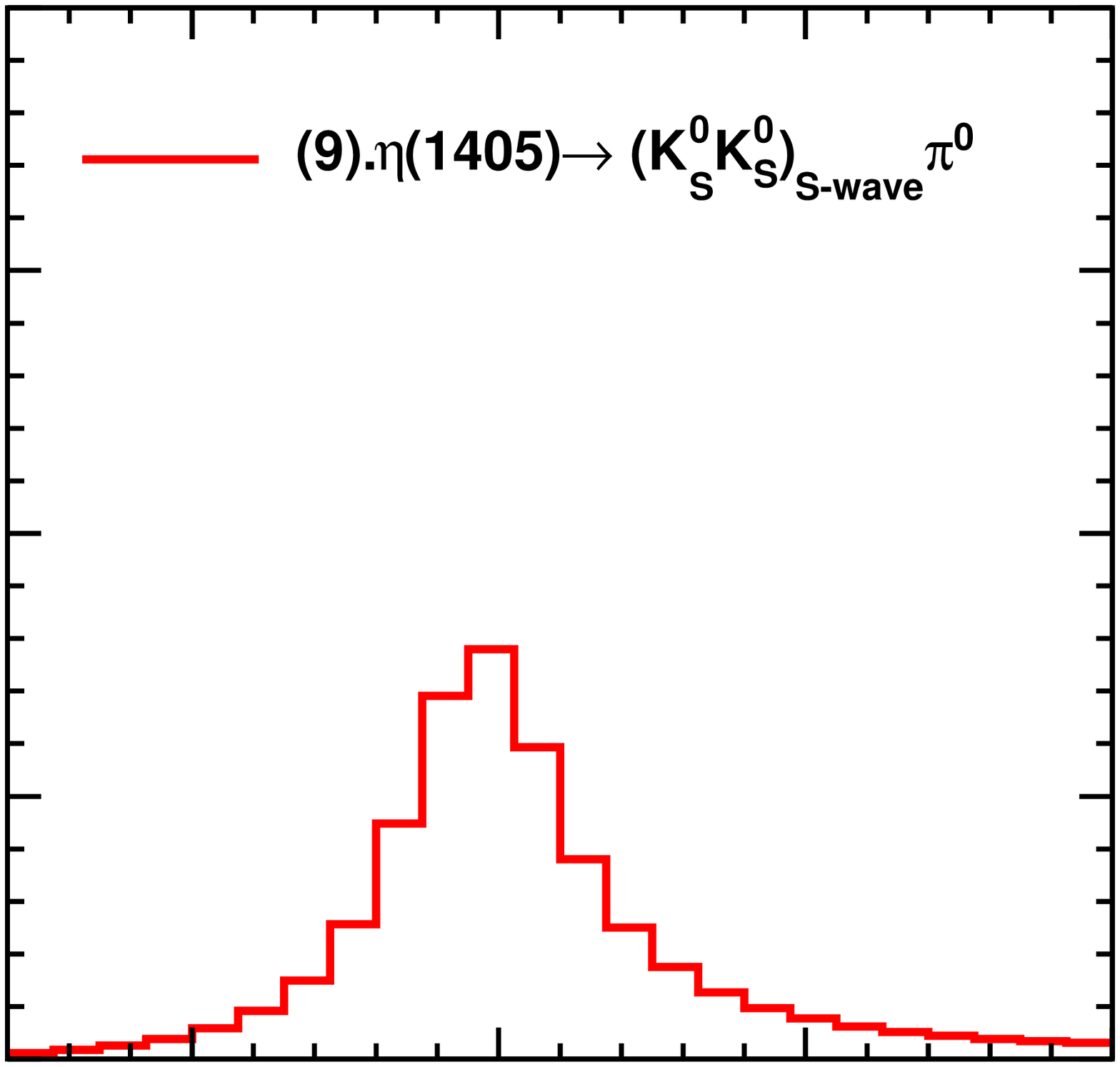}
		\put(18,42){$(i)$}
	\end{overpic}
	\begin{overpic}[width=0.36\textwidth]{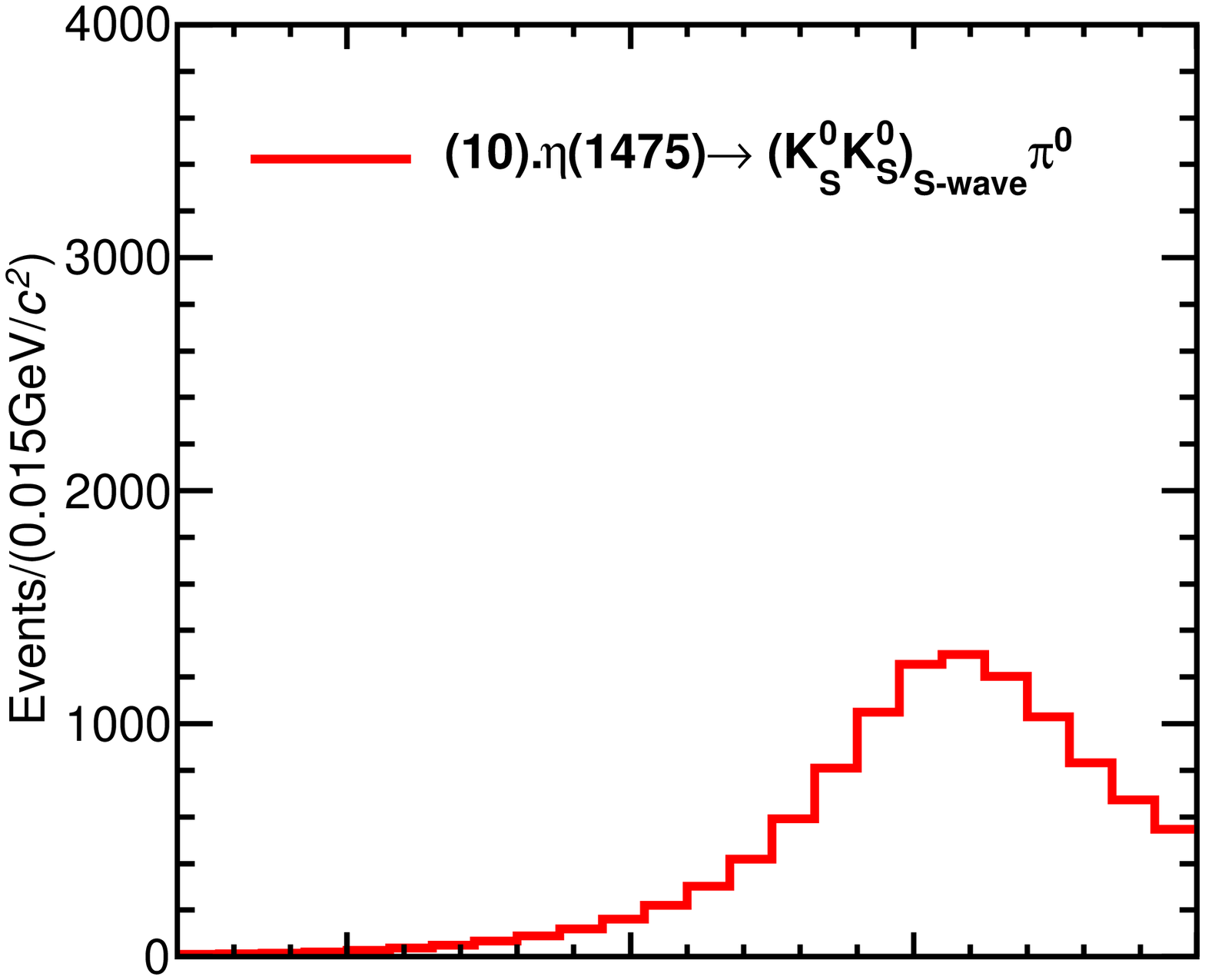}
		\put(28,34){$(j)$}
	\end{overpic}
	\begin{overpic}[width=0.30\textwidth]{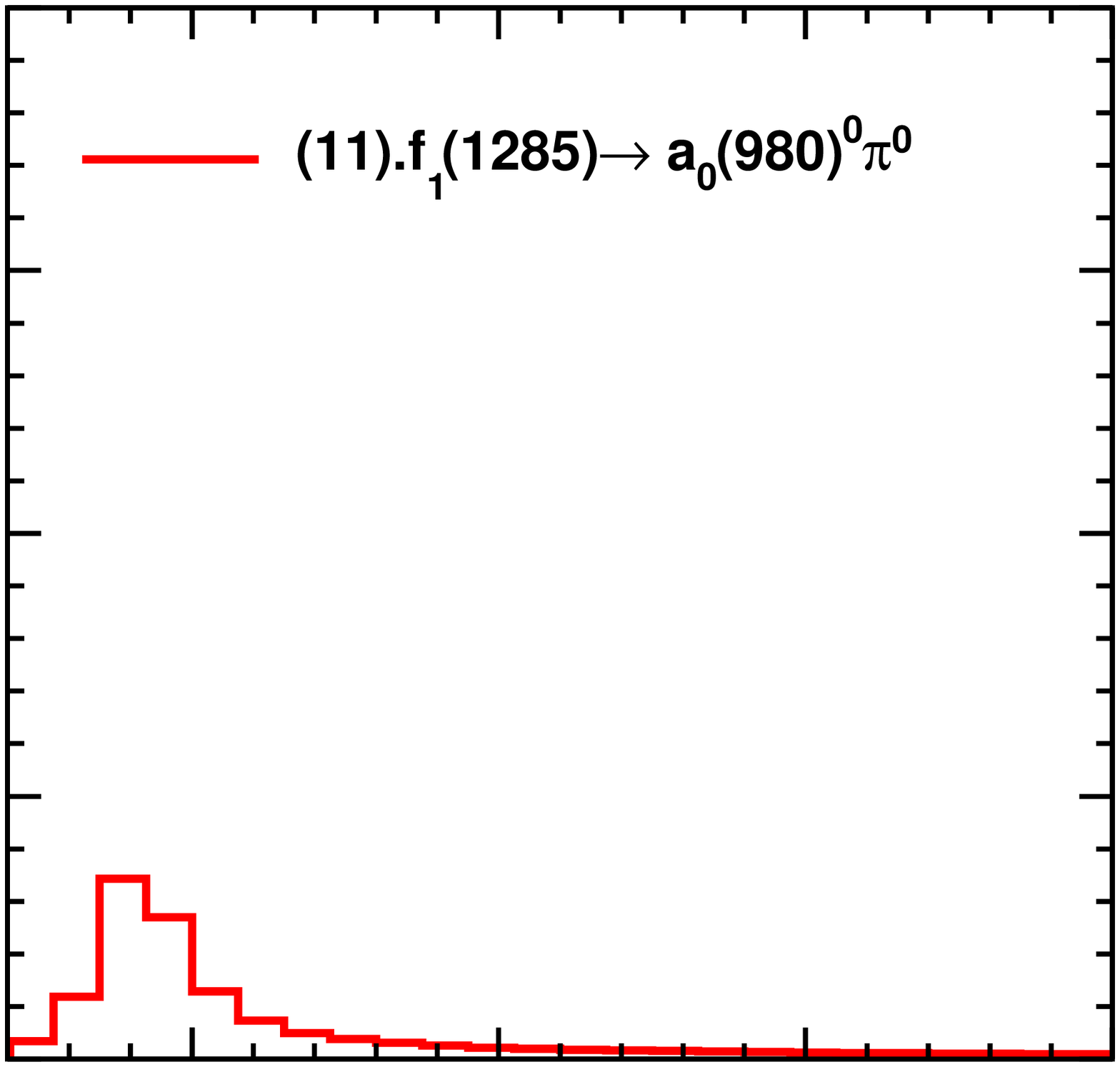}
		\put(18,42){$(k)$}
	\end{overpic}
	\begin{overpic}[width=0.30\textwidth]{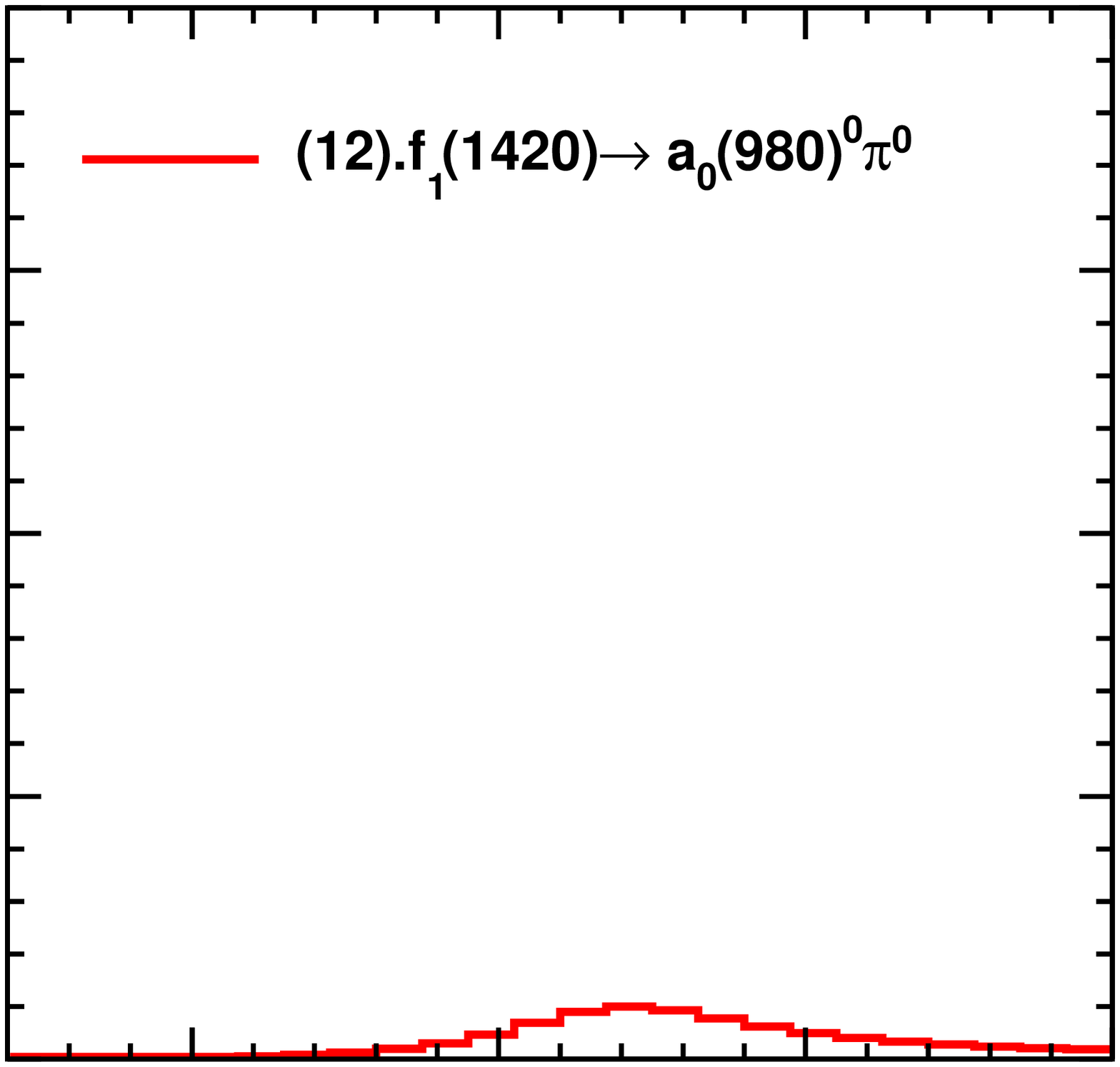}
		\put(18,42){$(l)$}
	\end{overpic}
	\begin{overpic}[width=0.36\textwidth]{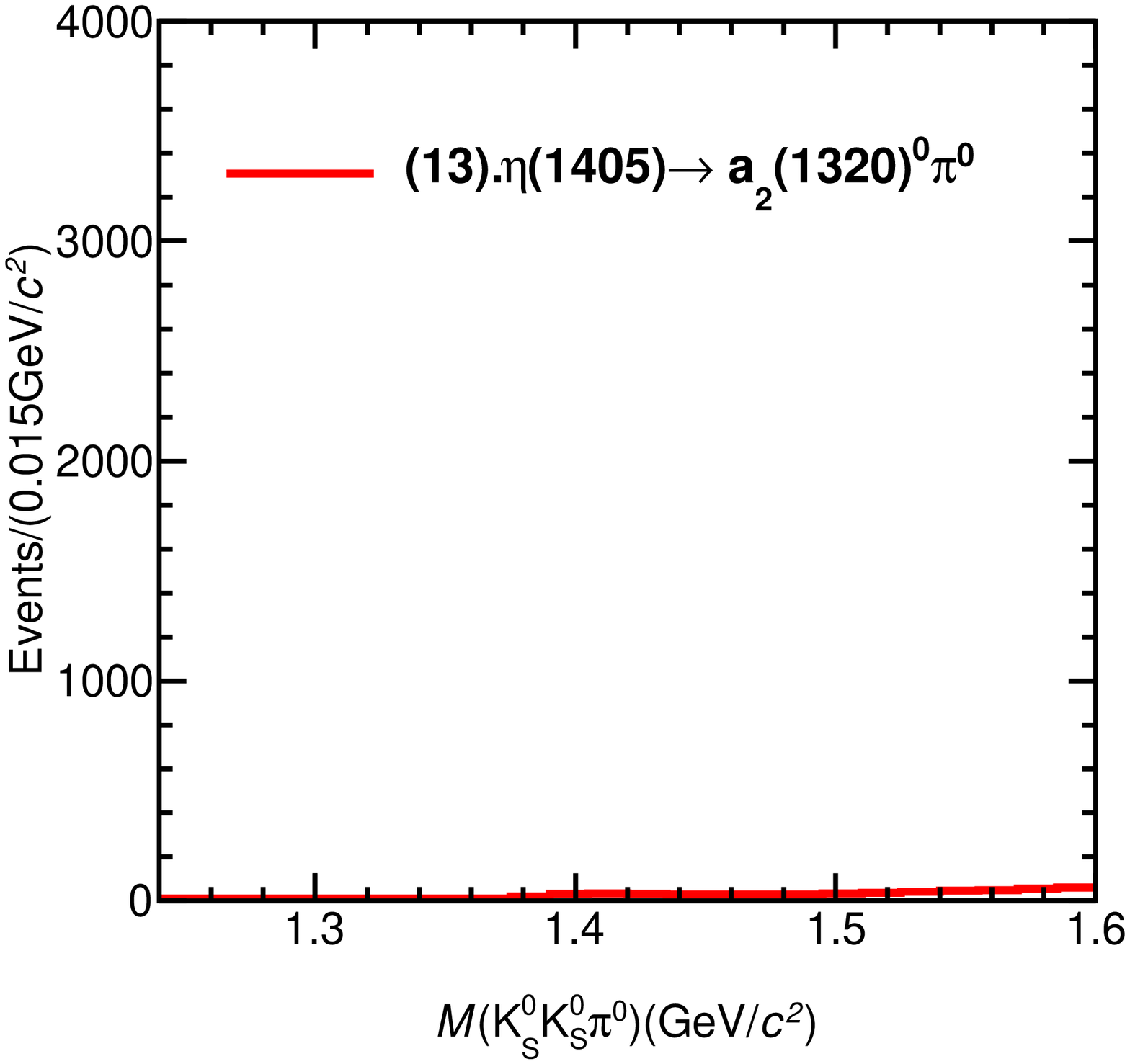}
		\put(28,51){$(m)$}
	\end{overpic}
	\begin{overpic}[width=0.30\textwidth]{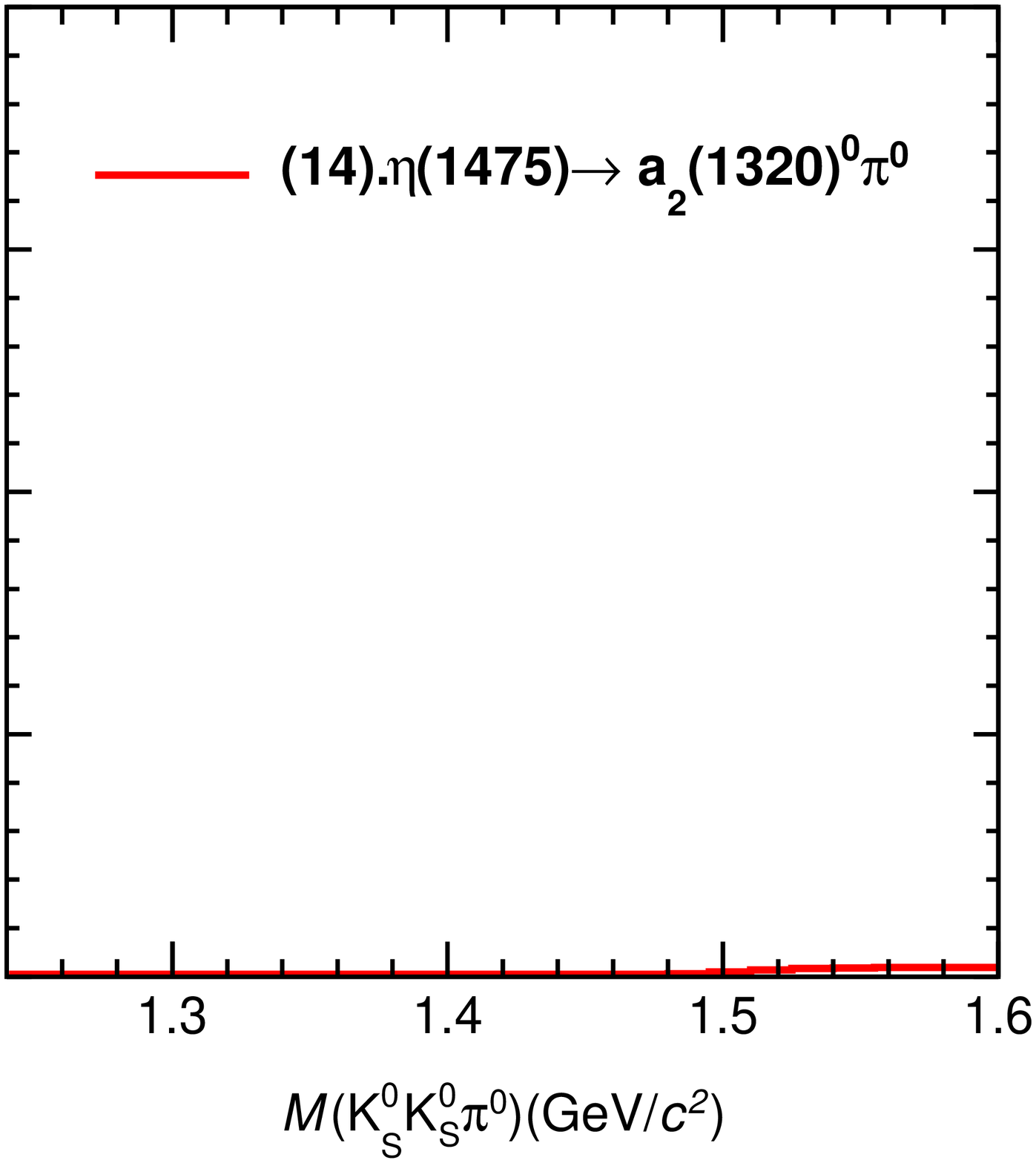}
		\put(16,52){$(n)$}
	\end{overpic}
	\caption{Invariant mass distribution of $\ksz\ksz\piz$ for each of the 14 individual components corresponding to table~\ref{t2} included in the MD PWA nominal solution.}	
	\label{f3_1}
\end{figure}

Figures~\ref{f3}(a)-(c) show distributions of the data and the MD PWA fit projections (PHSP MC sample weighted with the differential cross sections) on the invariant mass of $\ksz\ksz\piz$, $\ksz\ksz$ and $\ksz\piz$, individually, while figures~\ref{f3}(d)-(g) show various angular distributions. As a supplement for figure~\ref{f3}(a) in detail, the figure~\ref{f3_1} shows the $M(\ksz\ksz\piz)$ spectrum for each of the 14 individual components corresponding to table~\ref{t2} included in the MD PWA nominal solution. Additionally, the dips on the $\cos\theta$ distribution of $\gamma$ in $\jp$ helicity frame shown in figure~\ref{f3}(d) are caused by the acceptance of the EMC detector, and the dips on $\cos\theta$ distribution of $\piz$ in $\ksz\ksz\piz$ helicity frame shown in figure~\ref{f3}(f) are correlated with the veto of $\omega$ on the $\gamma\piz$ invariant mass spectrum which is applied on the event selection part. The pull projection of residual beneath each distribution is displayed to demonstrate the comparison between data and MC projection correspondingly, where the $\chi$ is defined as
\begin{equation}
\chi=\dfrac{n_{i}-\nu_{i}}{\sqrt{\nu_{i}}},
\end{equation}
where $n_{i}$ and $\nu_{i}$ are the number of entries for the data and the fit projections of the nominal solution in the $i^{\textrm{th}}$ bin, respectively.

\begin{figure}[htbp]
	\centering
	\begin{overpic}[width=0.66\textwidth]{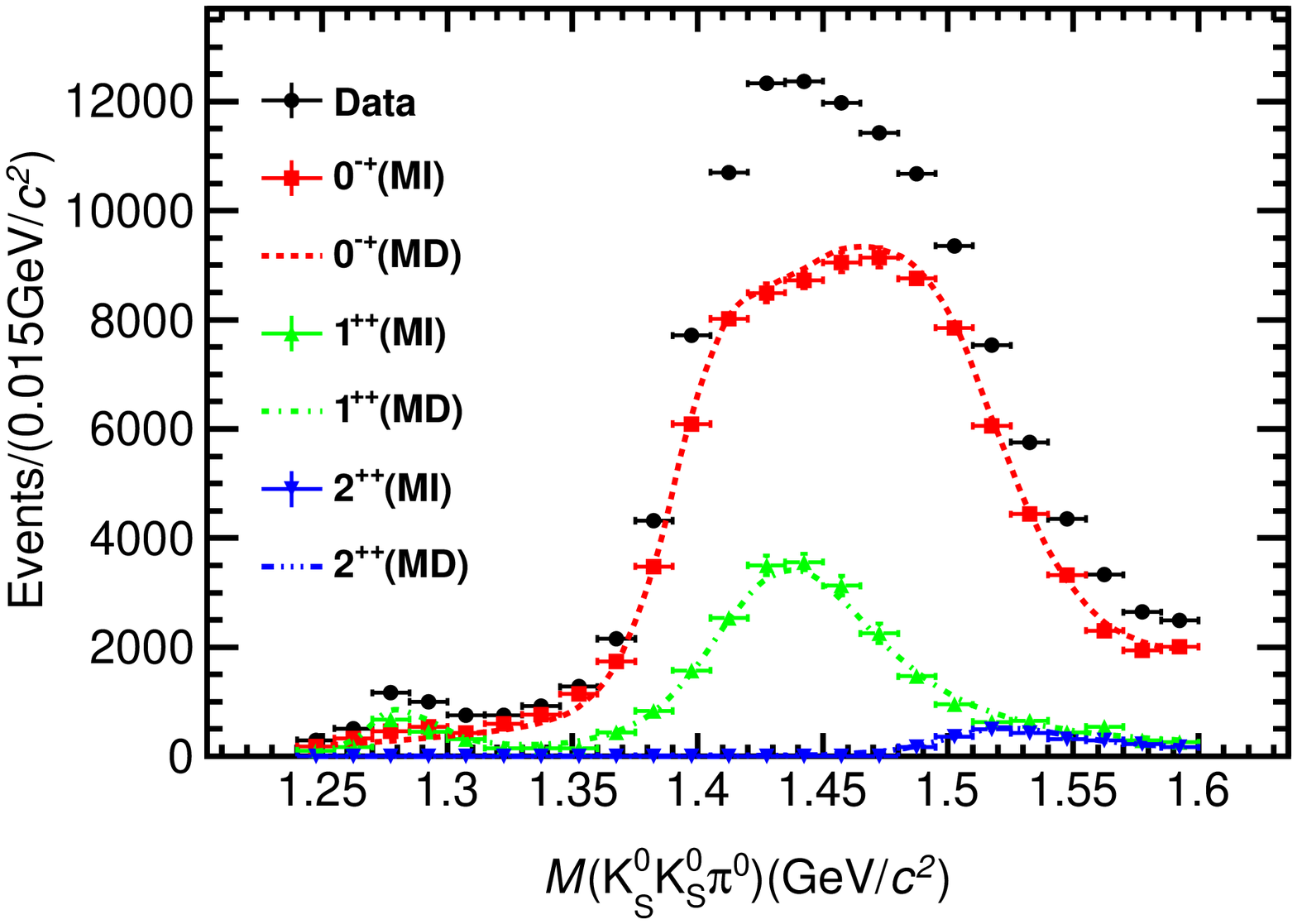}
		\put(80,60){$\bf (a)$}
	\end{overpic}
	\begin{overpic}[width=0.66\textwidth]{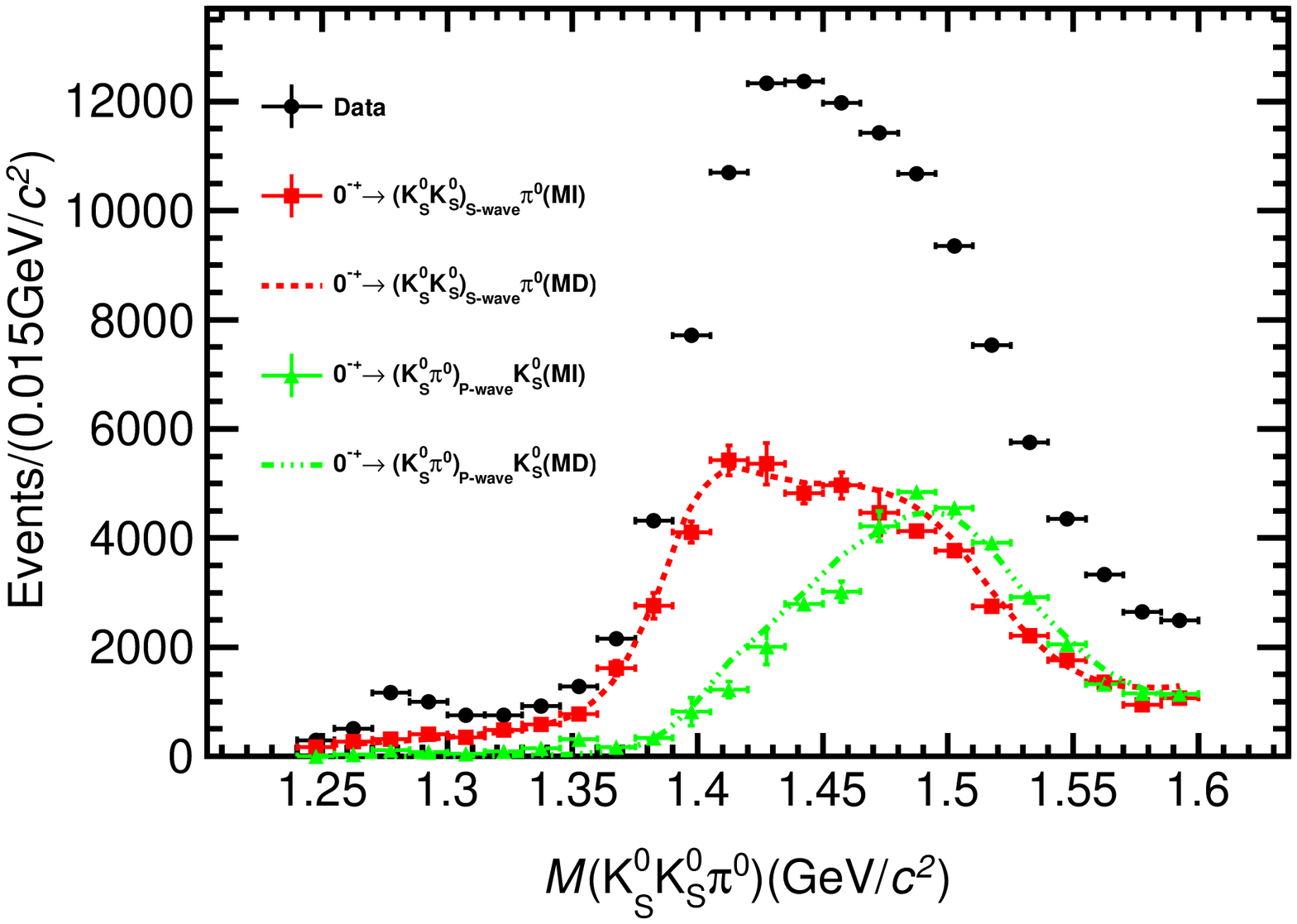}
		\put(80,60){$\bf (b)$}
	\end{overpic}
	\begin{overpic}[width=0.66\textwidth]{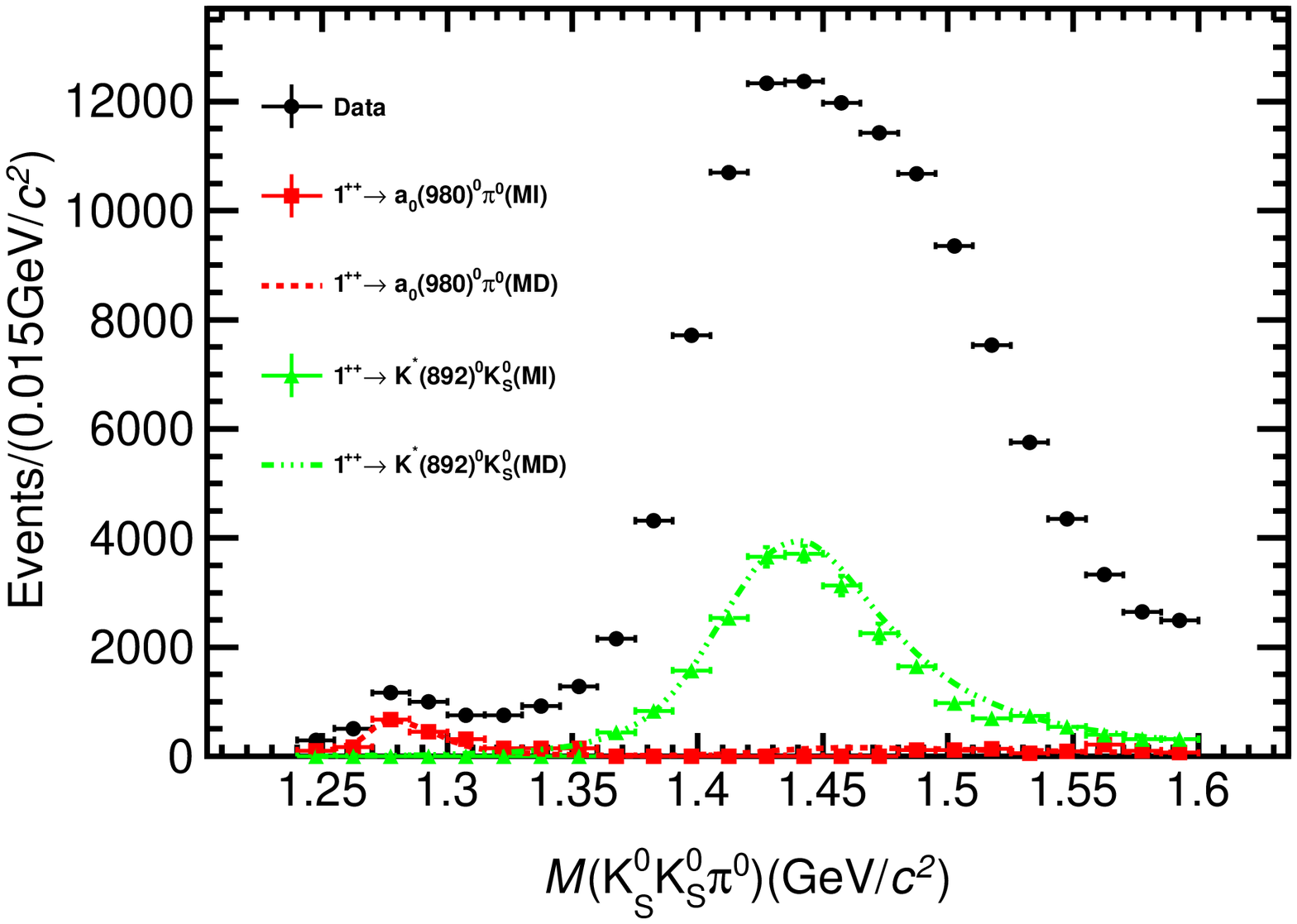}
		\put(80,60){$\bf (c)$}
	\end{overpic}
	\caption{The $\ksz\ksz\piz$ intensity spectra for the (a) dominant spin-parity components, (b) dominant decay modes for the pseudoscalar component, and (c) dominant decay modes for the axial vector component. The dots with error bars and the dotted lines are the intensity obtained from the MI PWA and MD PWA results, respectively. The uncertainty on the total data is statistical, and those of the others are obtained from the PWA in each bin.}	
	\label{f4}
\end{figure}

To validate the consistency between the MI and MD PWA results, comparisons of the $\ksz\ksz\piz$ intensity spectra for individual spin-parity components obtained by the two PWA approaches are shown in figure~\ref{f4}(a), where good agreement is observed. We also compare the distributions of the pseudoscalar component decaying into $(\ksz\ksz)_\textrm{S-wave}\piz$, $(\ksz\piz)_\textrm{P-wave}\ksz$ and the axial vector component decaying into $a_{0}(980)^{0}\piz$, $K^{*}(892)^{0}\ksz$ between the two approaches, individually, as shown in figures~\ref{f4}(b) and (c), where good agreement is also achieved. 

As described previously, the lineshape and dynamics of the pseudoscalar component are of great interest in this analysis. Therefore, different scenarios are tried in the MD analysis.  An alternative fit is performed by replacing the two pseudoscalar states $\eta(1405)$ and $\eta(1475)$ with one resonance with floating resonance parameters. The result shows the NLL value is worsen by 1564.4. The result from the TSM test by considering both $\eta(1440)$ and $f_{1}(1420)$ TSM contributions in the MD PWA, where the intermediate on-shell $K^{*}K+c.c$ can exchange an on-shell kaon or anti-kaon and then rescatter to $a_{0}(980)^{0}\piz$ in this process~\cite{tsm3,tsm4}, shows the NLL value is worse by 1169.4. The description of the $\ksz\ksz\piz$ invariant mass spectrum deteriorates significantly compared with the two states scenario.

\section{SYSTEMATIC UNCERTAINTIES}

\begin{table}[htbp]
	\normalsize
	\renewcommand\arraystretch{1.2}
	\begin{center}
		\begin{tabular}{cc}
			\hline
			\hline
			Source &Uncertainty(\%) \\
			\hline
			Photon detection &3.0 \\
			
			$\ksz$ reconstruction &2.0 \\
			
			Kinematic fit &1.6 \\
			
			Number of $\jp$ events &0.4 \\
			
			Intermediate branching fractions &0.2 \\
			\hline
			Total &4.0 \\
			\hline
			\hline
		\end{tabular}
	\end{center}
	\caption{Summary of systematic uncertainties related to event selection for the determination of the branching fraction (relative uncertainties).}
	\label{t4}
\end{table}	

In the MD PWA, the sources of systematic uncertainties can be classified into two categories, {\it i.e.}, those associated with event reconstruction and selection, and those related to the PWA.

The systematic uncertainties associated with the event reconstruction and selection are applied in the branching fraction measurement only, and mainly include those from the photon detection efficiency, $\ksz$ reconstruction efficiency, kinematic fit, the number of $\jpsi$ events, and the branching fractions of $\ksz\to\pip\pim$ and $\piz\to\gamma\gamma$. The systematic uncertainty of the photon detection efficiency is 1.0\% per photon, obtained by studying the control sample of $\jp\to\rho\pi$~\cite{47}. Therefore, 3\% is assigned as the corresponding uncertainty for three photons. The systematic uncertainty of $\ksz$ reconstruction efficiency is studied with the control samples of $\jp\to K^{*\pm}K^{*\mp}$ and $\jp\to\phi\ksz K^{\pm}\pi^{\mp}$, and is 1\% per $\ksz$ candidate~\cite{48}. Therefore, 2\% is assigned as the corresponding uncertainty for the two $\ksz$ candidates. The systematic uncertainty associated with the kinematic fit is 1.6\%, obtained by the track helix parameter correction method~\cite{49}. The total number of $\jpsi$ decays is obtained using inclusive hadron events, and the uncertainty is 0.4\%~\cite{njp}. The uncertainties associated with the decay branching fractions of $\ksz\to\pi^+\pi^-$ and $\piz\to\gamma\gamma$ are quoted from the PDG~\cite{13}. All the uncertainties associated with event reconstruction and selection are summarized in table~\ref{t4}, and the total uncertainty is 4\%, which is the quadrature sum of the individual values.

\begin{table}[htbp]
	\begin{center}
		\renewcommand\arraystretch{1.3}
		\small
		\begin{tabular}{ccccccccccc}
			\hline
			\hline
			\multirow{2}{*}{Source} &\multicolumn{2}{c}{$\eta(1405)$}	 &\multicolumn{2}{c}{$\eta(1475)$} &\multicolumn{2}{c}{$f_{1}(1285)$} &\multicolumn{2}{c}{$f_{1}(1420)$} &\multicolumn{2}{c}{$f_{2}(1525)$} 
			\\
			\cline{2-11}
			
			& $\Delta$M &$\Delta\Gamma$ & $\Delta$M & $\Delta\Gamma$ & $\Delta$M &$\Delta\Gamma$ & $\Delta$M &$\Delta\Gamma$ & $\Delta$M &$ \Delta\Gamma$ \\
			\hline
			Breit-Wigner formula &$+11.0$ &$_{-11.8}^{+5.2}$  &$_{-31.6}^{+15.5}$ &$+14.6$ &$_{-1.5}^{+0.8}$ &$_{-2.9}^{+0.6}$ &$_{-0.5}^{+27.9}$ &$_{-9.9}^{+13.3}$ &$_{-7.5}^{+2.6}$ &$_{-5.1}^{+2.0}$ \\
			
			Resonance parameters &$+2.5$ &$+1.8$  &$-5.6$ &$-10.8$ &$+0.3$ &$+1.8$ &$-0.2$ &$-2.4$ &$+1.6$ &$-3.0$ \\
			
			Extra components &$_{-0.3}^{+0.1}$ &$_{-2.0}^{+0.6}$  &$_{-2.2}^{+0.1}$ &$_{-1.8}^{+2.3}$ &$_{-0.2}^{+0.8}$ &$_{-0.1}^{+5.2}$ &$_{-0.5}^{+1.5}$ &$_{-3.9}^{+2.7}$ &$+1.1$ &$-1.4$ \\
			\hline
			Total &$_{-0.3}^{+11.3}$ &$_{-12.0}^{+5.5}$  &$_{-32.2}^{+15.5}$ &$_{-10.9}^{+14.8}$ &$_{-1.5}^{+1.2}$ &$_{-2.9}^{+5.5}$ &$_{-0.7}^{+27.9}$ &$_{-10.9}^{+13.6}$ &$_{-7.5}^{+3.2}$ &$_{-6.1}^{+2.0}$ \\
			\hline
			\hline								
		\end{tabular}
	\end{center}
	\caption{Summary of the systematic uncertainty sources and their corresponding contributions to the systematic uncertainties in masses ($\textrm{MeV}/{c}^2$) and widths ($\textrm{MeV}$) of $\eta(1405)$, $\eta(1475)$, $f_{1}(1285)$, $f_{1}(1420)$ and $f_{2}(1525)$, denoted as $\Delta$M and $\Delta\Gamma$, respectively.}
	\label{t5}
\end{table}

\begin{sidewaystable}[htbp]
	\begin{center}
		\normalsize
		\renewcommand\arraystretch{1.3}
		\resizebox{\textwidth}{28mm}{	
			\begin{tabular}{clcccc|c}
				\hline
				\hline
				Resonance &Decay Mode &Event selection &Breit-Wigner formula &Resonance parameters &Extra components &Total \\
				\hline
				\multirow{2}{*}{$\eta(1405)$}
				&$\jp\to\gamma\eta(1405)\to\gamma\ksz(\ksz\piz)_\textrm{P-wave}\to\gamma\ksz\ksz\piz$  &$\pm4.0$  &$_{-56.8}^{+28.1}$ &$+20.0$ &$_{-7.8}^{+2.8}$ &$_{-57.5}^{+34.8}$  \\
				\cline{2-7}
				&$\jp\to\gamma\eta(1405)\to\gamma(\ksz\ksz)_\textrm{S-wave}\piz\to\gamma\ksz\ksz\piz$  &$\pm4.0$  &$_{-7.5}^{+19.0}$ &$+53.1$ &$_{-10.0}^{+6.1}$ &$_{-13.1}^{+56.9}$  \\
				\hline
				
				\multirow{2}{*}{$\eta(1475)$} 
				&$\jp\to\gamma\eta(1475)\to\gamma\ksz(\ksz\piz)_\textrm{P-wave}\to\gamma\ksz\ksz\piz$  &$\pm4.0$  &$_{-42.5}^{+59.4}$ &$+8.9$ &$_{-3.1}^{+5.8}$ &$_{-42.8}^{+60.5}$ \\
				\cline{2-7}
				&$\jp\to\gamma\eta(1475)\to\gamma(\ksz\ksz)_\textrm{S-wave}\piz\to\gamma\ksz\ksz\piz$ &$\pm4.0$  &$_{-7.4}^{+5.5}$ &$-13.7$ &$_{-4.3}^{+7.8}$ &$_{-16.6}^{+10.3}$   \\
				\hline
				
				$f_{1}(1285)$ &$\jp\to\gamma f_{1}(1285)\to\gamma a_{0}(980)^{0}\piz\to\gamma\ksz\ksz\piz$  &$\pm4.0$  &$_{-11.2}^{+1.0}$ &$+4.3$ &$_{-2.5}^{+39.6}$ &$_{-12.2}^{+40.0}$  \\
				\hline
				
				\multirow{2}{*}{$f_{1}(1420)$} 
				&$\jp\to\gamma f_{1}(1420)\to\gamma K^{*}(892)^{0}\ksz\to\gamma\ksz\ksz\piz$  &$\pm4.0$  &$_{-16.8}^{+2.5}$ &$+6.4$ &$_{-1.2}^{+6.1}$ &$_{-17.3}^{+10.0}$  \\
				\cline{2-7}
				&$\jp\to\gamma f_{1}(1420)\to\gamma a_{0}(980)^{0}\piz\to\gamma\ksz\ksz\piz$  &$\pm4.0$  &$_{-41.8}^{+15.8}$ &$+7.8$ &$+47.8$ &$_{-42.0}^{+51.1}$ \\
				\hline
				
				$f_{2}(1525)$
				&$\jp\to\gamma f_{2}(1525)\to\gamma K^{*}(892)^{0}\ksz\to\gamma\ksz\ksz\piz$  &$\pm4.0$  &$_{-4.7}^{+15.3}$ &$-3.3$ &$_{-0.5}^{+1.9}$ &$_{-7.0}^{+15.9}$  \\
				\hline
				\hline		
		\end{tabular}}
	\end{center}
	\caption{Summary of the systematic error sources and their corresponding contributions to the branching fractions (relative uncertainties, in \%).}
	\label{t6}	
\end{sidewaystable}

The systematic uncertainties associated with the PWA, which are applied in both measurements of branching fractions and the resonance parameters in the MD approach, include:

\begin{itemize}
\item {\emph{Uncertainty from the resonance parametrization form.}}  In the nominal fit, the intermediate states are parameterized by a relativistic BW function with a constant width. Since the $\eta(1405)$, $\eta(1475)$ and $f_{1}(1420)$ are close to the threshold of $K^{*}(892)K$, an alternative BW function with mass-dependent width is used to parameterize the resonances,
\begin{equation}
BW(s)=\dfrac{1}{M^{2}-s-iM\Gamma(s)},
\end{equation}
with
\begin{equation}
\Gamma(s)=\Gamma(M^{2})\bigg{(}\dfrac{M}{\sqrt{s}}\bigg{)}\bigg{(}\dfrac{\rho(s)}{\rho(M^2)}\bigg{)}^{2l+1}B^{2}_{l}(\rho(s)),
\end{equation}
where $s$ is the invariant mass squared of the intermediate states, $M$ and $\Gamma(M^2)$ are the corresponding nominal mass and width, $\rho$ is the momentum of the daughter particle in its mother resonance rest frame calculated by the invariant mass squared of the mother resonance, $B_{l}(\rho(s))$ is the Blatt-Weisskopf barrier factor~\cite{50} and $l$ is the orbital angular momentum. 

\item {\emph{Uncertainty from resonance parameters.}} In the nominal fit, the resonant parameters of $a_{0}(980)^{0}$, $K^{*}(892)^{0}$ and $a_{2}(1320)^{0}$ states are fixed. The related uncertainties are evaluated with alternative fits in which these resonance parameters are varied within one standard deviation of the PDG values~\cite{13} or published results~\cite{41}.

\item {\emph{Uncertainty from extra components.}} Uncertainties from possible extra components are estimated by adding the processes $f_{1}(1510)\to K^{*}(892)^{0}\ksz$, $\eta(1295)\to a_{0}(980)^{0}\piz$, $f_{1}(1420)\to(\ksz\piz)_\textrm{P-phsp}\ksz$ and $f_{1}(1285)\to(\ksz\piz)_\textrm{P-phsp}\ksz$, 
which are  the four most significant extra components evaluated from data, into the nominal configuration individually.
\end{itemize}

For each alternative fit performed to estimate the systematic uncertainties from the PWA fit procedure, the changes of the measurements are taken as the one-sided systematic uncertainties. For each measurement, the individual uncertainties are assumed to be independent and are added in quadrature to obtain the total systematic uncertainty on the negative and positive sides, respectively. The sources of systematic uncertainties are applied to the measurements of masses and widths of intermediate states, and their contributions are summarized in in table~\ref{t5}. The relative systematic uncertainties relevant to the branching fraction measurements are summarized in table~\ref{t6}.

\section{SUMMARY}
In summary, mass independent and dependent PWAs have been performed on the decay $\jpsi\to\gamma\ksz\ksz\piz$ with $M(\ksz\ksz\piz)<1.6\gevcc$ based on $10.09\times10^{9}$ $\jpsi$ decays collected with the BESIII detector. The analyses show good consistency with each other in the intensity of individual components, and exhibit that the pseudoscalar and axial vector components are the dominant contributions. The mass dependent PWA shows that the pseudoscalar component can be well described with the two isoscalar states using relativistic Breit-Wigner model, {\it i.e.}, $\eta(1405)$ with a mass of $1391.7\pm0.7_{-0.3}^{+11.3}\mevcc$ and a width of $60.8\pm1.2_{-12.0}^{+5.5}\mev$ and $\eta(1475)$ with a mass of $1507.6\pm1.6_{-32.2}^{+15.5}\mevcc$ and a width of $115.8\pm2.4_{-11.0}^{+14.8}\mev$, and both decay into $(\ksz\ksz)_\textrm{S-wave}\piz$ and $(\ksz\piz)_\textrm{P-wave}\ksz$. The axial vector component can be well described with $f_{1}(1285)\to a_{0}(980)^{0}\piz$, $f_{1}(1420)\to a_{0}(980)^{0}\piz$ and $f_{1}(1420)\to K^{*}(892)^{0}\ksz$. The tensor component $f_{2}(1525)\to K^{*}(892)^{0}\ksz$ is observed for the first time in this process.

For the pseudoscalar component, the relatively flat lineshape between 1.4$\gevcc$ and 1.5$\gevcc$ rules out one standard resonance parameterization. In addition, the current TSM models~\cite{tsm1,tsm2,tsm3,tsm4} can not describe data well. This analysis provides a precision measurement of the pseudoscalar component, and is essential to study the dynamics of pseudoscalar structure in this mass region. More studies on the TSM and other phenomenological mechanisms are necessary to improve our understanding of the isoscalar pseudoscalar spectrum.

\acknowledgments
The BESIII collaboration thanks the staff of BEPCII and the IHEP computing center for their strong support. This work is supported in part by National Key R\&D Program of China under Contracts Nos. 2020YFA0406300, 2020YFA0406400; National Natural Science Foundation of China (NSFC) under Contracts Nos. 11335008, 11625523, 11635010, 11705192, 11735014, 11835012, 11922511, 11935015, 11935016, 11935018, 11950410506, 11961141012, 12022510, 12025502, 12035009, 12035013, 12061131003, 12105276, 12122509, 12192260, 12192261, 12192262, 12192263, 12192264, 12192265, 12235017; the Chinese Academy of Sciences (CAS) Large-Scale Scientific Facility Program; Joint Large-Scale Scientific Facility Funds of the NSFC and CAS under Contract No. U1732263, U1832103, U1832207, U2032111; the CAS Center for Excellence in Particle Physics (CCEPP); 100 Talents Program of CAS; The Institute of Nuclear and Particle Physics (INPAC) and Shanghai Key Laboratory for Particle Physics and Cosmology; ERC under Contract No. 758462; European Union's Horizon 2020 research and innovation programme under Marie Sklodowska-Curie grant agreement under Contract No. 894790; German Research Foundation DFG under Contracts Nos. 443159800, Collaborative Research Center CRC 1044, GRK 2149; Istituto Nazionale di Fisica Nucleare, Italy; Ministry of Development of Turkey under Contract No. DPT2006K-120470; National Science and Technology fund; National Science Research and Innovation Fund (NSRF) via the Program Management Unit for Human Resources \& Institutional Development, Research and Innovation under Contract No. B16F640076; STFC (United Kingdom); Suranaree University of Technology (SUT), Thailand Science Research and Innovation (TSRI), and National Science Research and Innovation Fund (NSRF) under Contract No. 160355; The Royal Society, UK under Contracts Nos. DH140054, DH160214; The Swedish Research Council; U. S. Department of Energy under Contract No. DE-FG02-05ER41374

\appendix
\section{Formulas of covariant tensor amplitudes}
As a supplement, we provide the formulas of covariant tensor amplitudes applied in PWA nominal solution referring to ref.~\cite{39}, where $\ksz$,$\ksz$, $\piz$ are denoted as particle 1, 2, 3. The correspinding intermediate resonances and decay modes are listed in the table~\ref{t2}.

For the $\jpsi\to\gamma 0^{-+}$ vertex, there is one independent coupling:
\begin{equation}
	\langle\gamma 0^{-+}|(K^{*}K)1\rangle =S_{\mu\nu}B_{1}(Q_{\psi\gamma X})f^{(0^{-+})}_{(123)}\bigg[\tilde{t}^{(1)}_{(K^{*}K)\lambda}\tilde{t}^{(1)\lambda}_{(13)}f^{(K^{*})}_{(13)}+\tilde{t}^{(1)}_{(K^{*}K)\lambda}\tilde{t}^{(1)\lambda}_{(23)}f^{(K^{*})}_{(23)}\bigg]
\end{equation}		
\begin{equation}
	\langle\gamma 0^{-+}|(a_{0}\piz)1\rangle =S_{\mu\nu}B_{1}(Q_{\psi\gamma X})f^{(0^{-+})}_{(123)}f^{(a_{0})}_{(12)}
\end{equation}		
\begin{equation}
	\langle\gamma 0^{-+}|(a_{2}\piz)1\rangle =S_{\mu\nu}B_{1}(Q_{\psi\gamma X})f^{(0^{-+})}_{(123)}f^{(a_{2})}_{(12)}\tilde{t}^{(2)}_{(a_{2}\piz)\gamma\delta}\tilde{t}^{(2)\gamma\delta}_{(12)} 
\end{equation}

For the $\jpsi\to\gamma 1^{++}$ vertex, there are two independent couplings:
\begin{equation}
	\langle\gamma 1^{++}|(K^{*}K)1\rangle =\epsilon_{\mu\nu\alpha\beta}p^{\alpha}_{\psi}f^{(1^{++})}_{(123)}\bigg[\tilde{t}^{(1)\beta}_{(13)}f^{(K^{*})}_{(13)}+\tilde{t}^{(1)\beta}_{(23)}f^{(K^{*})}_{(23)}\bigg]
\end{equation}		
\begin{equation}
	\langle\gamma 1^{++}|(K^{*}K)2\rangle =q_{\mu}S_{\nu\beta}B_2(Q_{\psi\gamma X})f^{(1^{++})}_{(123)}\bigg[\tilde{t}^{(1)\beta}_{(13)}f^{(K^{*})}_{(13)}+\tilde{t}^{(1)\beta}_{(23)}f^{(K^{*})}_{(23)}\bigg]
\end{equation}		
\begin{equation}
	\langle\gamma 1^{++}|(a_{0}\piz)1\rangle = \epsilon_{\mu\nu\alpha\beta}p^{\alpha}_{\psi}f^{(1^{++})}_{(123)}\tilde{t}^{(1)\beta}_{(a_{0}\piz)}f^{(a_{0})}_{(12)}
\end{equation}		
\begin{equation}
	\langle\gamma 1^{++}|(a_{0}\piz)2\rangle =q_{\mu}S_{\nu\beta}\tilde{t}^{(1)\beta}_{(a_{0}\piz)}B_2(Q_{\psi\gamma X})f^{(1^{++})}_{(123)}f^{(a_{0})}_{(12)} \\
\end{equation}		

For the $\jpsi\to\gamma 2^{++}$ vertex, there are three independent couplings:
\begin{equation}
	\langle\gamma 2^{++}|(K^{*}K)1\rangle =P^{(2)}_{\mu\nu\alpha\beta}(K)\epsilon^{\alpha\lambda\gamma\sigma}K_{\lambda}f^{(2^{++})}_{(123)}\bigg[\tilde{t}^{(2)\beta}_{(K^{*}K)\gamma}\tilde{t}^{(1)}_{(13)\sigma}f^{(K^{*})}_{(13)}+\tilde{t}^{(2)\beta}_{(K^{*}K)\gamma}\tilde{t}^{(1)}_{(23)\sigma}f^{(K^{*})}_{(23)}\bigg]
\end{equation}		
\begin{equation}
	\langle\gamma 2^{++}|(K^{*}K)2\rangle =g_{\mu\nu}p^{\delta}_{\psi}p^{\delta^{'}}_{\psi}P^{(2)}_{\delta\delta^{'}\alpha\beta}(K)\epsilon^{\alpha\lambda\gamma\sigma}K_{\lambda}f^{(2^{++})}_{(123)}\bigg[\tilde{t}^{(2)\beta}_{(K^{*}K)\gamma}\tilde{t}^{(1)}_{(13)\sigma}f^{(K^{*})}_{(13)}+\tilde{t}^{(2)\beta}_{(K^{*}K)\gamma}\tilde{t}^{(1)}_{(23)\sigma}f^{(K^{*})}_{(23)}\bigg]B_{2}(Q_{\psi\gamma X})
\end{equation}		
\begin{equation}
	\langle\gamma 2^{++}|(K^{*}K)3\rangle =q_{\mu}p^{\delta}_{\psi}P^{(2)}_{\nu\delta\alpha\beta}(K)\epsilon^{\alpha\lambda\gamma\sigma}K_{\lambda}f^{(2^{++})}_{(123)}\bigg[\tilde{t}^{(2)\beta}_{(K^{*}K)\gamma}\tilde{t}^{(1)}_{(13)\sigma}f^{(K^{*})}_{(13)}+\tilde{t}^{(2)\beta}_{(K^{*}K)\gamma}\tilde{t}^{(1)}_{(23)\sigma}f^{(K^{*})}_{(23)}\bigg]B_{2}(Q_{\psi\gamma X})
\end{equation}		

For the $\jpsi\to\gamma 2^{-+}$ vertex, there are three independent couplings:
\begin{equation}
	\langle\gamma 2^{-+}|(a_{0}\piz)1\rangle =\epsilon_{\mu\nu\alpha\beta}p^{\alpha}_{\psi}f^{(2^{-+})}_{(123)}\tilde{t}^{(2)\beta\gamma}_{(a_{0}\piz)}q_{\gamma}f^{(a_{0})}_{(12)}B_1(Q_{\psi\gamma X})
\end{equation}		
\begin{equation}
	\langle\gamma 2^{-+}|(a_{0}\piz)2\rangle =S_{\mu\nu}p_{\psi\gamma}p_{\psi\delta}f^{(2^{-+})}_{(123)}\tilde{t}^{(2)\gamma\delta}_{(a_{0}\piz)}f^{(a_{0})}_{(12)}B_3(Q_{\psi\gamma X})
\end{equation}		
\begin{equation}
	\langle\gamma 2^{-+}|(a_{0}\piz)3\rangle =q_{\mu}S_{\nu\gamma}f^{(2^{-+})}_{(123)}\tilde{t}^{(2)\gamma\delta}_{(a_{0}\piz)}p_{\psi\delta}f^{(a_{0})}_{(12)}B_3(Q_{\psi\gamma X})
\end{equation}
here the partial-wave amplitudes are constructed with the pure-orbital-angular-momentum covariant tensors $\tilde{t}^{(l)}_{\mu_{1}\cdots\mu_{l}}$ and the covariant spin wave functions $\phi_{\mu_{1}\cdots\mu_{l}}$ together with the operators $g^{\mu\nu}$ and $\epsilon^{\mu\nu\alpha\beta}$. Besides,  $P^{(S)}_{\mu_{1}\cdots\mu_{S}\mu^{'}_{1}\cdots\mu^{'}_{S}}(p_{a})$ is the spin projection operator of meson a, $B_{l}$ is the Blatt-Weisskopf barrier factor, the propagators of intermediate resonances are denoted as $f$ and the $S_{\mu\nu}$ is defined as
\begin{equation}
	S_{\mu\nu}=\epsilon_{\mu\nu\alpha\beta}p^{\alpha}_{\psi}q^{\beta}
\end{equation}
where $p^{\alpha}_{\psi}$ and $q^{\beta}$ are the momenta of $\jpsi$ and the radiative photon. More detailed derivation of the formulas, definitions of physical quantities and specific expressions have been given by the ref.~\cite{39}.

\newpage

\noindent$\Large\textbf{The BESIII Collaboration}$ \\
\\
\begin{small}
		M.~Ablikim$^{1}$, M.~N.~Achasov$^{11,b}$, P.~Adlarson$^{70}$, M.~Albrecht$^{4}$, R.~Aliberti$^{31}$, A.~Amoroso$^{69A,69C}$, M.~R.~An$^{35}$, Q.~An$^{66,53}$, X.~H.~Bai$^{61}$, Y.~Bai$^{52}$
% ...
		\\
		\\
	{\it
			$^{1}$ Institute of High Energy Physics, Beijing 100049, People's Republic of China\\
% ...
$^{4}$ Bochum Ruhr-University, D-44780 Bochum, Germany\\
% ...
$^{11}$ G.I. Budker Institute of Nuclear Physics SB RAS (BINP), Novosibirsk 630090, Russia\\
% ...
$^{31}$ Johannes Gutenberg University of Mainz, Johann-Joachim-Becher-Weg 45, D-55099 Mainz, Germany\\
% ...
$^{35}$ Liaoning Normal University, Dalian 116029, People's Republic of China\\
% ...
$^{52}$ Southeast University, Nanjing 211100, People's Republic of China\\
			$^{53}$ State Key Laboratory of Particle Detection and Electronics, Beijing 100049, Hefei 230026, People's Republic of China\\
% ...
$^{61}$ University of Jinan, Jinan 250022, People's Republic of China\\
% ...
$^{66}$ University of Science and Technology of China, Hefei 230026, People's Republic of China\\
% ...
$^{69}$ University of Turin and INFN, (A)University of Turin, I-10125, Turin, Italy; (B)University of Eastern Piedmont, I-15121, Alessandria, Italy; (C)INFN, I-10125, Turin, Italy\\
			$^{70}$ Uppsala University, Box 516, SE-75120 Uppsala, Sweden\\
% ...
$^{b}$ Also at the Novosibirsk State University, Novosibirsk, 630090, Russia\\
% ...
}
\end{small}
\newpage

\end{document}